\documentclass[twoside,twocolumn,english]{elsarticle}
\usepackage[T1]{fontenc}
\usepackage[latin9]{inputenc}
\usepackage{amsbsy}
\usepackage{graphicx}
\usepackage{makecell}

\usepackage{subfig}

\makeatletter
\usepackage[english]{babel}
\usepackage{lineno}

\usepackage{amsmath}
\usepackage{lineno}
\usepackage{hyperref}
\makeatother

\usepackage{babel}
\begin{document}
\twocolumn[   \begin{@twocolumnfalse}

\begin{frontmatter}{}

\title{\textbf{Control of fixation duration during visual search task execution} }

\author{A.Y.Vasilyev}

\address{Queen Mary University of London, Mile End Road, E1 4NS}
\begin{abstract}
We study the ability of human observer to control fixation duration during execution of visual search tasks. We conducted the eye-tracking experiments with natural and synthetic images and found the dependency of fixation duration on difficulty of the task and the lengths of preceding and succeeding saccades. In order to explain it, we developed the novel control model of human eye-movements that incorporates continuous-time decision making, observation and update of belief state.  This model is based on Partially Observable Markov Decision Process with lag in observation and saccade execution that accounts for a delay between eye and cortex.  We validated the computational model through comparison of statistical properties of simulated and experimental eye-movement trajectories.  \end{abstract}

\end{frontmatter}{}

\end{@twocolumnfalse} ]

\section{Introduction}
Eye-movements are essential actions of the human oculomotor system that overcome the limitation of acuity outside of foveal region. The saccades - high velocity gaze shifts - are used to direct high acuity foveal vision to the most informative locations of a scenes. The saccades are followed by fixations, during which the fixational eye-movements are generated in order to maintain high resolution perceptual input.  The implementation of these processes is crucial to the natural behavioural objectives of humans and animals. However, relatively little is known about the general scheme of control of fixation duration and location. Humans exhibit high variability in fixation duration, which depends on many factors including the visual task and the health conditions of observers. For example, the patients with Age-related Macular Degeneration demonstrate significantly longer fixation duration than the normal controls \cite{van2013macular}. Meanwhile, the normal controls also demonstrate the increase of average fixation duration with the increase of difficulty of visual search task \cite{hooge1996control,gould1973eye,moffitt1980evaluation}.  \par Previously, it was suggested that control of fixation duration is involuntary, and there is a mechanism that estimates the required time needed for the discrimination of target \cite{hooge1996control,hooge1998adjustment}.  Later it was argued \cite{engbert2002dynamical} that fixation duration is influenced both by global control that defines prior distribution of fixation duration and also a local control that inhibits the next saccade if visual processing is not completed.  This principle of mixed control was incorporated into saccade timing models, which successfully explain the statistics of fixation duration during observation of natural scenes \cite{nuthmann2010crisp}, visual search \cite{trukenbrod2014icat} and reading \cite{engbert2005swift}. Despite the remarkable achievements of saccade timing models in reproducing of eye-movement behaviour, they leave unexplained why the neural implementation of this complex framework is beneficial for human observers in the first place. These models perfectly capture the dynamic of saccade programming, but they don't reflect the process of task execution itself, because the status of task execution doesn't influence the model's internal variables.  For example, in the ICAT model \cite{trukenbrod2014icat} the visual search is not terminated when the target item is fixated and processed by foveal vision. Because perception is not incorporated into saccade timing models, the target objects are processed in the same way as distractors, therefore these models can't judge if the observer has located the target. We argue that saccade timing models don't explain the execution of visual search task, but instead describe dynamics of visual attention in array of distractors. 

\par Meanwhile, the control models of human eye-movements \cite{najemnik2005optimal, butko2008pomdp,vasilyev2017optimal} give interpretation of visual behavior from optimality point of view and have capability to incorporate mixed control of fixation duration as a mere consequence of their architecture. The global control of fixation duration is related to overall task requirement \cite{trukenbrod2014icat} and influences the fixation duration throughout the execution of the episode. The prior knowledge of properties of stimulus image and target object(s) in control models define the policy of observer, which governs it's eye-movement behaviour and remains unchanged during the task execution.  Therefore, the control models incorporate the global control through learning and implementation of the policy. In the meantime, the local control, which is related to decision making based solely on processing of visual input during the fixation \cite{engbert2002dynamical}, is implemented in control models as integration of visual input into belief state and output of policy (decision) based on updated belief state. The only component, which is absent in control models, but required for control of fixation duration, is operation at frequency higher than frequency of saccadic events.  
\par In order to study the adaptivity of fixation duration we extended the computational model presented in our previous work \cite{vasilyev2017optimal} by introduction of continuous observation and decision making. In such a model the observer is given an ability to terminate the current fixation at any moment of time. Therefore, the optimal observer is faced with the problem of trade-off between the quality of extracted visual information and time costs of fixation.   We expect that an extended computational model subject to temporal constraints will provide quantitative explanation of increase of average fixation duration in our psycho-physical experiment with decrease of detectability of target object. Furthermore, the noisy observation model \cite{najemnik2009simple} reformulated in continuous time is identical to evidence accumulation models such as LATEST \cite{tatler2017latest}, and we expect that this computational model will provide similar distribution of fixation duration. 
\par The modeling of decision making in continuous time will allow us to take into account the influence of saccade latency. It's well known that saccade programming is executed in two stages: labile and non-labile stage \cite{becker1979analysis,engbert2005swift}. During the first stage the observer can cancel the execution of a saccade according to initial decision, and initiate a saccade towards a new direction. Continuous perception and decision making can describe this process and take into account the influence of duration of non-labile stage.  
\par  Reformulation of the computational model of eye-movements in continuous time, in the future, will allow us to add fixational eye-movements during fixation intervals, and study their influence on decision making of the observer. Previously, the control models of eye-movements \cite{najemnik2009simple,eckstein2015optimal} didn't take into account the presence of drift and microsaccades during fixation intervals.  The microsaccade occurrence shortly before saccade onset results in increase of saccade reaction time \cite{rolfs2006shortening} in response task, which indicates the influence of fixational eye-movements on saccade programming. Furthermore, the primary role of saccades in counteracting of perceptual fading \cite{martinez2004role} indicates their importance in the process of extraction of visual information during the fixation. 

\section{Extension of computational model}

We developed the novel computational model of human eye-movements that incorporates continuous-time decision making, observation and update of belief state. We motivate this extension by experimental evidence of longer fixation duration in lower visibility condition \cite{van2013macular}. 
\subsection{Delayed interaction}
According to the literature, saccade programming is assumed to be a two-stage process that consists of labile and non-labile stages \cite{becker1979analysis,engbert2005swift}. The labile stage is the first stage of the saccade programming, during which the initial saccade command can be cancelled in a favour of saccade to another location. The saccade to the next location is executed after the non-labile stage. The visual input is active during both labile and non-labile stages and suppressed during the execution of a saccade.  Therefore, the visual input received during non-labile stage of saccade programming can be used for decision-making only during the next fixation.

\par We model this effect as a time lag in interaction $\tau$ between the observer and environment, which has the duration equal to the length of non-labile stage. 
In our model the interaction between observer and environment is modelled as exchange of "messages" with certain frequency $\Omega=1/ \Theta_{int}$, where $\Theta_{int}$ is called resolution interval. The resolution interval introduces discretization required for numerical solver and has no biological meaning. At time step $n$ observer receives observation $S_{n}$ and vector of actual fixation location $\mathbf{x}_{n}$, which are used for inference of belief state $\mathbf{b}_{n}$. In our model the belief state is probability distribution over potential target locations, and observation is stochastic noise signal at each location. The belief state is used for decision making - estimation of the next intended location to fixate: $\overline{\mathbf{x}}_{n+1} =
\pi(\mathbf{b}_{n}).$ However this decision to fixate is not executed immediately, but with delay, which defined by resolution interval and time lag in interaction: $lag=\lceil \tau/ \Theta_{int} \rceil$, where $\lceil \rceil$ is ceiling function.  Therefore, the environment receives the delayed message of decision $\overline{\mathbf{x}}_{n-lag+1}$ from the observer, which is used for execution of saccade:  $\mathbf{x}=\alpha(\overline{\mathbf{x}})$. At this stage the world state is updated, and the gaze is transferred to the new location. The environment emits the observation $S_{n+lag}$, which will be available to the observer only after a number of steps defined by $lag$. If at the moment environment was in the stage of execution of the previous saccade command $\overline{\mathbf{x}}_{n-lag}$, the new saccade signal is neglected, and the environment emits null observation. The time lag in interaction $\tau$ is chosen as $50$ ms, which is an experimental value for the delay between eye and V1 \cite{foxe2002flow,reichle2003ez}. 
\begin{figure}
\centering
\includegraphics[scale=1]{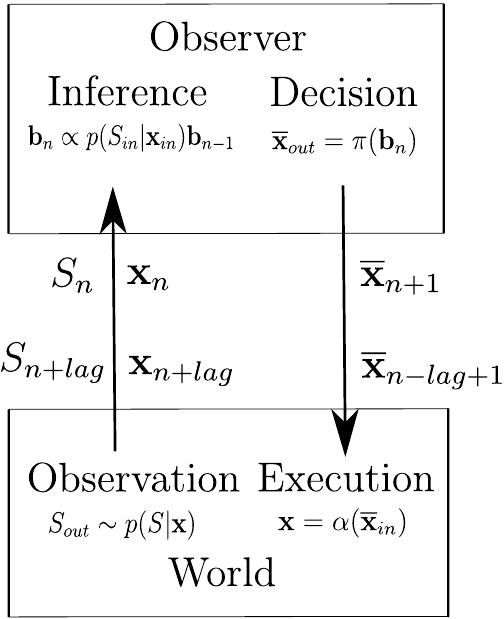}
\caption[Scheme of delayed interaction between observer and environment.]{At time step $n$ observer receives observation $S_{n}$ and vector of actual fixation location $\mathbf{x}_{n}$, which are used for inference of belief state $\mathbf{b}_{n}$. This belief state is used for decision making - estimation of the next intended location to fixate: $\overline{\mathbf{x}}_{n+1} =
\pi(\mathbf{b}_{n}).$  This decision is not executed immediately, by with delay defined as:  $lag=\lceil \tau/ \Theta_{int} \rceil$. On the same time step $n$,  environment receives the delayed message of decision $\overline{\mathbf{x}}_{n-lag+1}$ from the observer, which is used for execution of saccade:  $\mathbf{x}=\alpha(\overline{\mathbf{x}})$. At this stage the world state is updated, and the gaze is transferred to the new location. The environment emits the observation $S_{n+lag}$, which will be available to the observer only after a number of steps defined by $lag$. If at the moment environment was in the stage of execution of the previous saccade command $\overline{\mathbf{x}}_{n-lag}$, the new saccade signal is neglected, and the environment emits null observation.    }
\label{fig:delay}
\end{figure}

\subsection{Observation and inference}

\par We extend the previous computational model \cite{vasilyev2017optimal} by introduction of continuous time Gaussian white noise signal $ \dot{S}\bigl(t,\mathbf{x}_{t}, \, \mathbf{u}_\ell\bigr)$ at each location $\mathbf{u}_\ell$, each current fixation location $\mathbf{x}_{t}$ at time $t$ of world state, which satisfies following:
\begin{enumerate}
\item $E[\dot{S}\bigl(t,\mathbf{x}_{t}, \, \mathbf{u}_\ell\bigr)]=\delta\bigl(|\mathbf{u}_\ell - \mathbf{u}_\star|\bigr)/\Theta_{0}$
\item $E[\dot{S}\bigl(t_{1},\mathbf{x}_{t_{1}}, \, \mathbf{u}_{l_{1}}\bigr),\dot{S}\bigl(t_{2},\mathbf{x}_{t_{2}}, \, \mathbf{u}_{l_{2}}\bigr)]=\delta_{t_{1},t_{2}}\delta_{l_{1},l_{2}}/ ( \Theta_{0}  V^{2}(\mathbf{u}_\ell-\mathbf{x}_{t}))$
\end{enumerate}
where $\mathbf{u}_\star$ is location of target object, and $\Theta_{0}=250$ ms average fixation in double fixation task \cite{najemnik2005optimal}. Next, we assume that observer receives the observations as a sequence of integrated continuous time signals: $ S_{n}=S(t_{n},\mathbf{x}_{t_{n}}, \, \mathbf{u}_\ell)=\int_{t_{n}}^{t_{n}+\Delta t} \dot{S}\bigl(t,\mathbf{x}_{t}, \, \mathbf{u}_\ell\bigr) dt_{n}$ with finite duration $\Delta t$. These signals, according to conditions, will be Gaussian random variables with the following means and standard deviations:
\begin{equation}
E[S_{n}]=\Delta t \delta\bigl(|\mathbf{u}_\ell - \mathbf{u}_\star|\bigr)/\Theta_{0} \label{eq:mean}
\end{equation}
\begin{equation}
\sigma[S_{n}]=\frac{ \sqrt{\Delta t}}{\sqrt{\Theta_{0}} V(\mathbf{u}_\ell-\mathbf{x}_{n})} \label{eq:std}
\end{equation}
After receiving of observation signal $ S_{n}$ Bayesian inference is used to update the belief state:
\begin{equation}
b_{n,\ell} \propto
p\bigl(S_{n}\, |\, \mathbf{x}_{t_{n}}\bigr) \,
b_{n-1,\mspace{1mu}\ell}.\label{eq:inference2}
\end{equation}Our next assumption is that observer is effectively blind during execution of saccades, therefore, the signal from time intervals, which contain saccades should be more noisy or be neglected. For each time interval we estimate the total time $\Delta t_{s}$ the gaze is fixed. Then we use equations \ref{eq:mean} and \ref{eq:std} to generate the observation and equation \ref{eq:inference2} to estimate the belief state after receiving of observation from this time interval. If the saccade's duration is longer than the whole time step, the environment doesn't yield the observation. 
\subsection{Decision making and action execution}
 
\par
\begin{figure}
\centering
\includegraphics[scale=0.4]{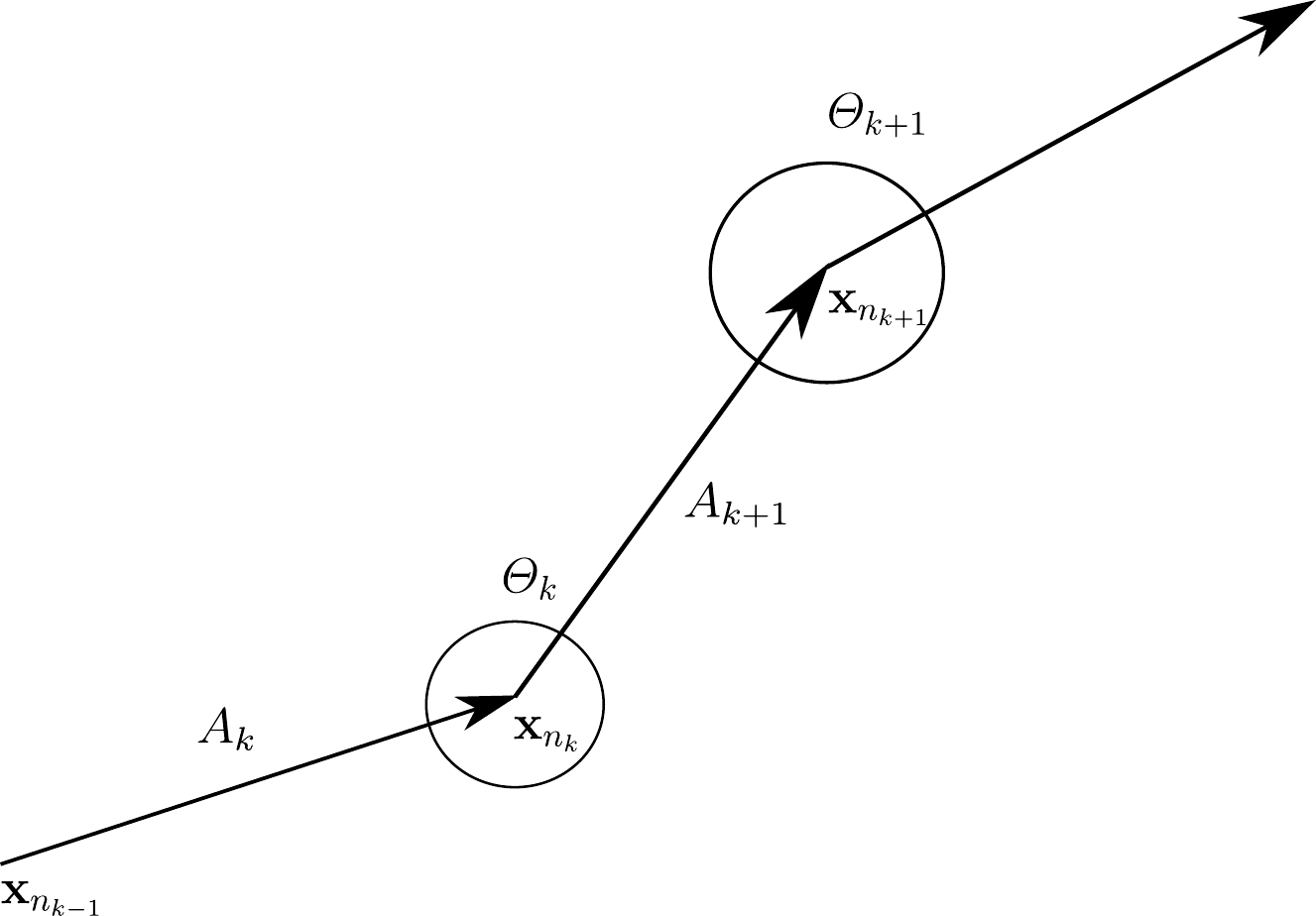} \caption{The sequence of saccade with endpoints $(\mathbf{x}_{n_{k-1}},\mathbf{x}_{n_{k}},\mathbf{x}_{n_{k+1}})$ and corresponding fixation durations $(\varTheta_{k}, \varTheta_{k+1})$. }
\label{fig:deagsac}
\end{figure}

  After the inference of belief state $\mathbf{b}_{n}$ observer makes decision where to fixate next according to the policy:
\begin{equation} 
\overline{\mathbf{x}}_{n+1} \leftarrow
\pi(\mathbf{b}_{n}). \label{eq:decisionmake}
\end{equation}
The command of execution of the next saccade is received by the environment with delay defined by $lag=\lceil \tau/ \Theta_{int} \rceil$. If at the moment of arrival of command the environment was not executing the previous saccade, the next saccade (or fixation) is executed as:
\begin{equation} \mathbf{x}_{n+1}=\begin{cases}
\overline{\mathbf{x}}_{n+1}, &\mathrm{ if }\ \overline{\mathbf{x}}_{n+1}=\mathbf{x}_{n} \\
\alpha(\overline{\mathbf{x}}_{n+1}), & \mathrm{otherwise}
\end{cases} \label{eq:decisionmake}
\end{equation}
here $\alpha(D)$ is the execution function: 
\begin{equation}
\mathbf{x}_{n+1}=\alpha(\overline{\mathbf{x}}_{n+1})=\overline{\mathbf{x}}_{n+1}+\zeta_{n}
\end{equation}
where $\zeta_{n}$ is Gaussian-distributed random error with zero mean
and standard deviation $\nu$ defined in \cite{engbert2005swift}:

\begin{equation}
\nu=\zeta_{0}+\zeta_{1}\left\Vert\overline{\mathbf{x}}_{n+1}-\mathbf{x}_{n}\right\Vert \label{eq:saccadevar-1}
\end{equation}
the error of saccade execution is proportional to intended saccade
amplitude $\left\Vert\overline{\mathbf{x}}_{n+1}-\mathbf{x}_{n}\right\Vert $ given in degrees,
the value of parameters: $\zeta_{0}=0.87\deg$, $\zeta_{1}=0.084$  (from \cite{engbert2005swift}).  The time $t$ of world state is updated in the following way:
 \begin{equation}
 t_{n+1}=t_{n}+\Theta_{int}
 \end{equation}
 The position of fixation is updated after the saccade command is executed. The time required to execute the saccade command $\mathbf{x}_{n}$ is defined as: 
\begin{equation} \varTheta_{sc}(n)=\varTheta_{mv}(n)+\varTheta_{st}(n) \label{eq:decisionmake}
\end{equation}
where $\varTheta_{mv}(n)$ is time required to execute the movement:
\begin{equation}
\varTheta_{mv}(n)=\tau_{mv}\left\Vert \mathbf{x}_{n}-\mathbf{x}_{n-1}\right\Vert ^{0.4}\label{eq:duration_sac}
\end{equation}
and $\varTheta_{st}(n)$ is time required to mechanically stop the eye-movement:
\begin{equation}
\varTheta_{st}(n)= \tau_{st} \left\Vert  \mathbf{x}_{n}-\mathbf{x}_{n-1}\right\Vert\label{eq:duration_sac}
\end{equation}The constants $\tau_{mv}=21 ms\cdot \deg^{-0.4}$,  $\tau_{st}=6 ms/\deg$   come from eye-tracking data from experiments of \cite{bartz1962eyemovement,salthouse1980determinants,rayner1978eye} and \cite{greene2006control,unema2005time} correspondingly. In the case if $\varTheta_{st}(n)+\varTheta_{mv}(n)>m\Theta_{int}$, where $m$ is an integer, the next $m$ saccade commands will be cancelled. Furthermore, the eye-movement is triggered only if decision where to fixate next doesn't match with current location:  $ \overline{\mathbf{x}}_{n+1}\neq\mathbf{x}_{n}$, according to equation \ref{eq:decisionmake}. This means that saccades don't start at every time step, and the $k$-th saccade will start at time step $n_{k}\geq k$. Therefore the length of $k$-th saccade equals: 
\begin{equation}
A_{k}=\left\Vert  \mathbf{x}_{n_{k}}-\mathbf{x}_{n_{k-1}}\right\Vert
\end{equation}
And the corresponding duration of $k$-th duration equals:
\begin{equation}
\varTheta_{k}=t_{n_{k+1}}-t_{n_{k}}-\varTheta_{sc}(n_{k})
\end{equation}which means that $k$-th fixation starts after $k$-th saccade is completely finished. 
The diagram \ref{fig:deagsac} explains  the notations in this section. 

\subsection{Learning an optimal policy}

Given the initial world state $S_{0}$ we define the value function for policy $\mu$ as an expectation of a return
$R$: 

\begin{equation}
V_{\mu}(S_{0})=E\left[R|\mu,S_{0}\right]\label{eq:value_function}
\end{equation}
The random variable $R$ denotes the return and is defined by:

\begin{equation}
R\equiv - \sum_{n=0}^{N}\Theta_{int}=N\Theta_{int}\label{eq:reward}
\end{equation}
where $N$ is total number of steps in episode. We start with representation of the stochastic policy $\mu$ \cite{butko2008pomdp}:

\begin{equation}
\mu(D,p)=\frac{\exp(f(D,p))}{\underset{l}{\sum}\exp(f(l,p))}\label{eq:soft_max}
\end{equation}
where $f(D,p)$ is the decision preference function, which indicates a tendency to choose 
the decision $D$ in the belief state $p$. In this study we limit
the search of $f(D,p)$ to convolution
of belief state $p$ due to shift-invariance of visual search process \cite{butko2008pomdp}:

\begin{equation}
f(D,p)=\underset{l}{\sum}K(D-l)p(l)\label{eq:gainfunction}
\end{equation}
our task is the search of kernel function $K$, which
corresponds to the policy that maximizes the value function $V_{\mu}$:

\begin{equation}
K^{*}=\underset{K}{\arg\max}\:V_{\mu(K)}(S_{0})\label{eq:problem}
\end{equation}
for any starting world state $\forall S_{0}$. The policy $\mu(K^{*})$
is called optimal policy of gaze allocation. We approach problem \ref{eq:problem} with algorithm of Policy Gradient Parameter Exploration \cite{sehnke2010parameter}.
The table \ref{table:parameters} shows the values of all parameters used in simulations of computational model. 
\begin{table*}
\centering 
\begin{tabular}{|c | c | c | c|} 
\hline 
\thead{Meaning} & \thead{Symbol} & \thead{Value} & \thead{Source} \\ [1ex] 
\hline 
RMS contrast of background noise & $e_{n}$ & $(0.1,0.15,0.2,0.25)$& Set  \\ 
RMS contrast of target & $e_{t}$ & 0.2 & Set\\
error of saccade execution: intercept & $\zeta_{0}$& 0.87 deg  &\cite{engbert2005swift}\\
error of saccade execution: slope  & $\zeta_{1}$ & 0.084  &\cite{engbert2005swift}\\
average fixation duration in double fixation task   & $\Theta_{0}$ & $250$ ms  & \cite{najemnik2005optimal}\\
saccade duration coefficient & $\tau_{sac}$ & 21 $ms\cdot \deg^{-0.4}$  & \cite{lebedev1996square}\\ 
time to stop the saccade: slope   & $\varTheta_{0,fix}$ &  $6 ms/\deg$  & \cite{greene2006control} \\
resolution interval   & $\Theta_{int}$ &  $(128, 64, 32, 16, 8)ms$  & Set\\
size of stimulus image  & S &  $15 \times 15 \deg^{2} $   & Set\\
dimensionality of computational grid & $N\times N$& $24 \times 24$ & Set\\
[1ex]
\hline 
\end{tabular}\\
\vspace*{1ex}
\caption{The parameters of computational model used in our simulations, including those that were set experimentally.}
\label{table:parameters}
\end{table*}
\section{Psychophysical experiments}
In this section we describe the eye-tracking experiments of visual search with natural and synthetic images.

\subsection{Participants}

The group of five participants with normal to corrected-to-normal vision
participated in the experiment. All participants were postgraduate students of Queen Mary University of London. The participants was aware of the experimental settings and passed 10 minutes of training sessions with four different experimental conditions, which correspond to the certain value of the RMS contrast of background noise. The experiments were approved
by the ethics committee of Queen Mary University of London and informed
consent was obtained.
\subsection{Equipment}
We used DELL P2210 22'' LCD monitor (resolution $1680\times1050$,
refresh rate 60 Hz) driven by a Dell Precision laptop for all experiments.
The eye movements of the right eye were registered using Eye Tracker device
SMI-500 with a sampling frequency of $120$ Hz. The Eye tracker device
was mounted on the monitor. Matlab Psychtoolbox \cite{psychtool} was used to run the experiments
and generate the stimulus images. The saccades and fixations were detected automatically by software provided by SMI with dispersion based algorithm I-DT \cite{salvucci2000identifying}. The dispersion-based algorithm associates the gaze samples with the same fixation if the samples are located within a region of size $0.5\deg$ for the minimal fixation duration of $80$ ms. 

\subsection{Stimulus}
\begin{figure*}
 \centering
      \subfloat[\small Stimulus images corresponding to the first band of d-prime.]{\includegraphics[scale=0.1]{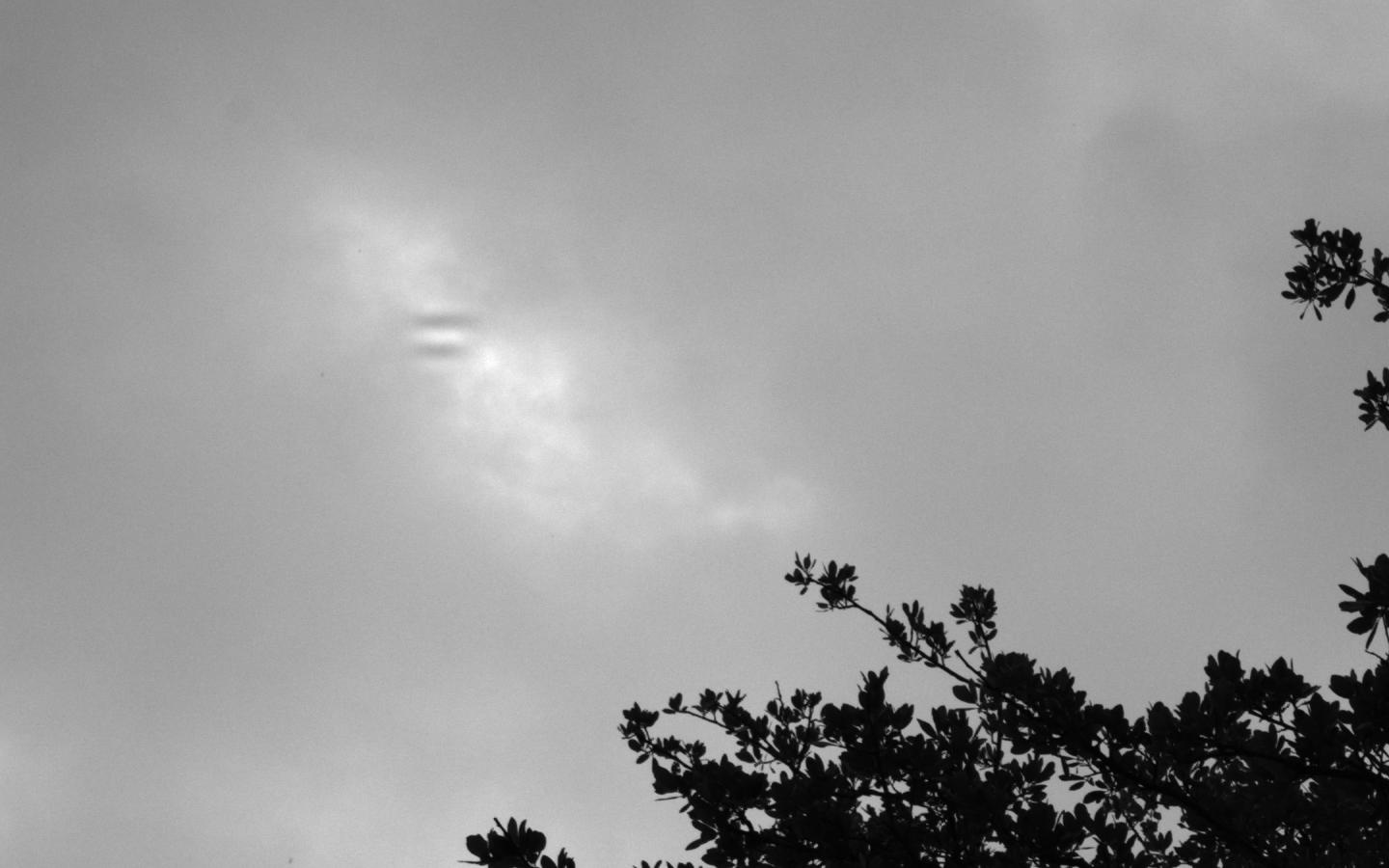} \includegraphics[scale=0.1]{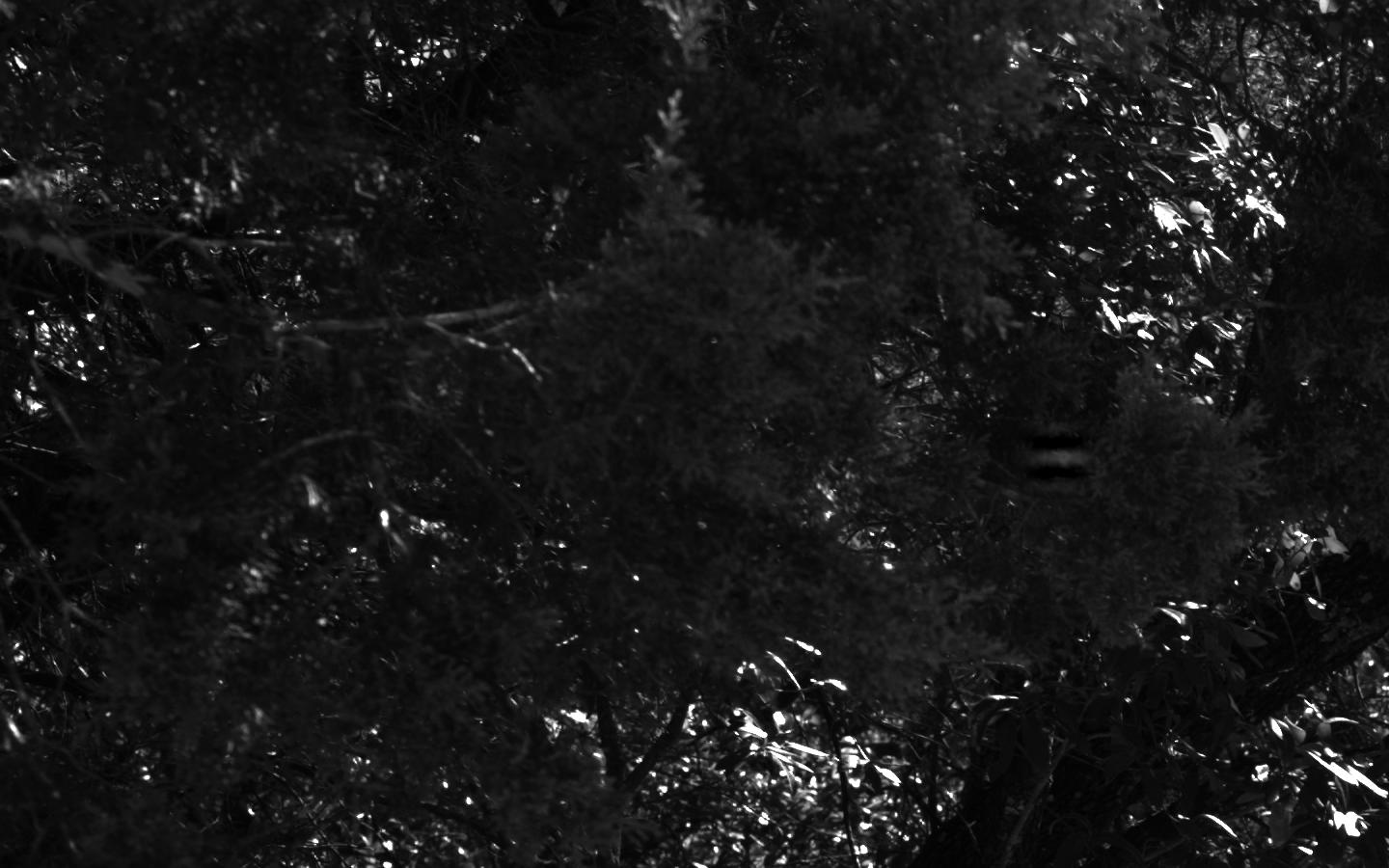}}

      \subfloat[\small Stimulus images corresponding to the fifth band of d-prime.]{ \includegraphics[scale=0.1]{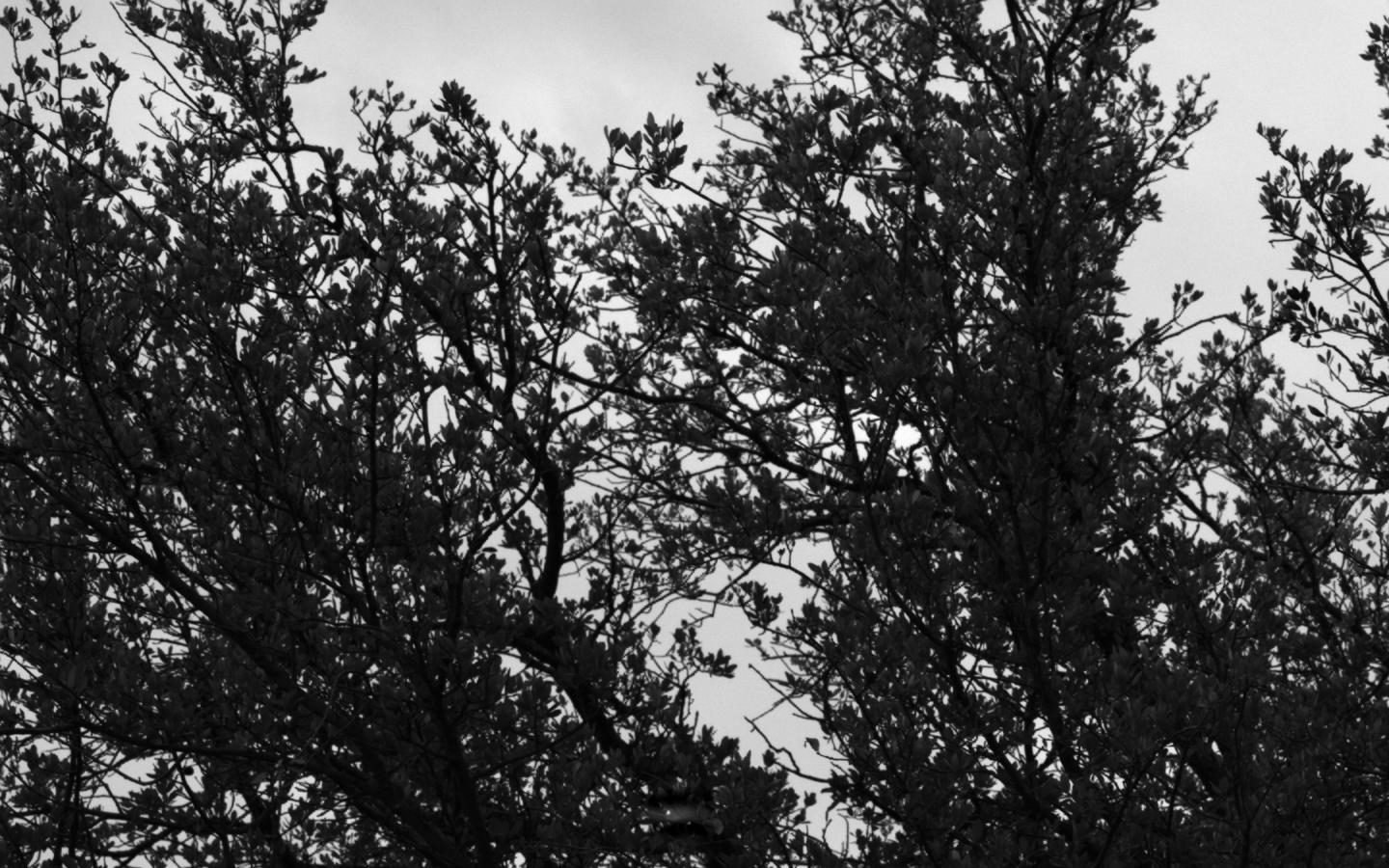} \includegraphics[scale=0.1]{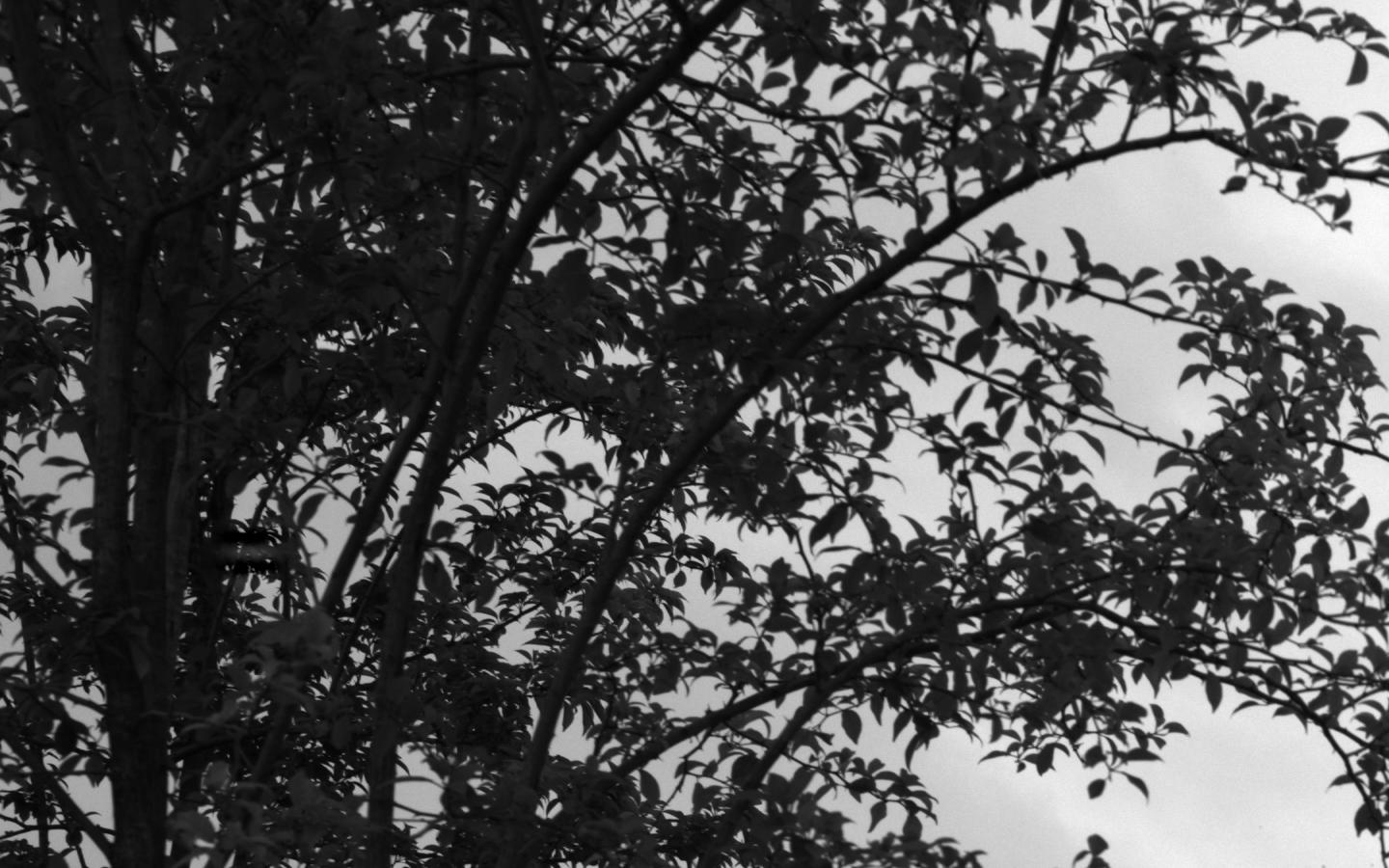}}

\caption[The natural images used in visual search experiments]{
\label{fig:highband}%
We used the database of natural images from \cite{geisler2011statistics} to create the stimulus images for visual search experiment. The target object was 4 cycles per degree windowed sine-wave grating. We estimated a mean of d-primes of  $10^{3}$ randomly placed 4 cpd sine-wave gratings for each image and sorted all 415 images according to the mean value.   On the next step the sorted array of images was divided on 5 equal ($20\%$) bands of d-prime. The target object on stimulus images from the first band (top) is the least difficult to detect. Whereas the fifth band consists of the most difficult stimulus images in experiment. }
\end{figure*}

We used the database of 1204 natural images from \cite{geisler2011statistics} and the procedure \cite{sebastian2017constrained} of estimation of d-prime - value of visibility map in the center of visual field - for  4 cycles per degree windowed sine-wave grating with a width of $0.8\deg$, which was a target object in our visual search experiments. We estimated a mean of d-primes of  $10^{3}$ randomly placed 4 cpd sine-wave gratings for each image and sorted all 415 images according to the mean value. On the next step the sorted array of images was divided on 5 equal ($20\%$) bands of d-prime. Therefore, we sorted our images by the level of detectability of target object. Next, we computed average d-prime for each band. For each image of the band we selected the location of target, whose estimate of d-prime is the closest to average one in the band, and generated the image with this target location. So, we formed 5 bands of difficulty with images that have close values of d-prime. The values of average d-prime for 5 bands are: 
{8.53, 3.98, 2.89, 2.24, 1.57}. 
The Figures \ref{fig:highband} demonstrates stimulus images for the first and the fifth bands of difficulty.
The second type of stimuli were  1/f noise images described in the original experiment \cite{najemnik2005optimal}.  The 1/f noise was generated on a square
region on the screen, which spans the visual angle of
$15\times15 \deg$. The target was 
sine grating $6$ $\deg^{-1}$
framed by a symmetric raised cosine. The target appeared randomly at
any possible location on the stimuli image within the square region. The
experiments were provided for one level of RMS contrast of target
$e_{t}=0.2$ and several levels of 1/f noise RMS contrast $e_{n}\in(0.1,0.15,0.2,0.25)$.
\subsection{Procedure}
\par Five participants with normal or corrected to normal vision performed visual search task on generated images. We used the SMI eye-tracker to record the eye-movements of participants. The visual search experiment consisted of five blocks corresponding to five bands with 2 minutes of rest between blocks. Each trial of search started with displaying of the fixation cross in the center of the screen for 1 second. 
The participants were instructed to find the target object on the image as soon as possible, after the stimulus image was displayed on the screen. The participants were asked to press "Next" button once they stopped their eyes on the target location. If the participants stopped their eyes further than 2 degrees from the target location, we considered this trial as unsuccessful and excluded it from further analysis. The number of unsuccessful trials grows with the difficulty of the band:  {3.1\%, 5.3\%, 8.3\%, 11.2\%, 17.1 \%} for corresponding five bands of d-prime. 
\par Psychophysical experiments of visual search \cite{vasilyev2017optimal} were repeated for constant target contrast $\epsilon_{t}=0.2$ and noise RMS contrast $\epsilon_{n}={0.1, 0.15, 0.2, 0.25}$. 5 normal controls were performing 240 trials of visual search task.

\section{Statistics of fixation duration}
In this section we discuss the statistics of fixation duration in human eye-movements and trajectories simulated with learned policy. Both learning of policy and simulations were conducted with parameters presented in the table \ref{table:parameters} and resolution interval $8$ ms.

\subsection{Dependence on difficulty }

\par
\begin{figure*}
\centering
\includegraphics[scale=0.4]{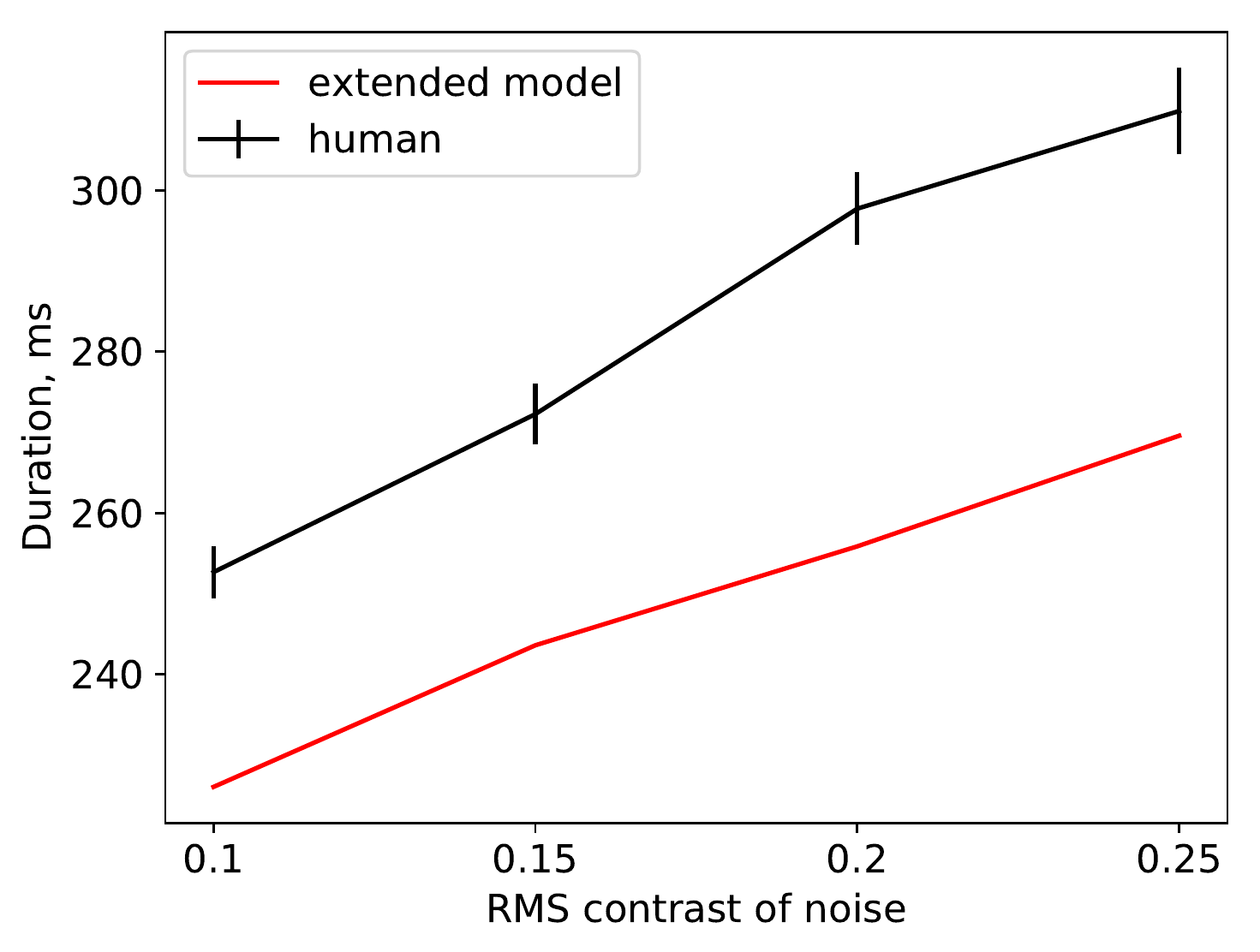}
\includegraphics[scale=0.4]{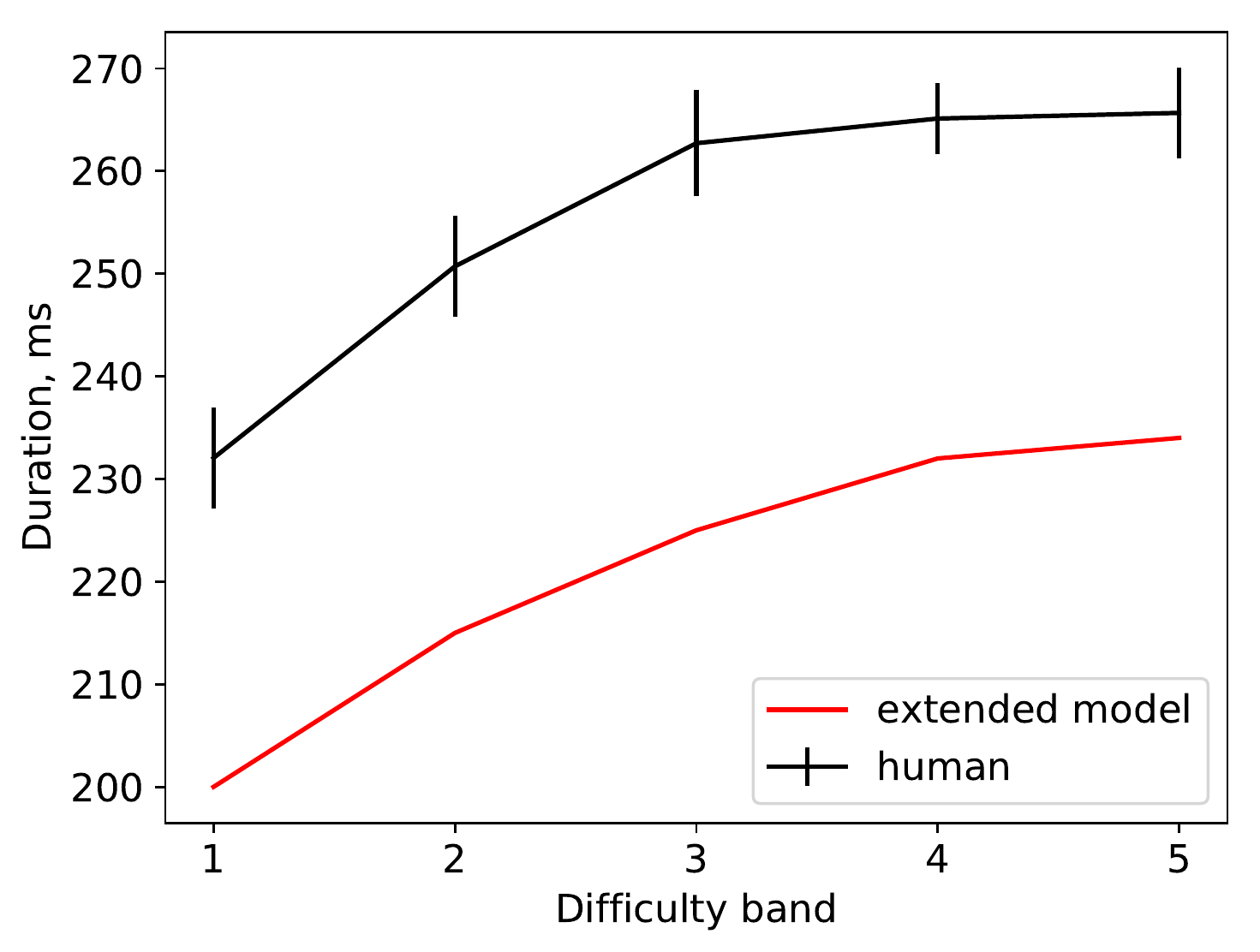}
\caption[The fixation duration in simulated eye-movement trajectories and experimental data.]{The fixation duration in simulated eye-movement trajectories and experimental data. The computational model significantly underestimates the values of fixation duration in humans for both cases of synthetic images (left) and natural images (right). } \label{fig:durdifnew}
\end{figure*}

 The figure \ref{fig:durdifnew} shows the change of fixation duration with difficulty for human observer and simulated agent in the case of synthetic images (left) and natural images (right). In the case of experimental values, we discovered significant difference between means of fixation durations for different values of RMS contrast of background noise in the case of synthetic images.  Using Bonferroni method for multiple comparison we found  mean fixation duration for all difficulty levels differ across all levels  ($p< 0.05$) for psychophysical experiments for both types of stimulus.
\begin{figure*}
\centering
\includegraphics[scale=0.4]{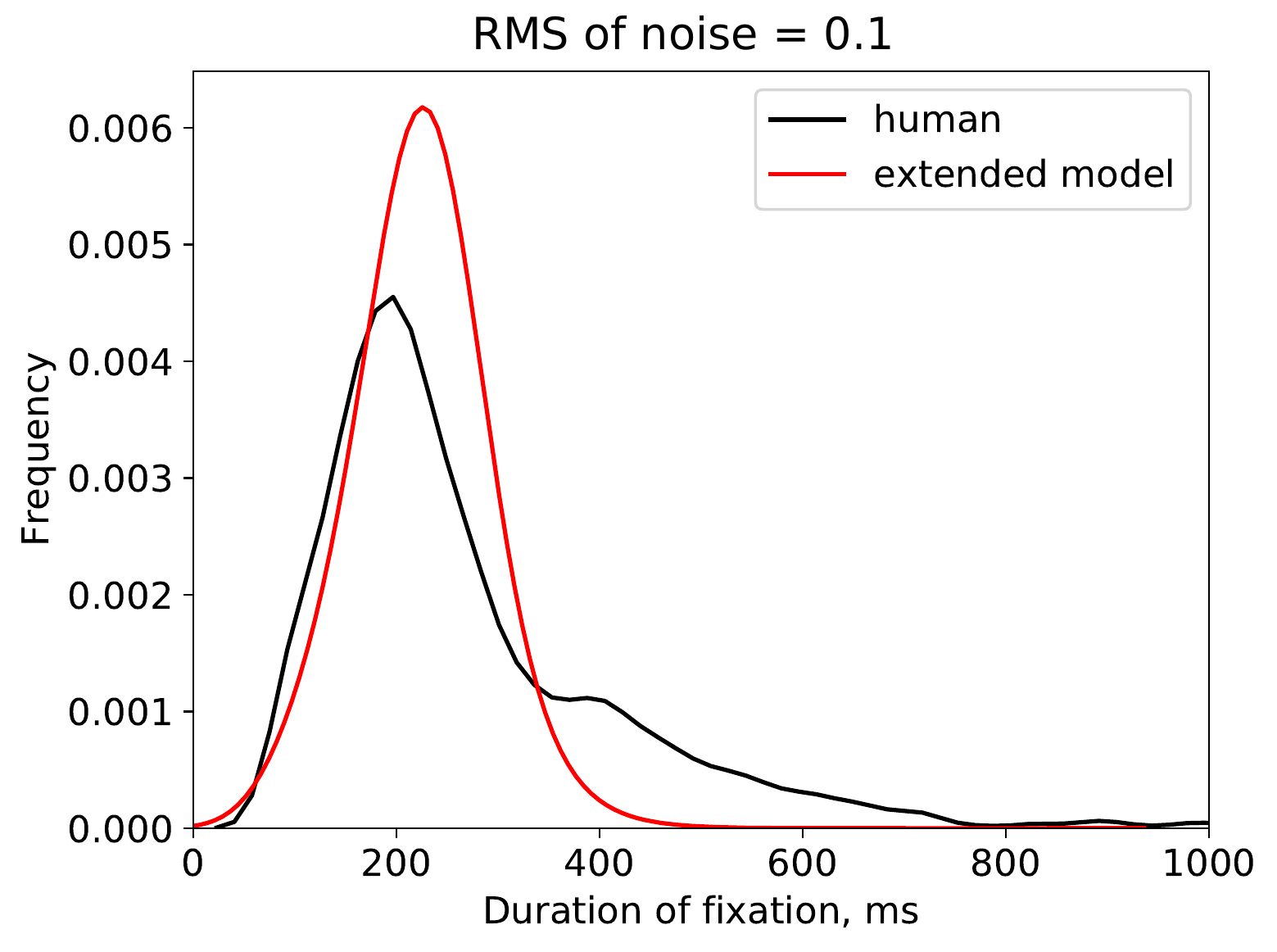}
\includegraphics[scale=0.4]{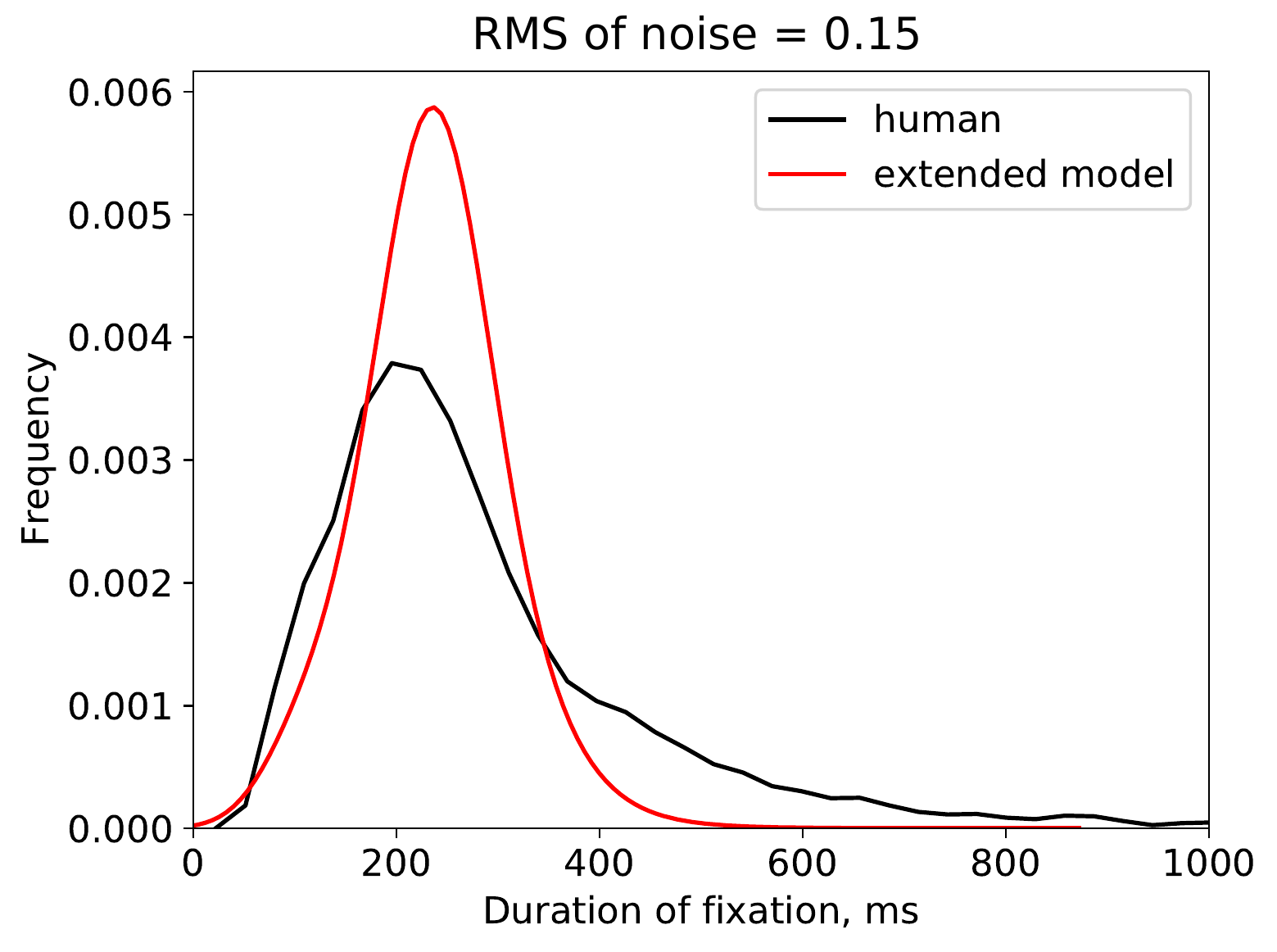}
\includegraphics[scale=0.4]{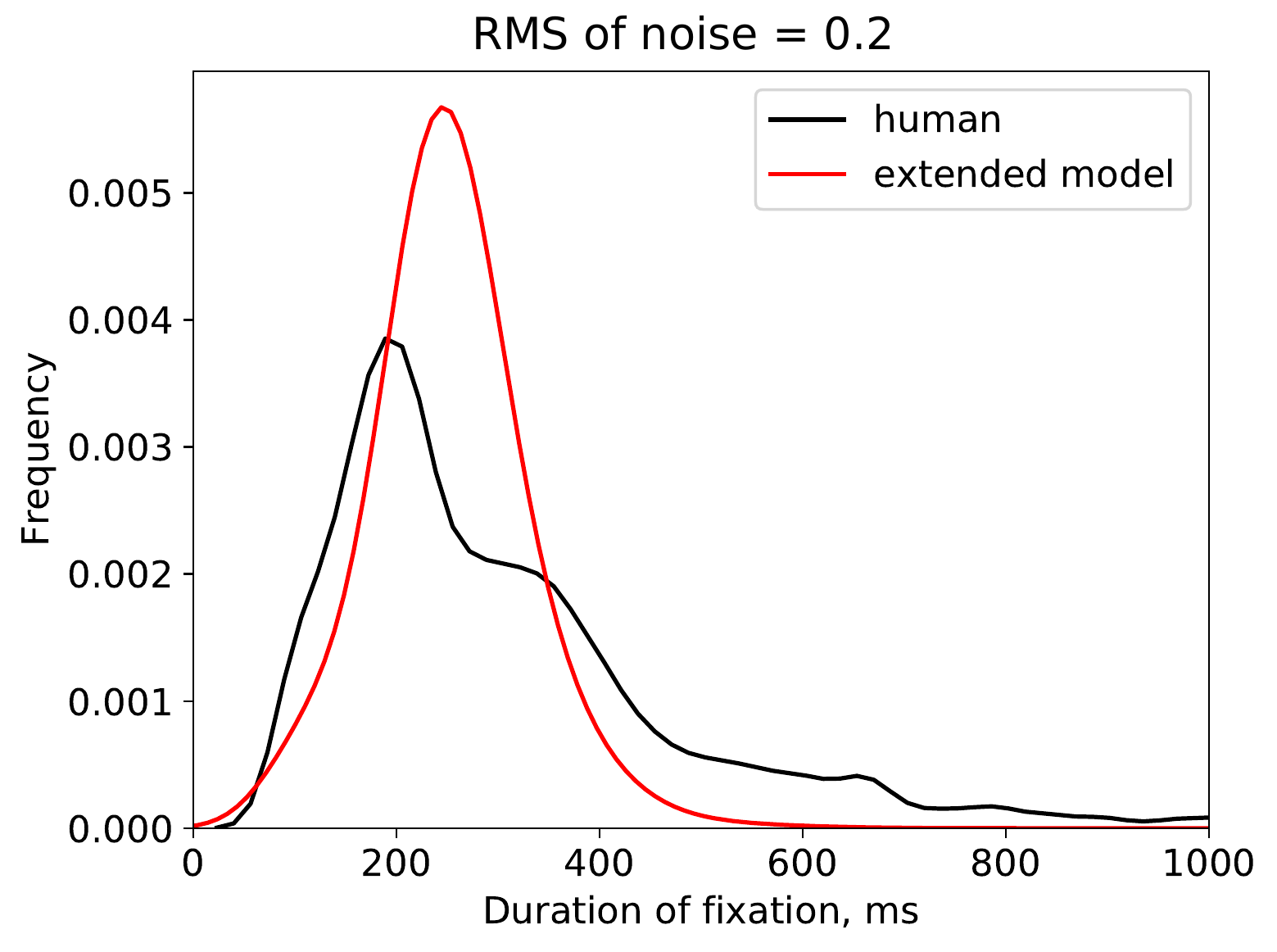}
\includegraphics[scale=0.4]{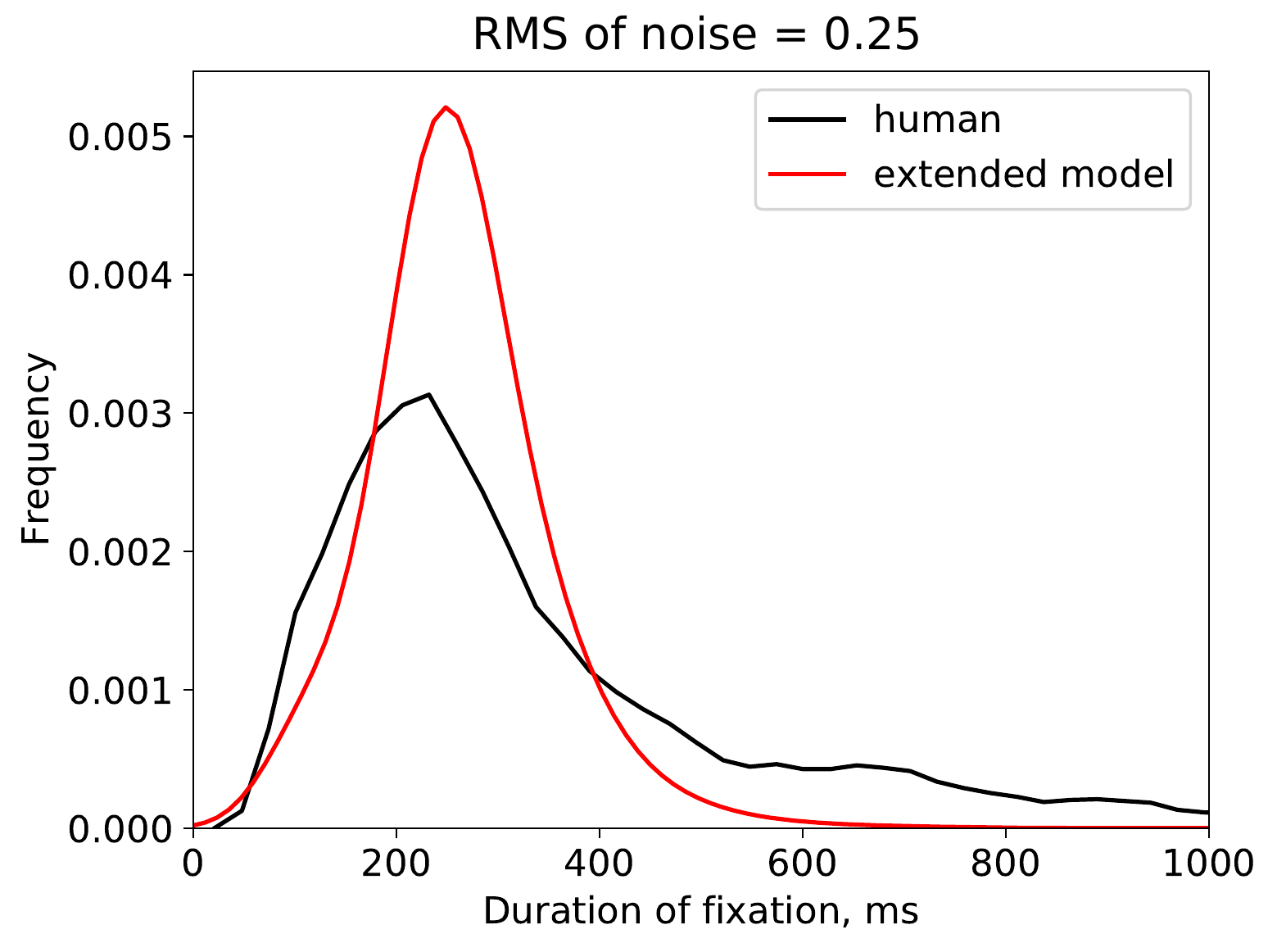}
\caption[The distributions of fixation duration in experiment on synthetic images and simulations of extended model. ]{The distributions of fixation duration in experiment on synthetic images and simulations of extended model. The computational model can't explain the heavy tail in the experimental distribution for any case of difficulty, and we assume that the reason of this mismatch is that our model doesn't take into account the fixational eye-movements (FEMs), which favour longer fixation duration.}
\label{fig:durdifcases}
\end{figure*}
\par We found that there is significant difference between the values of fixation duration predicted by the model and measured in experiments. The computational model can't explain the heavy tail in the experimental distribution for any case of difficulty  (see Figures \ref{fig:durdifcases} and \ref{fig:durdifnat}). We assume that the reason of this mismatch is that our model doesn't take into account the fixational eye-movements (FEMs): drift and microsaccades. The FEMs change the position of gaze during the fixation, which allows to extract additional visual information without initiation of saccade. This favors longer fixations and lower saccade frequency. However, the model correctly predicts the change of fixation duration with difficulty. Assuming that the model can't explain 35 ms of difference between simulated and experimental fixation duration, if the difference is subtracted from experimental values, they will not be significantly different from simulated values. Another reason of mismatch between model simulations and human behaviour is assumed preference of saccade initiation towards certain mean rate \cite{nuthmann2010crisp,engbert2002dynamical,engbert2005swift}, which originates from rhythmic activity in central neural system \cite{mcauley1999human}.  Because our approach is phenomenological, it uses Bayesian framework \cite{geisler2003bayesian} to model the processing within central neural system neglecting its temporal aspects.  More detailed theoretical model of neural system is required to account for preference in saccade initiation rate.

\subsection{Interaction with saccade length}
\par Another piece evidence of control of fixation duration during visual search process is correlation between the time series of lengths of both preceding and succeeding saccades and fixation duration. We estimated the Pearson correlation coefficients and p-values for both cases of psycho-physical experiments with natural and synthetic images and presented them in corresponding tables \ref{table:spearman_synt} and \ref{table:spaerman_nat}. The time series of duration of fixation (succeeding) preceding the saccades are (positively) negatively correlated to time series of lengths of (incoming) outgoing saccades. Meanwhile, in the case of synthetic images the p-value exceeds the significance level $0.05$ only for the RMS contrast value of $0.1$, and in the case of natural images p-value exceeds the significance level  for the first difficulty band. In the lower difficulty cases human observers execute the visual search task faster \cite{najemnik2009simple}, which results in smaller number of eye-movements.  Lower number of sample points didn't allow us to confirm the hypothesis on significance level 0.05.
\begin{table}
\centering
\begin{tabular}{|c | c |c | c | c | } 
\hline 
\multicolumn{5}{|c|}{Visual search on synthetic images} \\ \hline
\multicolumn{1}{|c|}{Case} & \multicolumn{2}{|c|}{Preceding fixation } & \multicolumn{2}{|c|}{Preceding saccade} \\
\hline
\thead {noise \\ contrast} &\thead{R}& \thead{p-value} & \thead{R} & \thead{p-value} \\
\hline 
0.1 & -0.057 & 0.08 & 0.111&0.07 \\ \hline
0.15  & -0.032 &0.03 & 0.073 &0.04 \\ \hline
0.2 & -0.067 &0.05 & 0.089  & 0.04 \\ \hline
0.25 & -0.045& 0.04 & 0.09  &0.03 \\
[1ex]
\hline 
\end{tabular}
\caption{Pearson correlation and p-values for the case of visual search task on synthetic images. The time series of duration of fixation (succeeding) preceding the saccades are (positively) negatively correlated to time series of lengths of (incoming) outgoing saccades. }
\label{table:spearman_synt}
\end{table}
\begin{table}
\centering
\begin{tabular}{|c | c |c | c | c | } 
\hline 
\multicolumn{5}{|c|}{Visual search on natural images} \\ \hline
\multicolumn{1}{|c|}{Case} & \multicolumn{2}{|c|}{Preceding fixation } & \multicolumn{2}{|c|}{Preceding saccade} \\
\hline
\thead {difficulty \\ band} &\thead{R}& \thead{p-value} & \thead{R} & \thead{p-value} \\
\hline 
1 & -0.045 & 0.09 & 0.08&0.08 \\ \hline
2  & -0.036 &0.05 & 0.05 &0.05 \\ \hline
3 & -0.023 &0.05 & 0.06  & 0.04 \\ \hline
4 & -0.045& 0.04 & 0.089  &0.03 \\ \hline
5 & -0.03 & 0.05 & 0.045 & 0.034 \\
[1ex]
\hline 

\end{tabular}
\caption{Pearson correlation and p-values for the case of visual search task on natural images. The time series of duration of fixation (succeeding) preceding the saccades are (positively) negatively correlated to time series of lengths of (incoming) outgoing saccades.  }
\label{table:spaerman_nat}
\end{table}
\par Our next goal is to compare the dependency of fixation duration on the lengths preceding and succeding of saccades between simulations and experiment.  Before the regression analysis, we find an optimal transformation of data with multivariate Box Cox transformation \cite{box1964analysis,cook2009applied}. The Box Cox transformation of a variable $x$ is defined as:
\begin{equation} x^{(\lambda)}=\begin{cases}
\frac{x^{\lambda}-1}{\lambda}\ , &\mathrm{ if } \ \lambda \neq 0 \\
\log(x), &\mathrm{ if } \ \lambda =0
\end{cases} 
\end{equation}The aim of multivariate Box Cox transformation is to find such parameter $\lambda$ that the transformed observations of $x^{\lambda}$ satisfy full normality assumption, e.g. the independent and normally distributed. We apply Box Cox transformation to the experimental observations $(\varTheta^{\lambda_{1}},A^{\lambda_{2}})$ to find such parameters $\lambda_{1}$ and $\lambda_{2}$ that observations satisfy the multivariate assumption:  $(\varTheta^{\lambda_{1}},A^{\lambda_{2}}) \sim\mathcal{N}(\mathbf{\mu},\,\Sigma)$, where $\mu$ and $\Sigma$ are mean and covariance matrix correspondingly. We present the result of the analysis in the tables \ref{table:Box Cox} and \ref{table:BoxCoxNat}. We apply the log-log transformation of data for both variables due to proximity of inferred parameters to zero. 

Next, we make an assumption of power-law relationship between saccade length and fixation:  $\varTheta=pA^{q}$ due to previously reported non-linear relationship between main saccadic characteristics \cite{lebedev1996square}. 
In the log-transformed data coefficients $p$ and $q$ correspond to intercept and slope of the linear relationship: $\log  \varTheta =q\log A+\log p$. We perform the log-transformation due to convenience of s
\par We log-transformed our data and formed the vectors $\log (\varTheta_{i}), \log(\Delta A_{i})$ of  fixation durations and outgoing saccade lengths  and $\log(\varTheta_{i+1}), \log(\Delta A_{i})$ of fixation durations and incoming saccade lengths.  We computed coefficients of linear robust fit using RANSAC method \cite{derpanis2010overview} for log-transformed vectors together with principal components. RANSAC method was used due to its robustness to outliers. The least square distance was chosen a measure of error. The maximal residual for a data point to be classified as an inlier was chosen as the median absolute deviation of the target values.  On each iteration of RANSAC the linear model was fitted to data points identified as inliers. The iterations of RANSAC stop if the number of iterations reaches the maximal one ($N$=1000). The figures \ref{fig:corrchart} and \ref{fig:corrchart1} show scatter-plots for vectors $\log(\varTheta_{i}),\log( \Delta A_{i})$ ( $\log(\varTheta_{i+1})$, $\log(\Delta A_{i})$) that corresponds to relation between preceding (succeeding) fixation duration and succeding (preceding) saccade length in the case of synthetic images. The figures \ref{fig:corrchartsym} and \ref{fig:corrchartsym1} demonstrate the same relation for saccade lengths and fixation duration in simulated data. We can see that in both cases of experimental and simulated data the regression line of robust fit is close to horizontal and coincides with the first principal component, which explains $65 \%$ of variability in experimental data and $57\%$ of variability on average in simulated data (see figures \ref{fig:corrchartnat} and \ref{fig:corrchartnat1} for the case of natural images). 

\begin{figure*}
\centering
\includegraphics[scale=0.4]{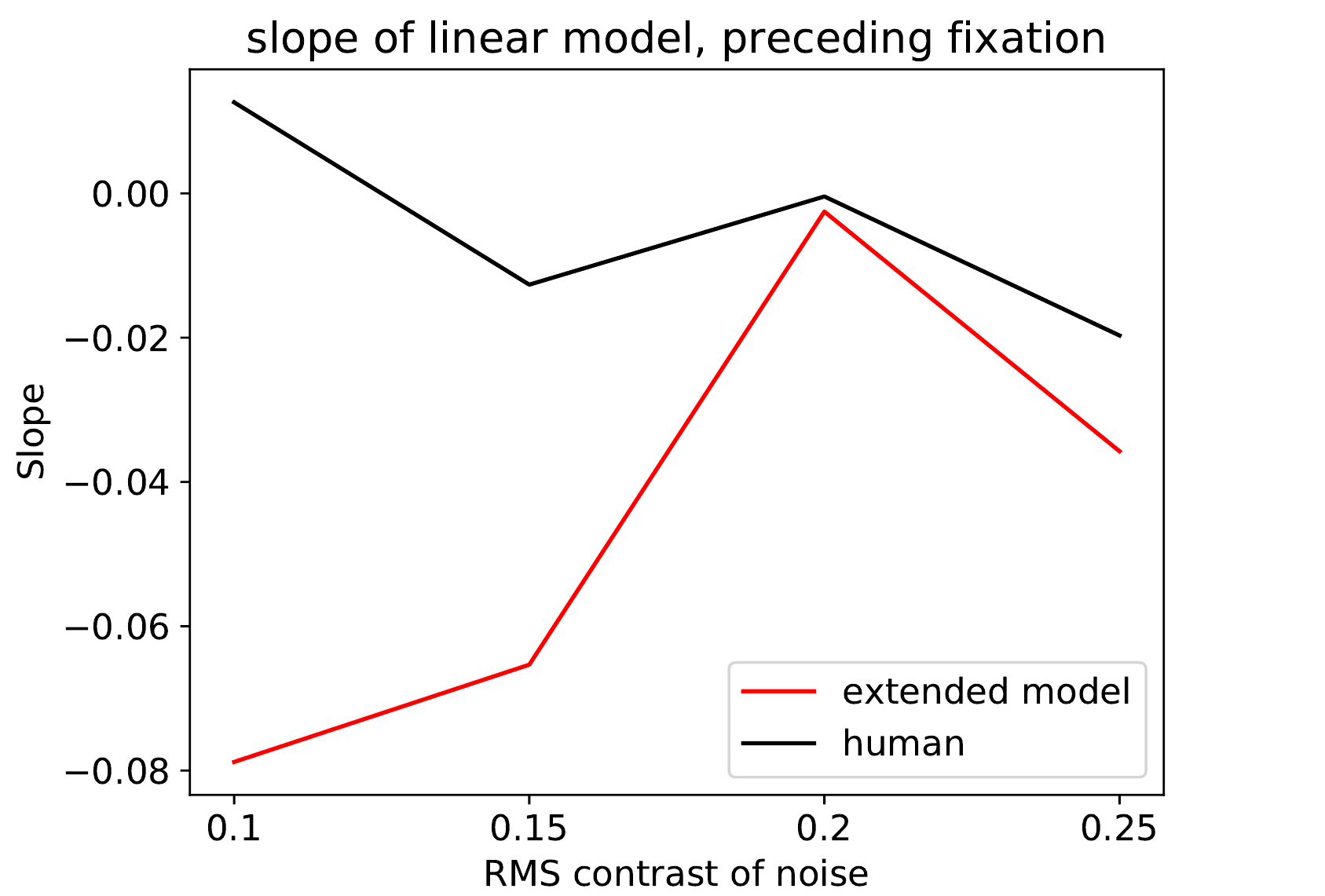}
\includegraphics[scale=0.4]{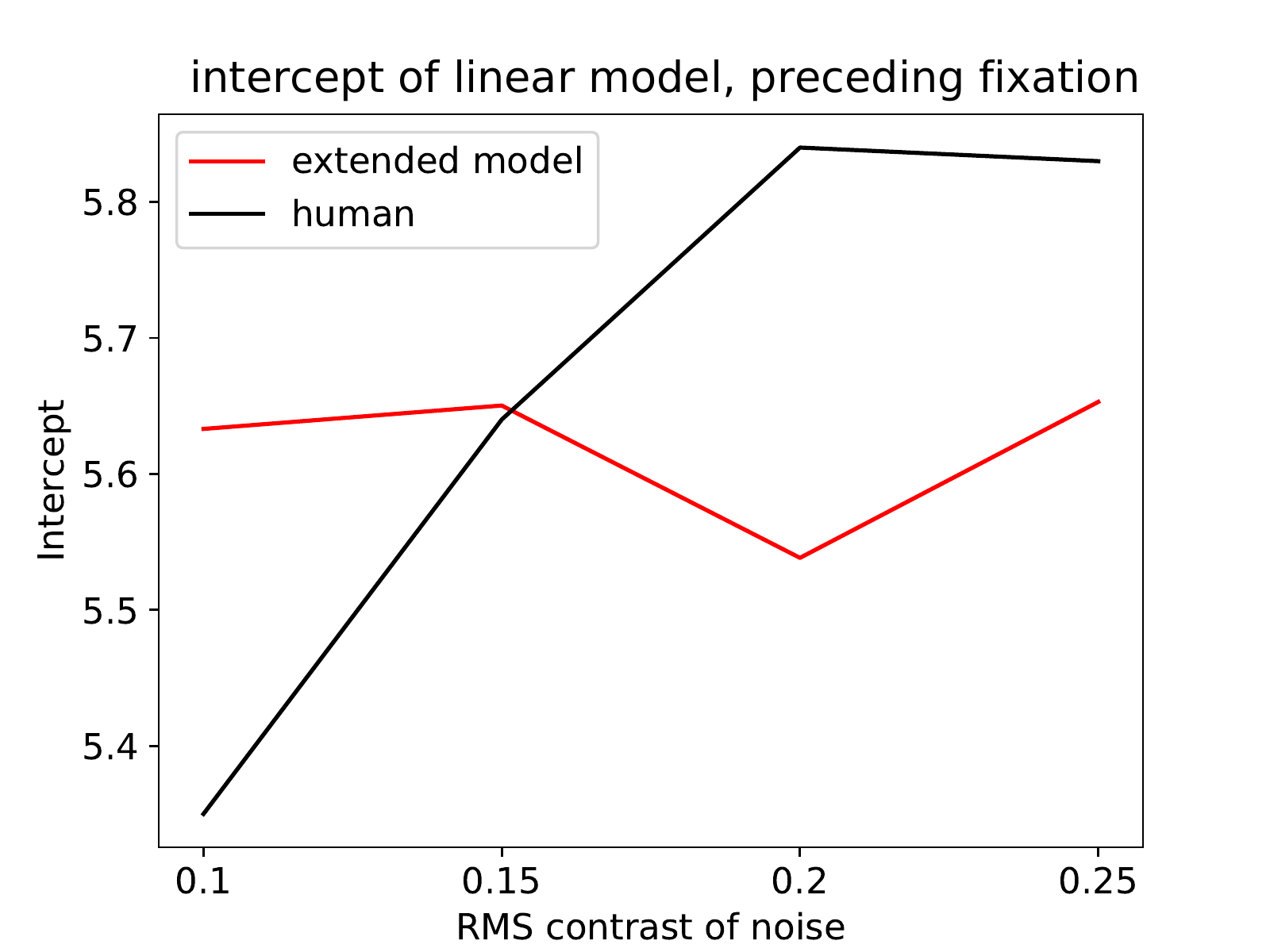}
\includegraphics[scale=0.4]{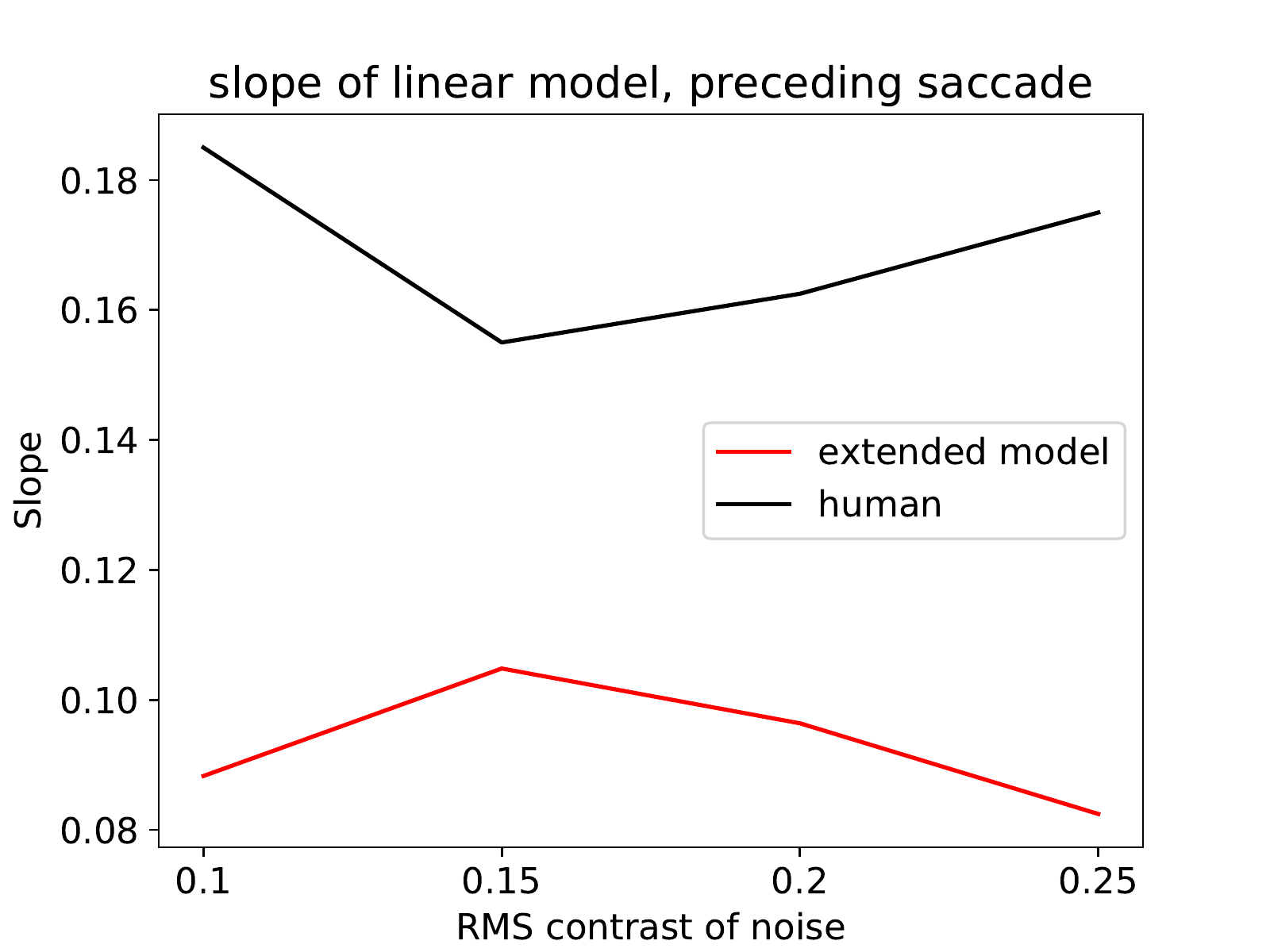}
\includegraphics[scale=0.4]{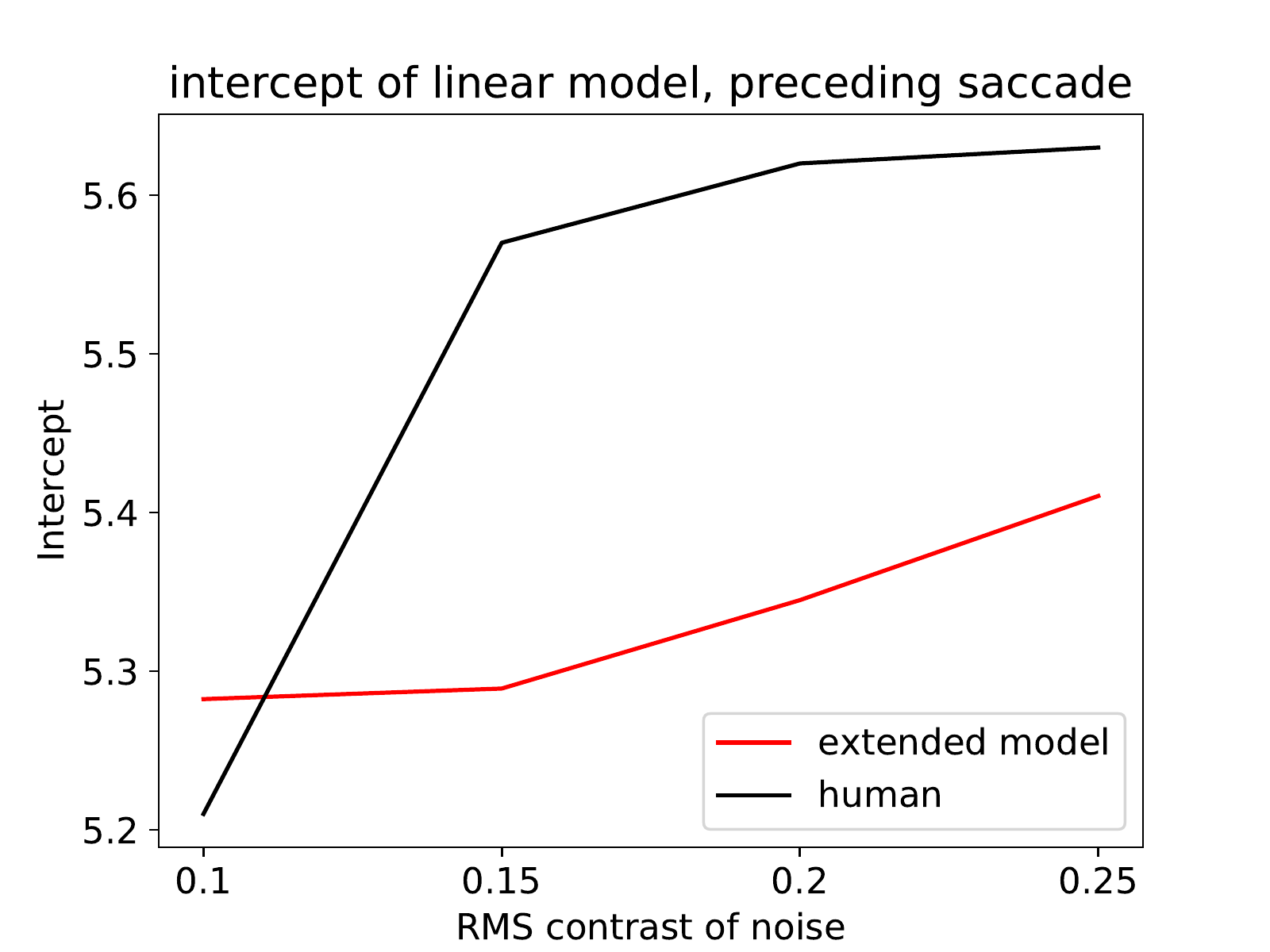}
\caption[The slopes and intercepts of linear approximation of log-transformed data of fixation duration and saccade lengths in the case of synthetic images]{The slopes and intercepts of linear approximation of log-transformed data of fixation duration and saccade lengths time-series in the case of synthetic images. We can see that slopes of linear dependency between fixation duration and preceding saccades are positive both for simulated and experimental eye-movements in the case of synthetic images, which is consistent with literature. In the other hand we can see that the values of slope of linear dependency between fixation duration and succeding saccade length obtained through RANSAC method are negative, which was not previously discussed in the literature.}
\label{fig:slopessynt}
\end{figure*}
\begin{figure*}
\centering
\includegraphics[scale=0.4]{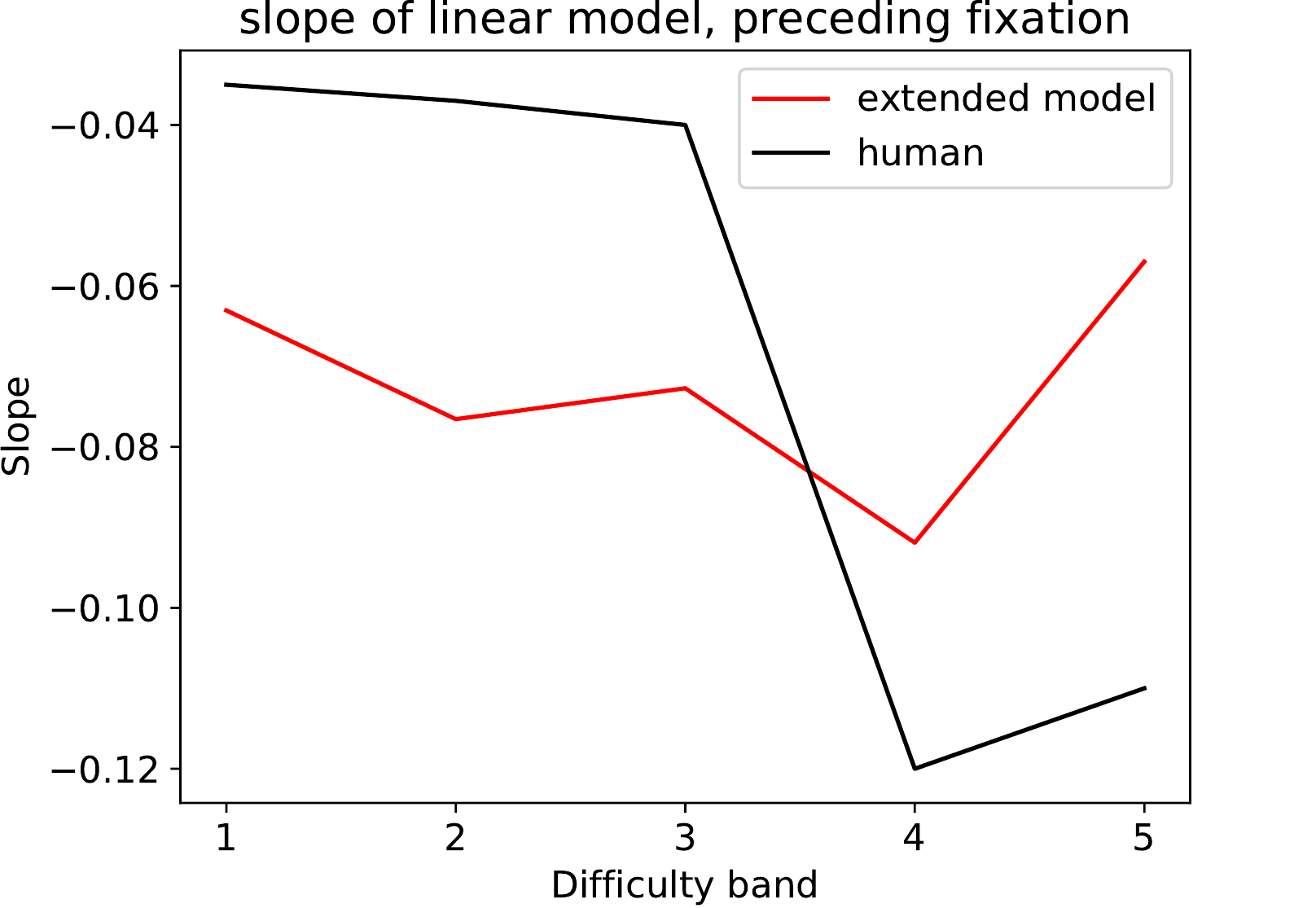}
\includegraphics[scale=0.4]{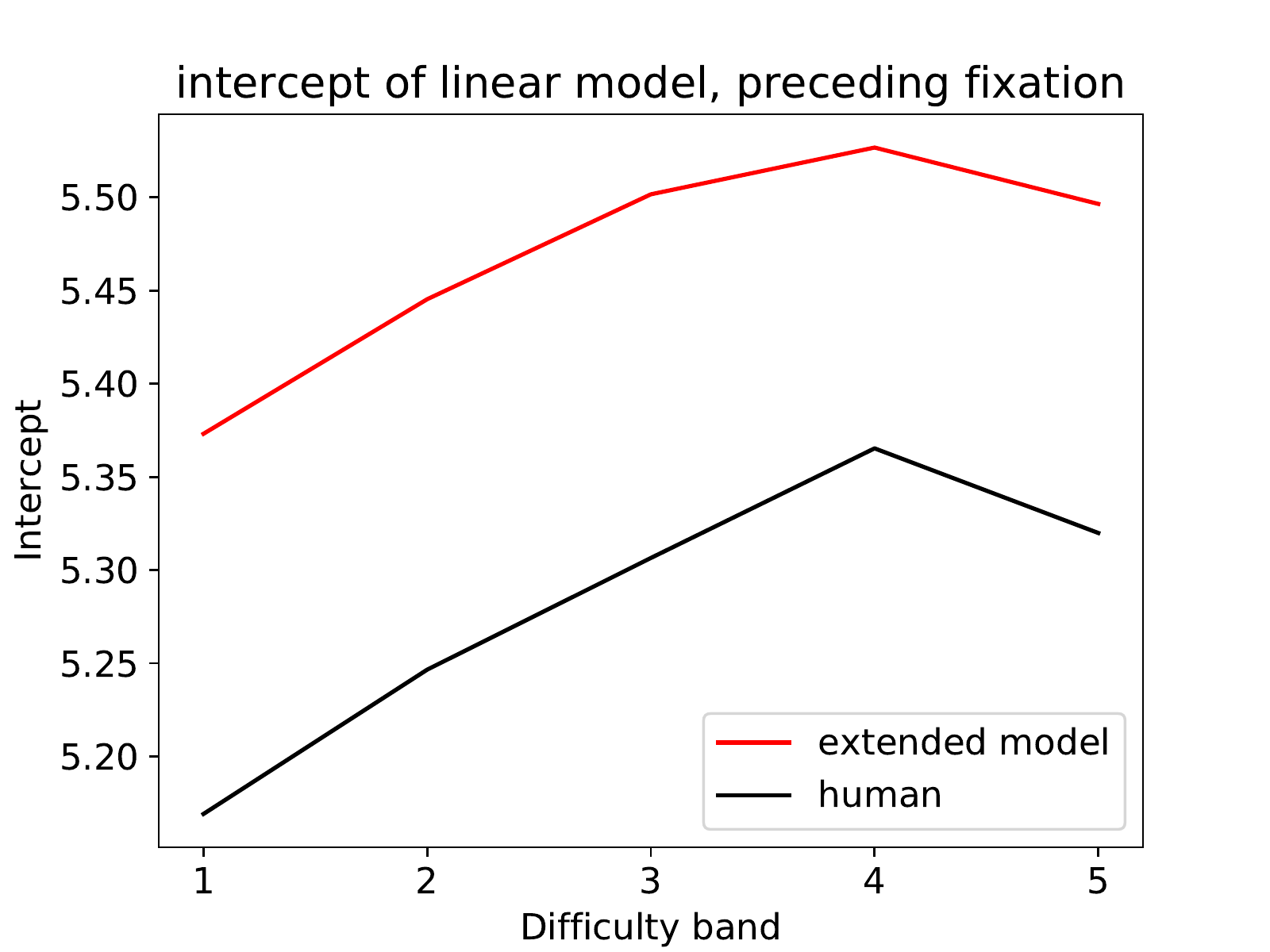}
\includegraphics[scale=0.4]{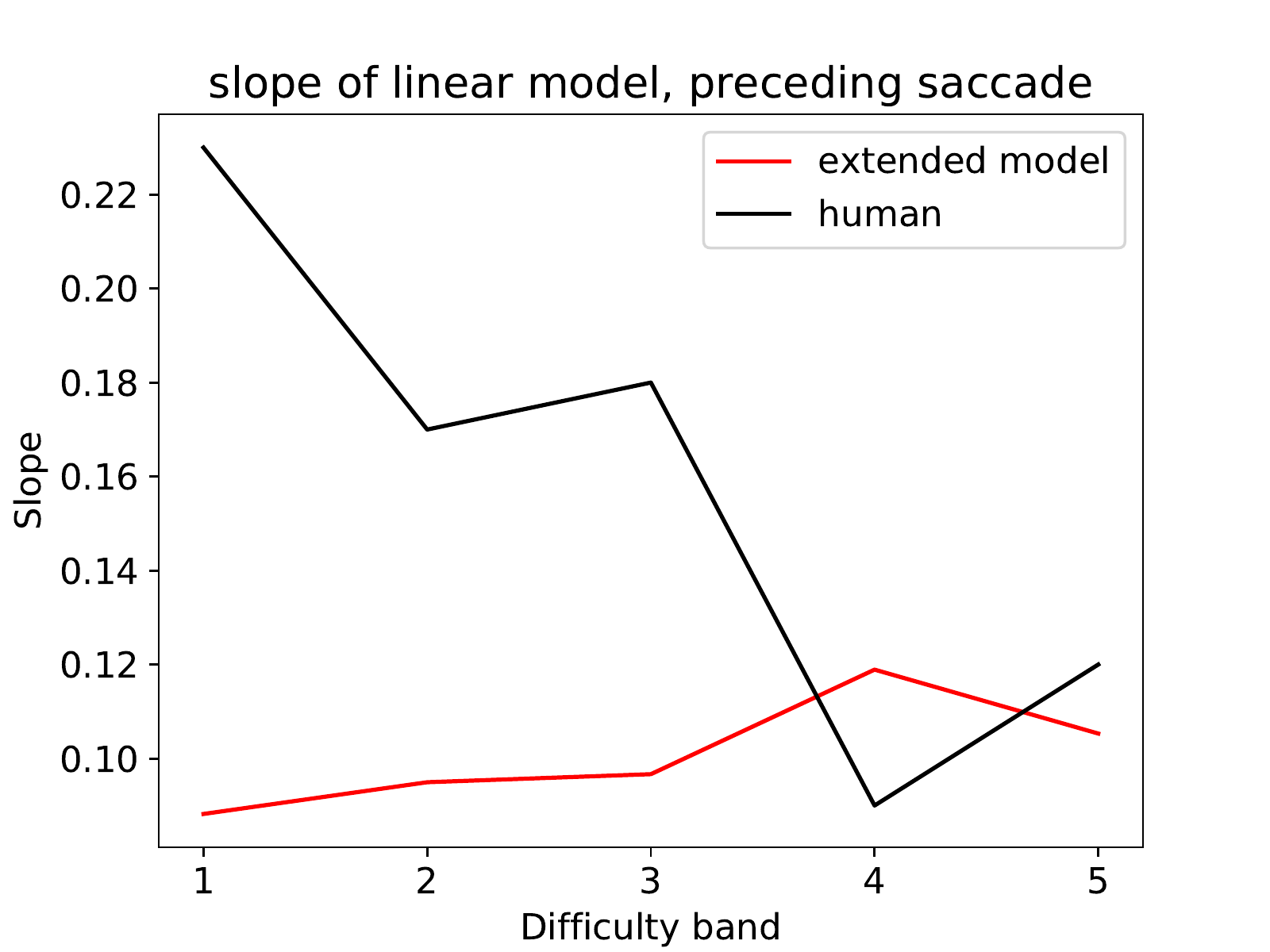}
\includegraphics[scale=0.4]{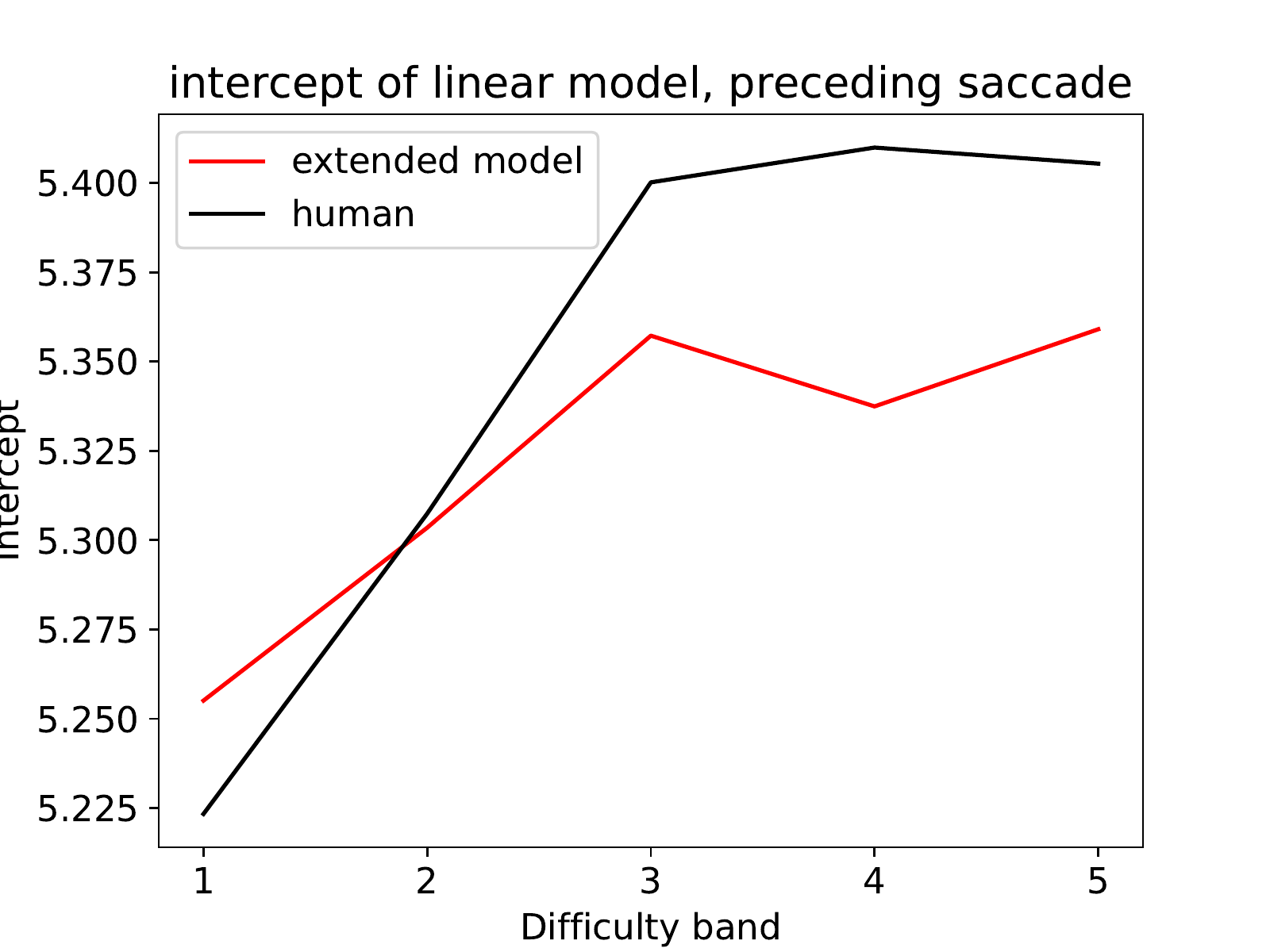}
\caption[The slopes and intercepts of linear approximation of log-transformed data of fixation duration and saccade lengths time-series in the case of natural images.]{The slopes and intercepts of linear approximation of log-transformed data of fixation duration and saccade lengths time-series in the case of natural images. We can see that slopes of linear dependency between fixation duration and preceding saccades are positive both for simulated and experimental eye-movements in the case of natural images, which is consistent with literature. In the other hand we can see that the values of slope of linear dependency between fixation duration and succeding saccade length obtained through RANSAC method are negative, which was not previously discussed in the literature.}
\label{fig:slopesnatural}
\end{figure*}

The red (black) line on Figures \ref{fig:slopessynt} and \ref{fig:slopesnatural} show  the slope and intercept of linear dependency between log-transformed length of saccade and log-transformed fixation duration on RMS contrast of noise for simulated (experimental) eye-movements. We can see that slopes of linear dependency between fixation duration and preceding saccades are positive both for simulated and experimental eye-movements in both cases of natural and synthetic images, which is consistent with literature. Previously it was shown that fixation duration is correlated with preceding saccade length during execution of fixation tasks. Salthouse et al. \cite{salthouse1980determinants} hypothesized that the eyes can't be stopped immediately after execution of saccade, and faster was the preceding eye-movement - a longer time interval is required to stop it. 
\par On the other hand we can see in Figures \ref{fig:slopessynt} and \ref{fig:slopesnatural} that the values of slope of linear dependency between fixation duration and succeding saccade length obtained through RANSAC method are negative in most cases both for simulated and experimental eye-movements. This effect was not previously discussed in the literature. In the experiment 6 of Salthouse et al. \cite{salthouse1980determinants}, which is a simple fixation task, they didn't find any evidence that fixation duration and amplitude of succeeding saccade are correlated. But that doesn't mean that this effect can't take place during execution of other visual tasks. Unlike fixation task, visual search requires integration of visual information extracted from previous fixations and information that is currently being extracted. There is always a trade-off between quality of extracted information and time spent on fixation, because it can be spent, alternatively, on moving to and fixating other locations, which may result in  negative dependency between fixation duration and succeding saccade length.

\section{Conclusion}

Our goal was to study the ability of human observer to control  fixation duration during the execution of  visual search task. 
We conducted the eye-tracking experiments with natural and synthetic images and found the following consistent patterns in visual behavior: a steady increase in average fixation duration with difficulty and dependency of fixation duration on the length preceding and succeding saccades. In order to explain these effects, we presented the extension of eye-movements model by introduction of continuous-time observation and decision making. The simulations of the model were made with discretized time steps of progressively decreasing duration in order to demonstrate the convergence of statistics of trajectories with resolutions. We have shown that basic characteristic of simulated time-series converge with increase of resolution, which means that decision making process is invariant to time step after reaching resolution of 16 ms. This allows us to perform reinforcement learning for different visibility maps corresponding to various difficulty conditions. 
\par We used the model to simulate the eye-movement time series and to compare them with experimental ones. We found significant difference between predicted and experimental values of fixation duration. However, the model is capable to correctly predict the increase in values with difficulty.  We assume that the reason of the  mismatch is that our model doesn't take into account the fixational eye-movements (FEMs): drift and microsaccades, which favour longer fixation duration. Despite the addition of FEMs to the model is not challenging itself, we require finer computational grid  since the amplitude of FEMs is much lower (within the single time step) than the one of saccade. To investigate the effect of FEMs on extraction of visual information during the fixation we may require the size of grid cell $0.01 \deg$ and time step of $1$ ms \cite{engbert2004microsaccades}. Since our current simulations with grid size of $0.625$ deg  \ref{table:parameters} and time step $8$ ms require several hours to learn the optimal policy, it is not feasible to introduce the addition of FEMs. The second reason of mismatch between model simulations and human behaviour is limitations of current approach in modelling of processes within central neural system. Our computational model is formulated in terms of \textsc{Po-Mdp}, which relies on Bayesian framework \cite{geisler2003bayesian} for inference of world states. Our approach is agnostic about temporal aspects of neural processing, and, therefore, it neglects the influence of rhythmic activity in central neural system on dynamics of visual attention and eye-movements \cite{mcauley1999human,fiebelkorn2019rhythmic}. We expect the scenario of bias of saccade rate towards theta-waves frequency \cite{vanrullen2013visual}, which can't be explained by phenomenological control models of visual behaviour.
\par We studied the dependency between fixation duration and the length of preceding and succeding saccades. The time series of fixation duration and length of saccades were log-transformed due to their skewed distribution. We used RANSAC method to find the coefficients of linear model, and found that the slope of the linear model is positive for the case of saccades preceding to the fixation. This effect was previously found in fixation task by Salthouse et al. \cite{salthouse1980determinants}, and has simple mechanical explanation. In the other, we found that the slope is negative for the case of outgoing saccades, which was discussed in literature previously. This effect is specific for visual search task, during observer is faced with problem of distribution of time on current fixation or the next saccade. 
\par Our future work includes a development of maximum likelihood algorithm for estimation of visibility maps of observer, in the same way it was  previously done for Infomax model of eye-movements \cite{vasilyev2019spatial}. This will allow us to accurately model the eye-movements of every individual participant without experimental measurements of its visibility map. The main obstacle of usage of maximum likelihood schemes is dependence of optimal policy on parameters of visibility maps. In other words, we have to run reinforcement learning algorithm every time for estimation of optimization function or its gradient, which is required for most maximum likelihood schemes.   Instead of learning of policy each time for new parameter values, it was shown that it's possible to learn the family of policies each of which is trained for a particular environment, which is called the Universal Policy (UP)\cite{yu2018policy}. This method deviates from the standard reinforcement learning approach by training the UP, which is explicitly conditioned both on model parameters and policy parameters. This method was recently developed and tested for control of biped \cite{yu2018policy} and quadruped \cite{yu2019sim} locomotion. The Universal Policy is capable to control the robots with various conditions, unlike the single policy, which is trained only for specific conditions. In our case UP will be parametrized by parameters of visibility map and saccade execution function.
\par   Our conclusion is that optimal control of eye-movements is one of the reasons behind the variability of fixation duration. The extension of computational model of eye-movements allowed us to quantitatively explain the consistent patterns in visual behaviour of human observers. As it was mentioned before, the model can't predict the asymptotic behaviour of fixation duration distribution, and it would require  further improvements including addition FEMs and optimization of policy learning algorithm.

\bibliographystyle{elsarticle-num-names}
\addcontentsline{toc}{section}{\refname}\bibliography{proplib}

%




\appendix

\section{Convergence of basic characteristics}
\label{sec:converge}
\par First of all, we demonstrate how the values of three main characteristics: length of saccade, fixation duration and response time - change with time resolution.  We consider the case $e_{n}=0.15, e_{t}=0.2$ and estimate the visibility map according to \cite{najemnik2005optimal}. We learn the policy for several resolution intervals $\Theta_{int}=(128, 64, 32, 16, 8)$ms with PGPE in order to demonstrate the convergence of basic characteristics with increase of resolution. The procedure of learning with PGPE is described in our previous work and the current implementation has the same values of hyper-parameters \cite{vasilyev2017optimal}.
\par
\begin{figure*}
\centering
\includegraphics[scale=0.37]{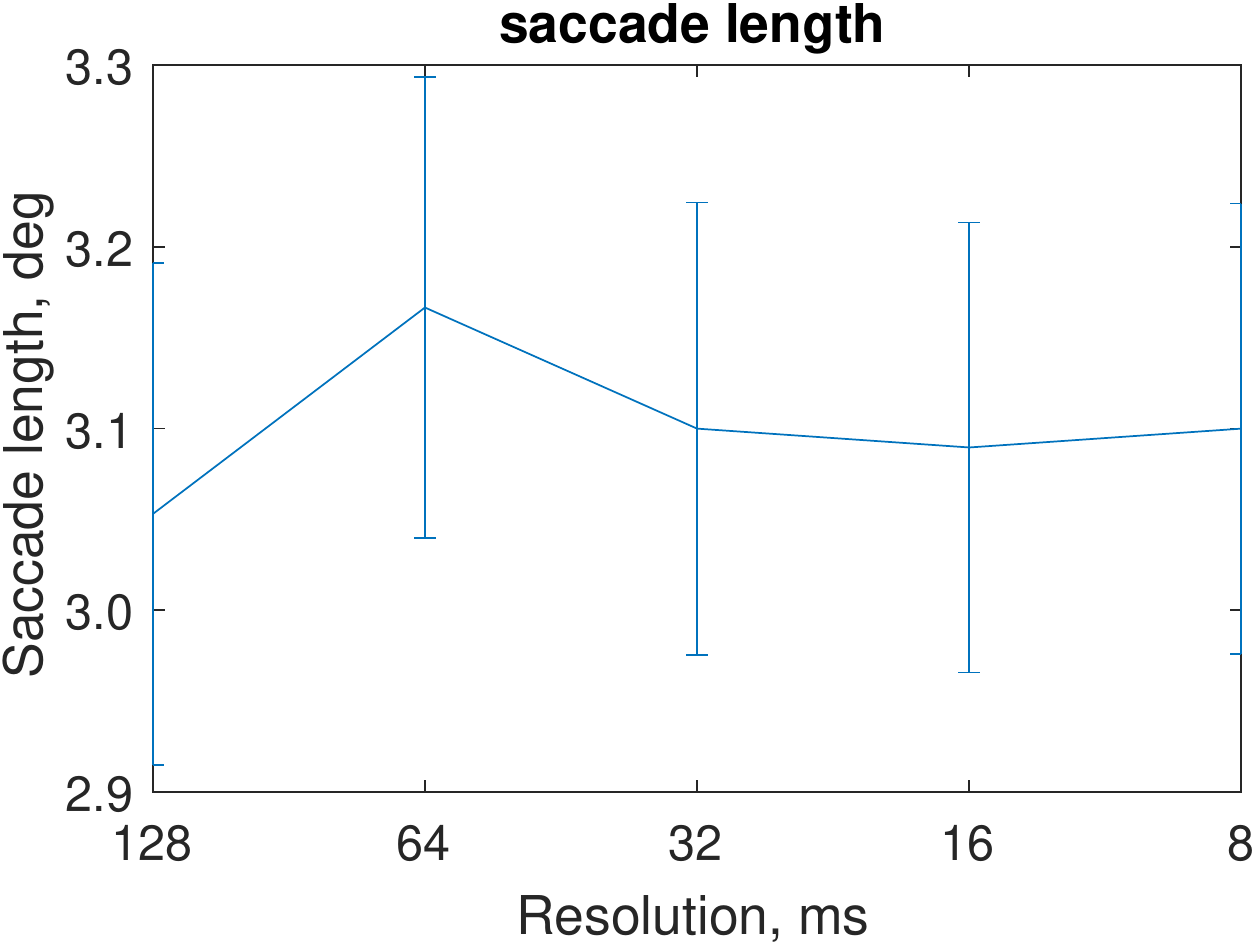}
\includegraphics[scale=0.37]{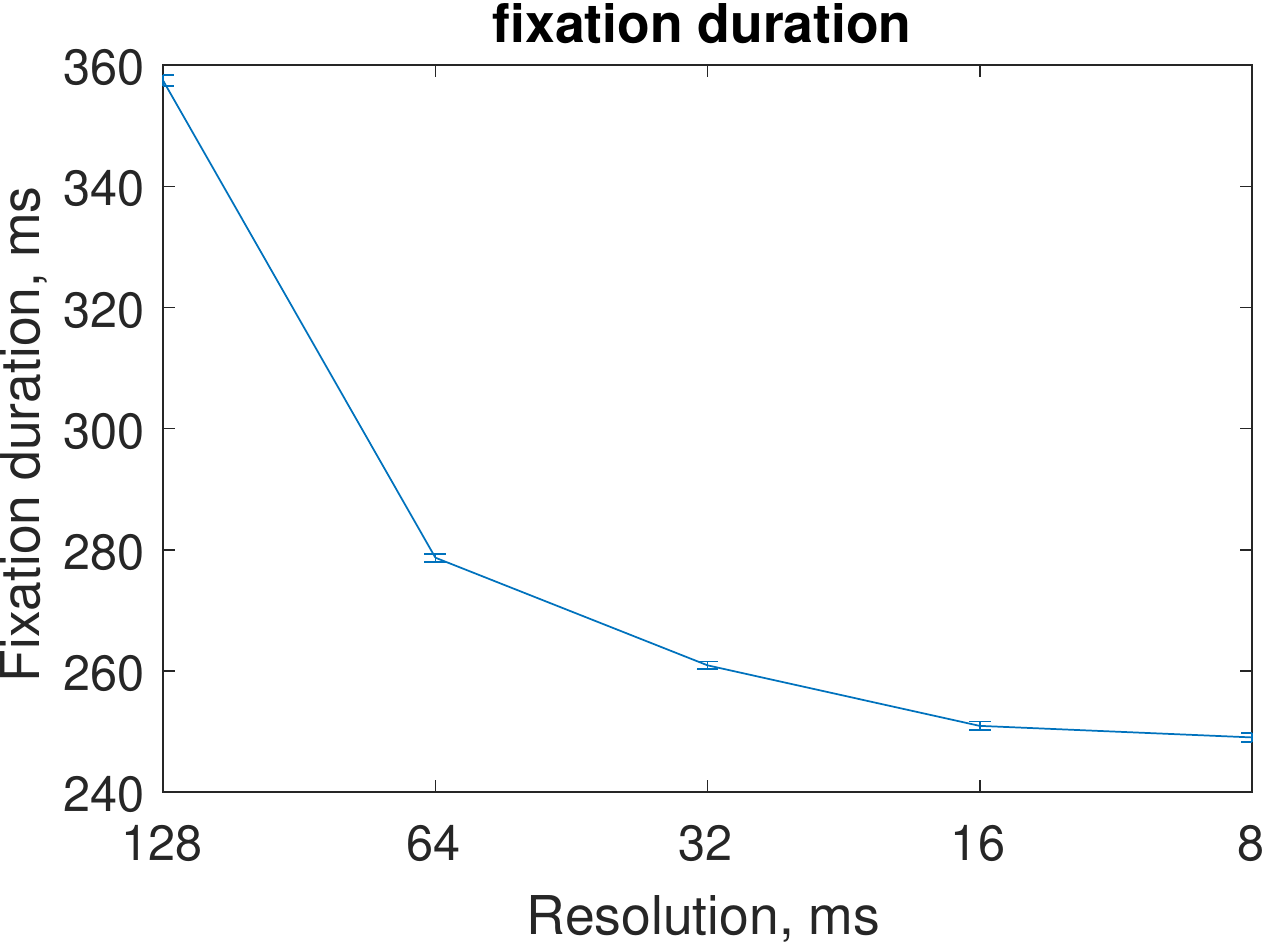}
\includegraphics[scale=0.37]{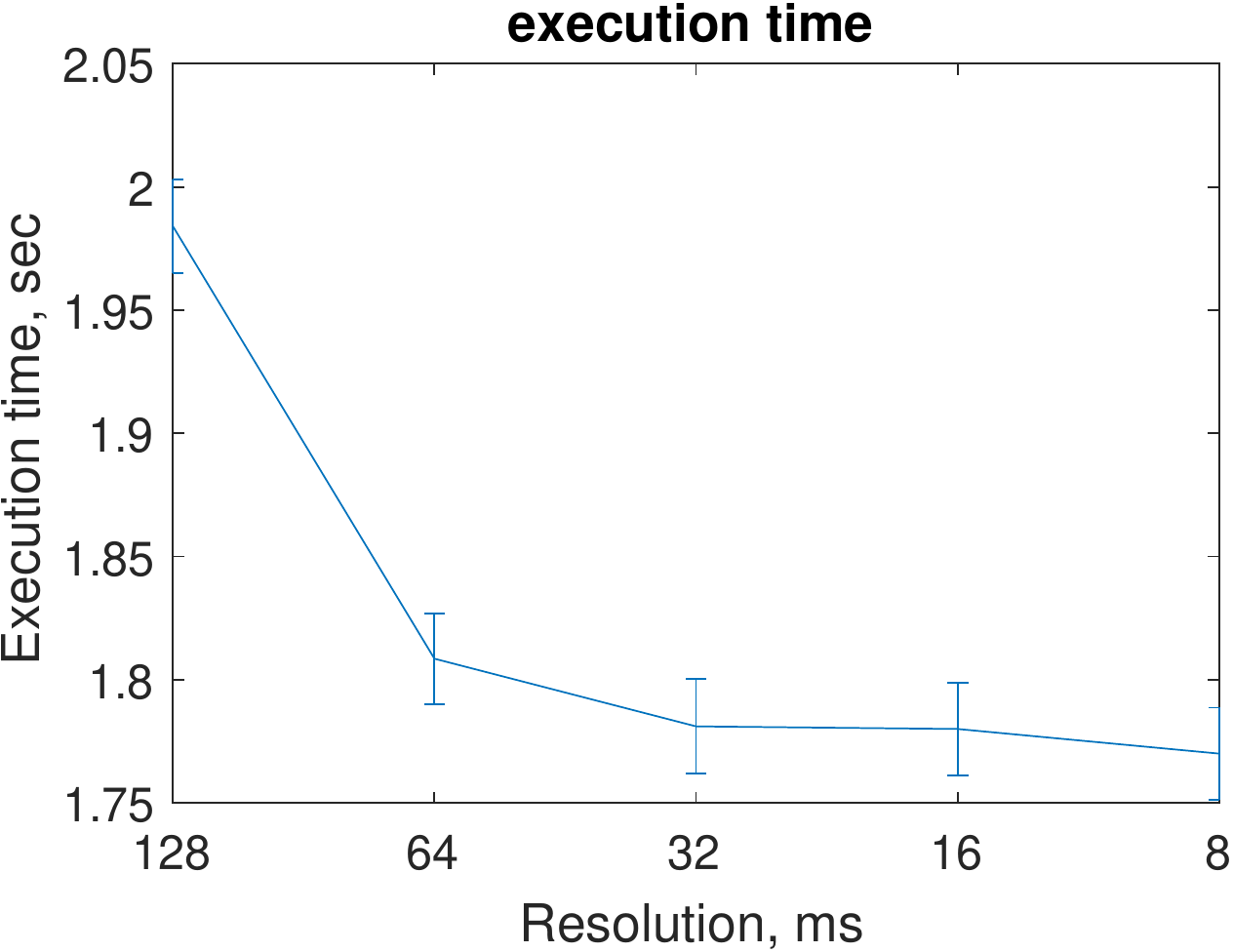}
\caption[Convergence of basic characteristics]{Convergence of basic characteristics. We simulated $10^{4}$ episodes of PO-MDP for each time resolution and corresponding policy to compute mean and standard error of characteristics. A change of saccade with resolution is not significant (left). Fixation duration decreases with resolution until it reaches $180 ms$, and then the change is not significant. Execution time significantly decreases from $2$ sec to $1.755$ sec, and after reaching resolution of $32$ ms there is no significant change.    }
\label{fig:converge}
\end{figure*}

 We simulated $10^{3}$ episodes of \textsc{Po-Mdp} for each time resolution and corresponding policy to compute mean and standard error of characteristics.  The figure \ref{fig:converge} shows that a difference of saccade length with resolution is not significant after resultion of $32$ ms (Student's t-test $p>0.05$).  At the same time, fixation duration decreases with resolution until it reaches $240$ ms, and then the change is not significant ($p>0.05$).  Execution time significantly decreases from $0.7$ sec to $0.5$ sec, and after reaching resolution of $32$ ms there is no significant change ($p>0.05$). We observed convergence of three basic characteristics for all experimental conditions.

\section{Distributions of fixation duration in experiments on natural images}
\begin{figure*}
\centering
\includegraphics[scale=0.4]{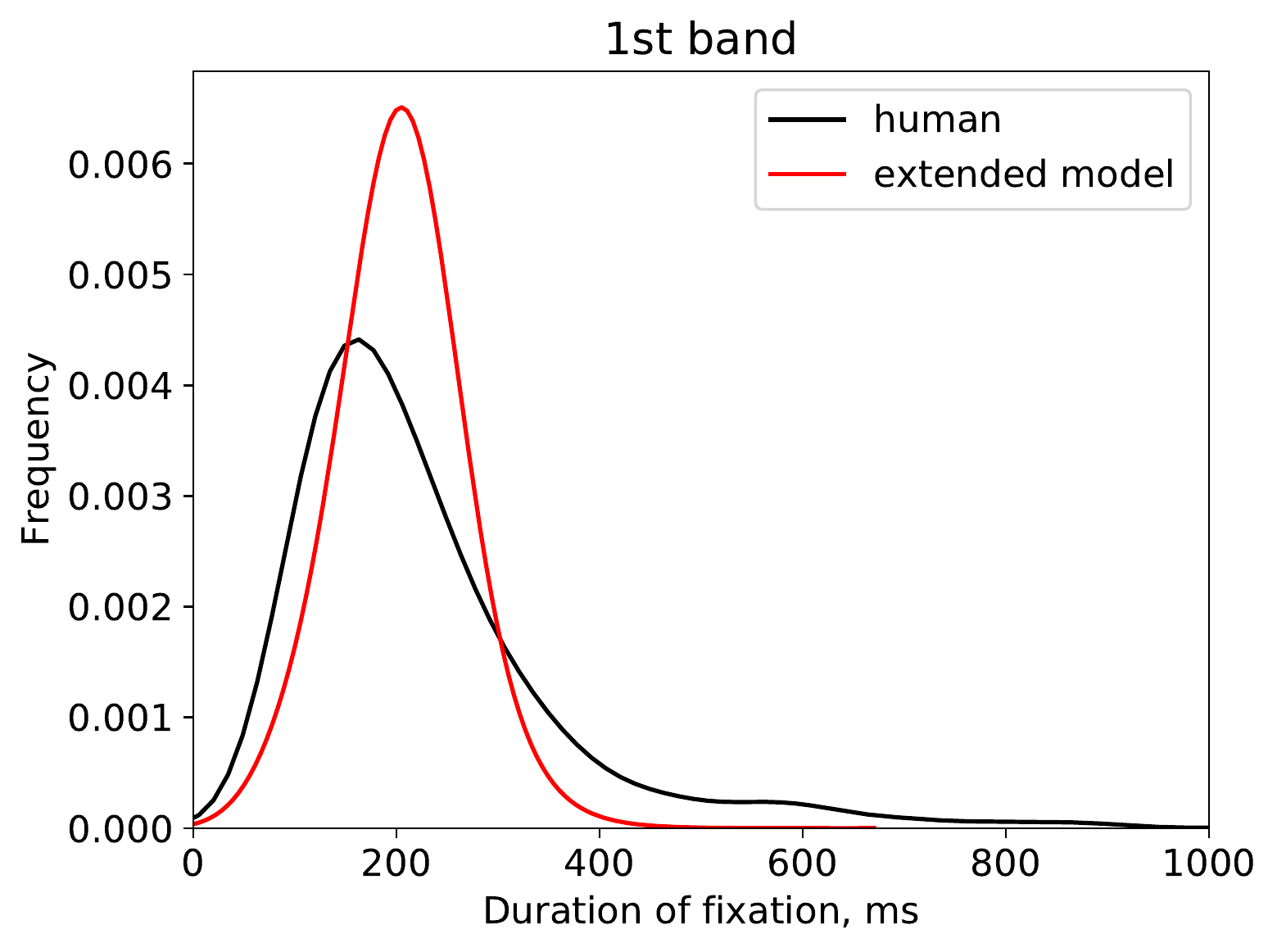}
\includegraphics[scale=0.4]{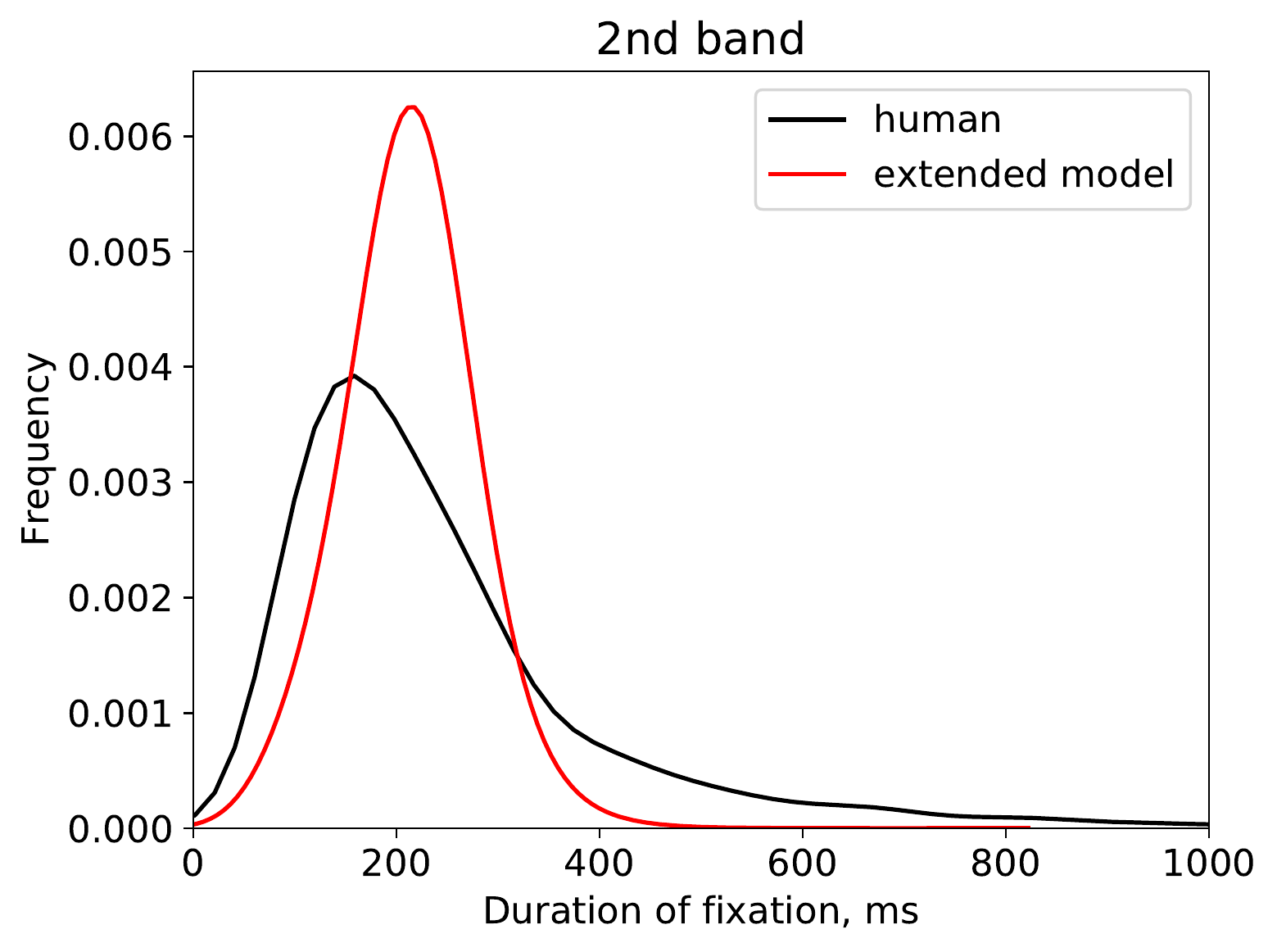}
\includegraphics[scale=0.4]{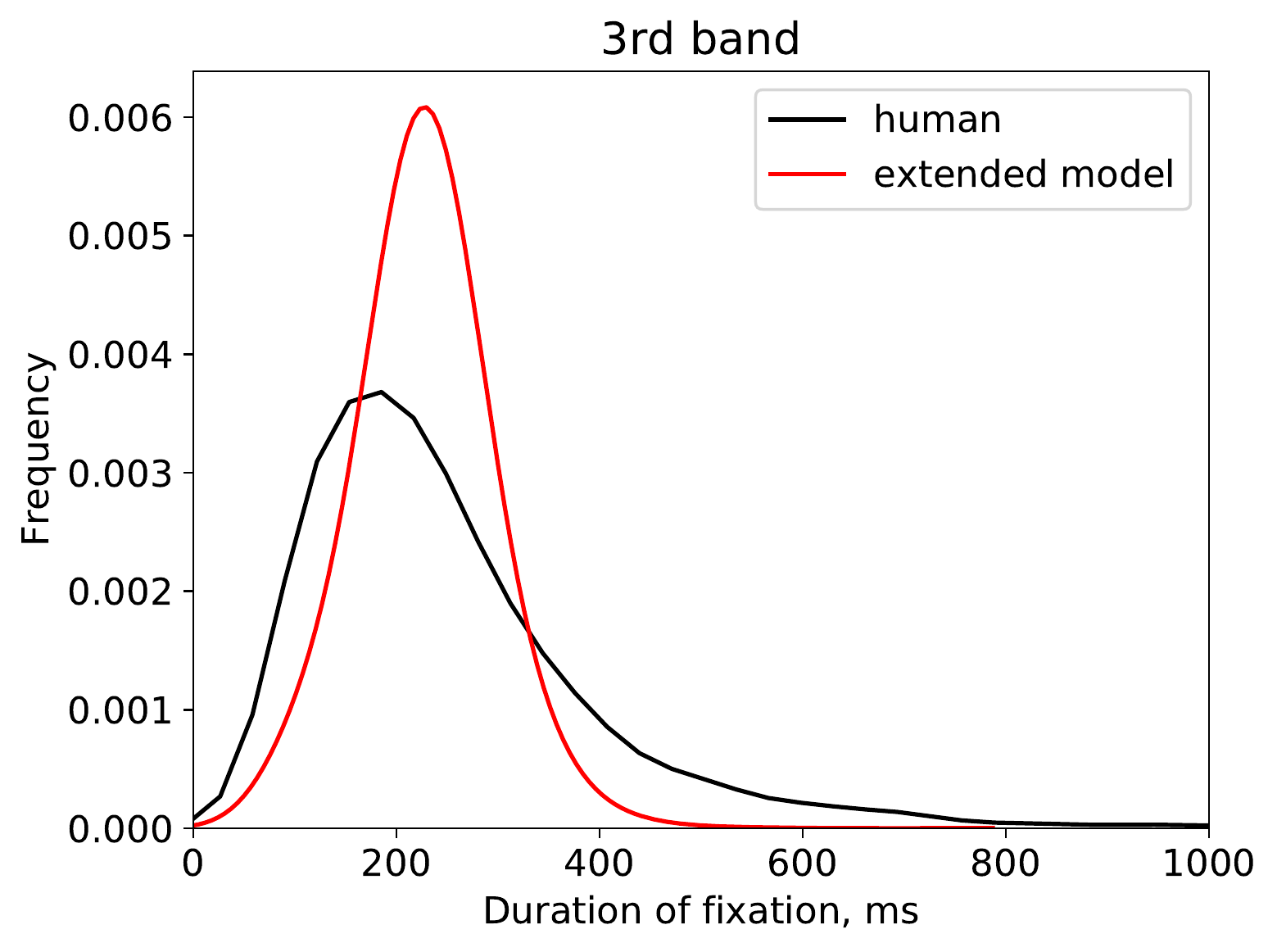}
\includegraphics[scale=0.4]{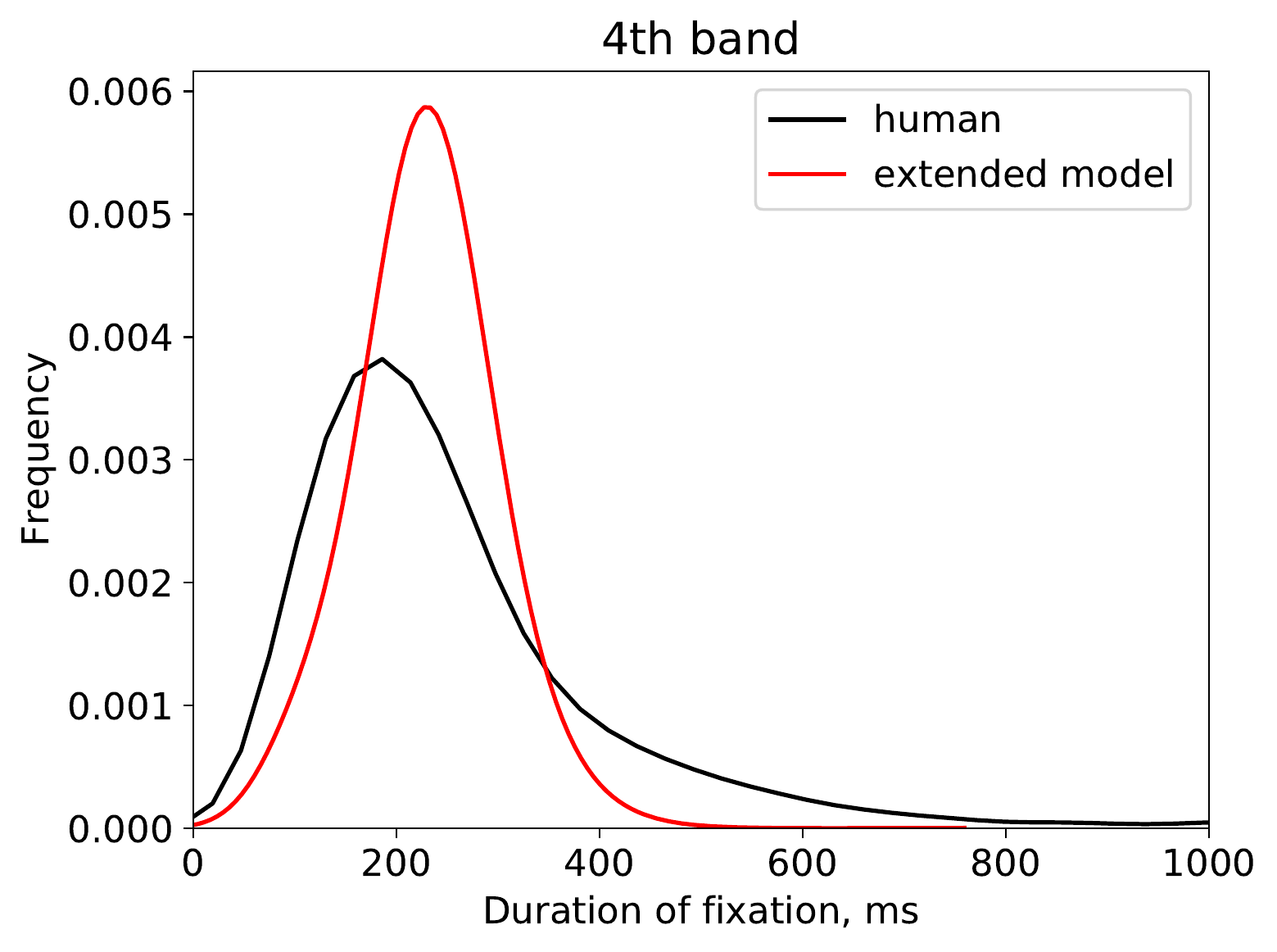}
\includegraphics[scale=0.4]{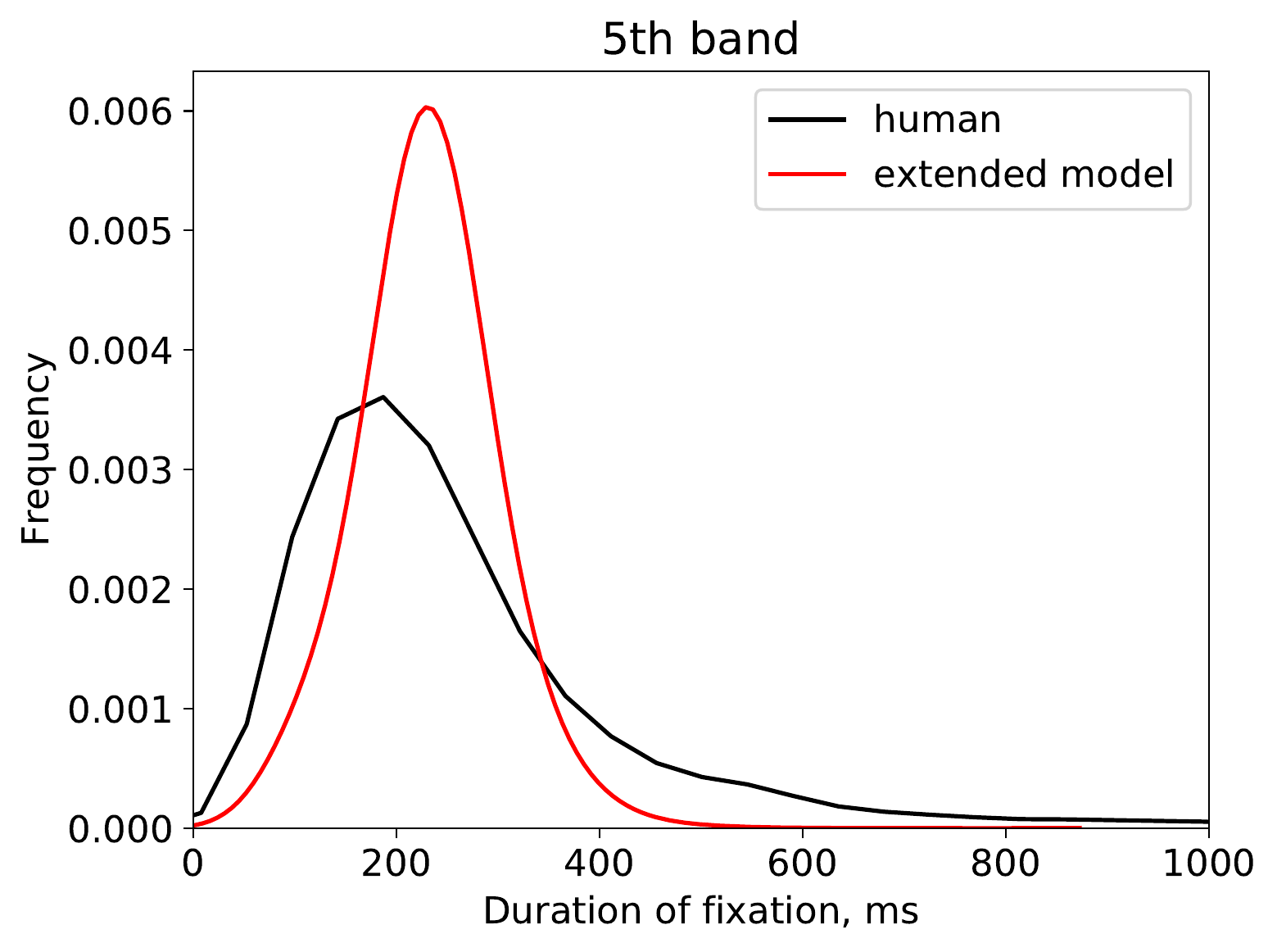}
\caption[The distribution of fixation duration in experiment on natural images and corresponding simulations of extended model. ]{The distribution of fixation duration in experiment on natural images and corresponding simulations of extended model. As well as in the case of synthetic images, the computational model can't explain the heavy tail in the experimental distributions, because fixational eye-movements were taken into account.  }
\label{fig:durdifnat}
\end{figure*}
\section{Coefficients of Box-Cox transformation}

\begin{table*}
\centering
\begin{tabular}{|c | c |c | c | c | } 
\hline 
\multicolumn{5}{|c|}{Visual search on synthetic images} \\ \hline
\multicolumn{1}{|c|}{Case} & \multicolumn{2}{|c|}{Preceding fixation } & \multicolumn{2}{|c|}{Preceding saccade} \\
\hline
\thead {noise \\ contrast} &\thead{Lambdas}& \thead{Standard errors} & \thead{Lambdas} & \thead{Standard errors} \\
\hline 
0.1 & (-0.015,0.061) & (0.026,0.036) & (0.013,0.045) &(0.056,0.076)  \\ \hline
0.15  & (0.024,-0.064) &(0.080,0.021) & (0.036,0.036)  &(0.076,0.023) \\ \hline
0.2 &  (-0.023,0.013) &(0.090,0.098)  & (0.057,0.068)  & (0.034,0.046)  \\ \hline
0.25 &  (0.058,0.043) & (0.023,0.037)  & (0.024,0.045)  & (0.077,0.034)  \\
[1ex]
\hline 
\end{tabular}
\caption{Predicted coefficients of Box Cox transformation and corresponding standard errors for the experimental data in the case of synthetic images. }
\label{table:Box Cox}
\end{table*}

\begin{table*}
\centering
\begin{tabular}{|c | c |c | c | c | } 
\hline 
\multicolumn{5}{|c|}{Visual search on real images} \\ \hline
\multicolumn{1}{|c|}{Case} & \multicolumn{2}{|c|}{Preceding fixation } & \multicolumn{2}{|c|}{Preceding saccade} \\
\hline
\thead {difficulty \\ band} &\thead{Lambdas}& \thead{Standard Error} & \thead{Lambdas} & \thead{Standard Error} \\
\hline 
1 & (0.011,-0.031) & (0.056,0.034) & (0.015,-0.035) &(0.056,0.076)  \\ \hline
2  & (0.023,0.084) &(0.089,0.078) & (0.056,0.076)  &(0.076,0.023) \\ \hline
3 & (-0.043,0.023) &(0.078,0.089)  & (-0.067,0.098)  & (0.034,0.046)  \\ \hline
4 & (0.046,-0.048) & (0.084,0.067)  & (0.084,-0.034)   &(0.055,0.078)  \\ \hline
5 & (0.078,0.023) & (0.088,0.067)  & (0.12,-0.045)  & (0.077,0.034)  \\
[1ex]
\hline 

\end{tabular}
\caption{Predicted coefficients of Box Cox transformation and corresponding standard errors for the experimental data in the case of natural images.  }
\label{table:BoxCoxNat}
\end{table*}
\begin{figure*}
\centering
\includegraphics[scale=0.98]{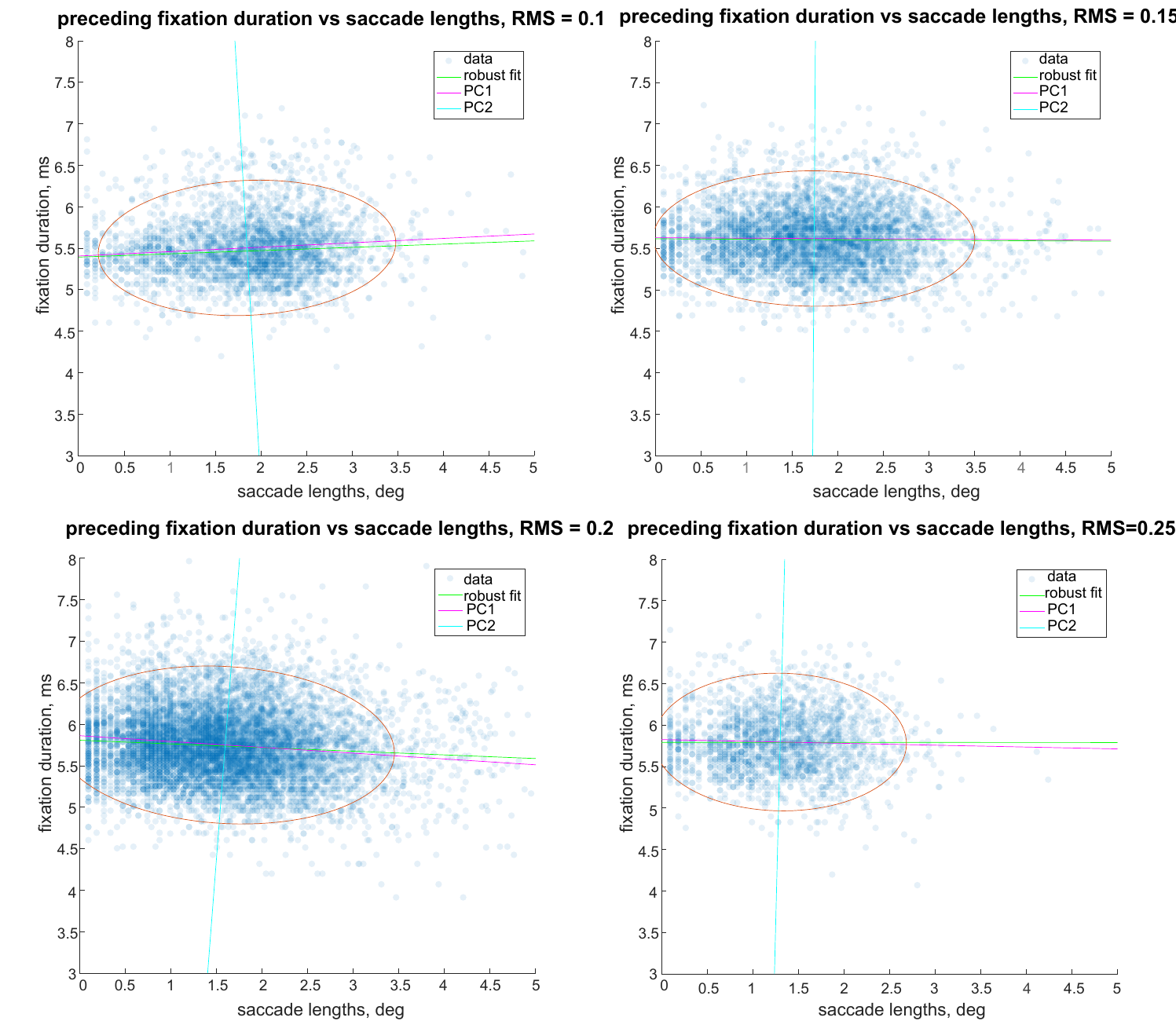}
\caption[Scatter plots of log-transformed data in logarithmic scale for preceding saccade in the case of experiments on synthetic images. ]{Scatter plots of log-transformed data: preceding fixation duration (vertical axis) and saccade lengths (horizontal axis) in logarithmic scale, principal components and linear approximation by robust fit in the case of experiments on synthetic images. The regression line of robust fit is close to horizontal and coincides with the first principal component, which explains $68 \%$ of variability in experimental data for the case of synthetic images and preceding fixation.   }
\label{fig:corrchart} 
\end{figure*}

\section{Scatter plots of log-transformed data}
\begin{figure*}
\centering
\includegraphics[scale=0.98]{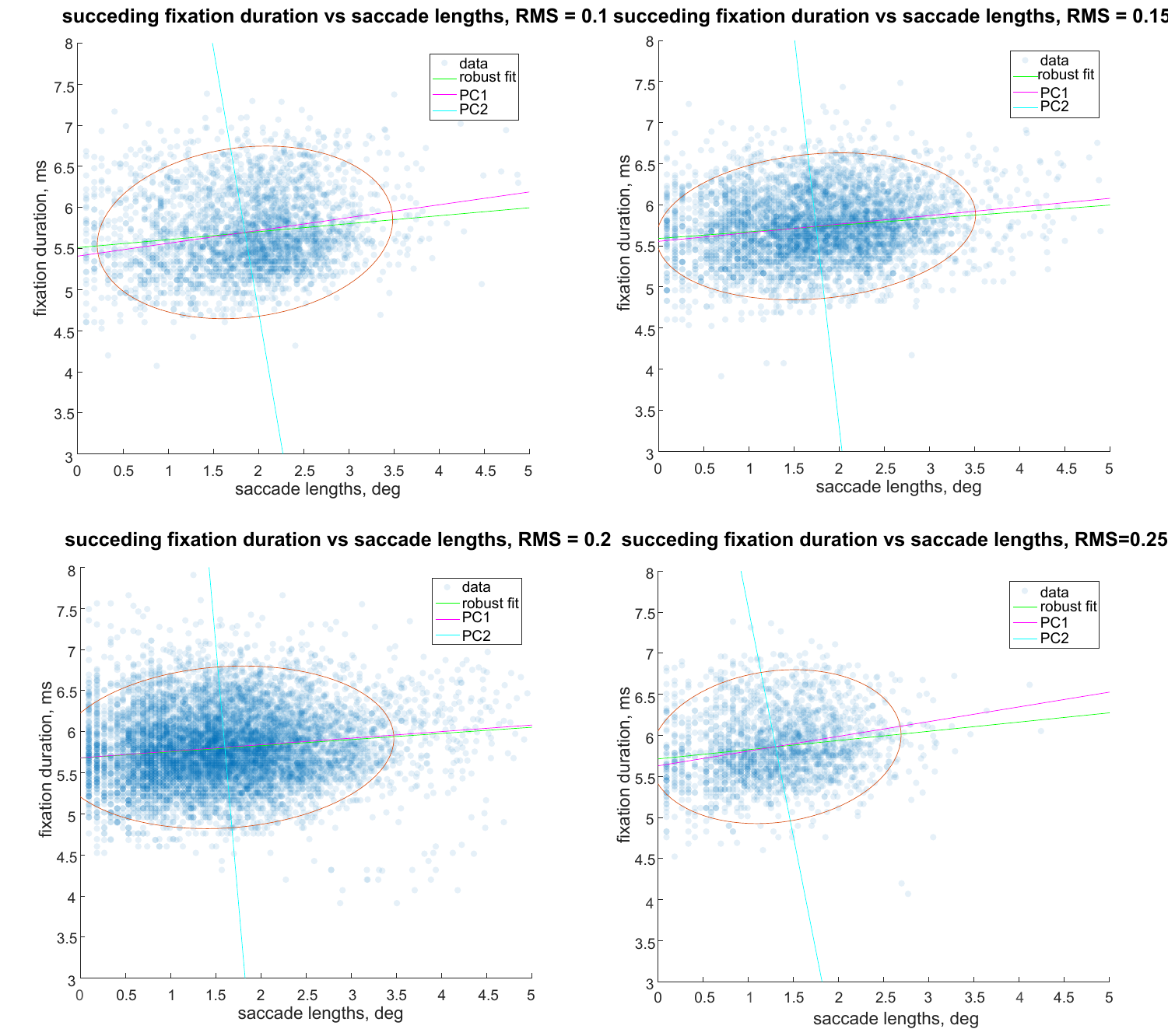}
\caption[Scatter plots of log-transformed data in logarithmic scale for preceding saccade in the case of experiments on synthetic images. ]{Scatter plots of log-transformed data: preceding fixation duration (vertical axis) and saccade lengths (horizontal axis) on logarithmic scale, principal components and linear approximation by robust fit for preceding fixation in the case of experiments on synthetic images. The regression line of robust fit is close to horizontal and coincides with the first principal component, which explains $63 \%$ of variability in experimental data for the case of synthetic images and preceding fixation. }
\label{fig:corrchart1}
\end{figure*}
\begin{figure*}
\centering
\includegraphics[scale=0.39]{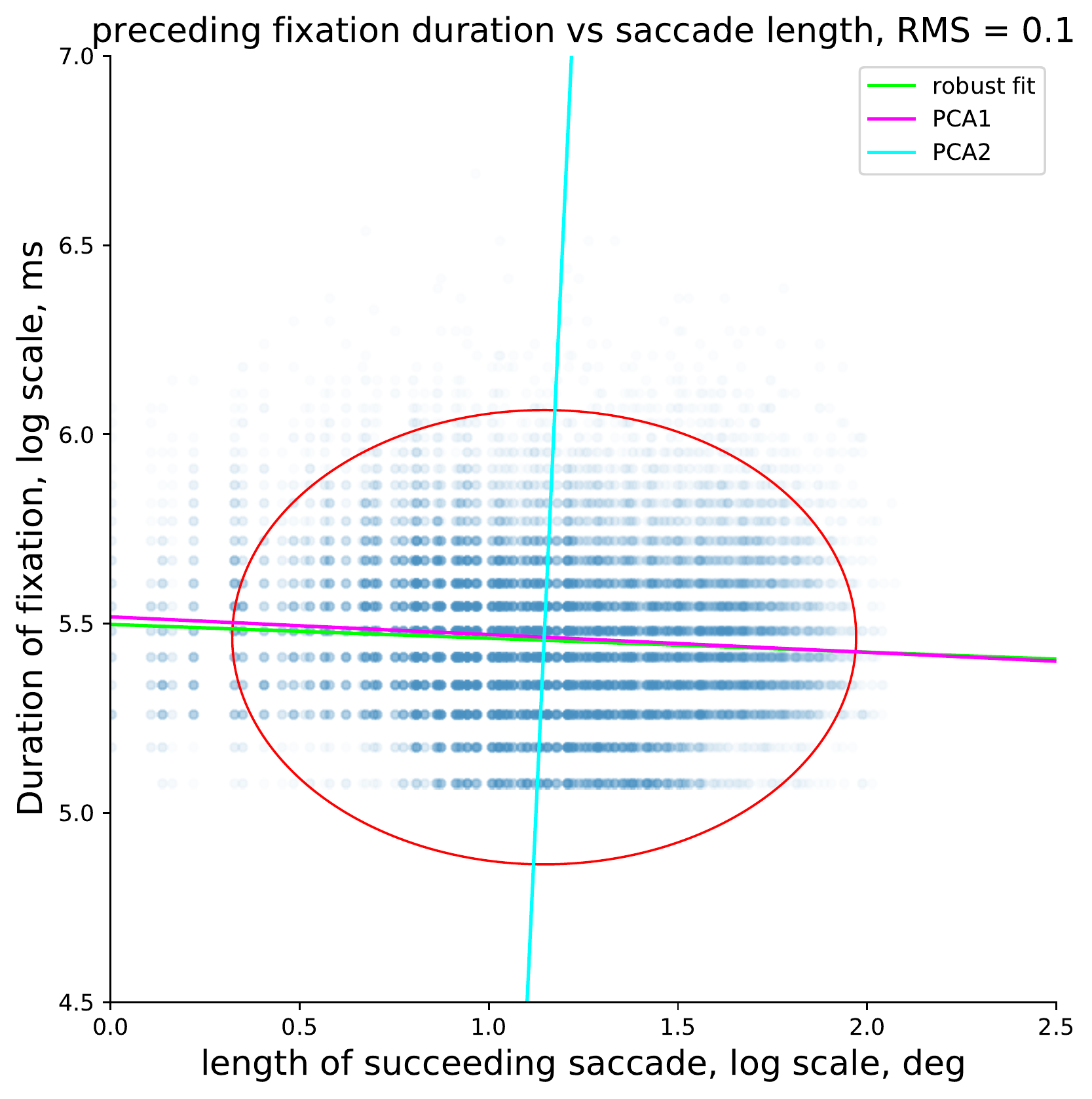}
\includegraphics[scale=0.39]{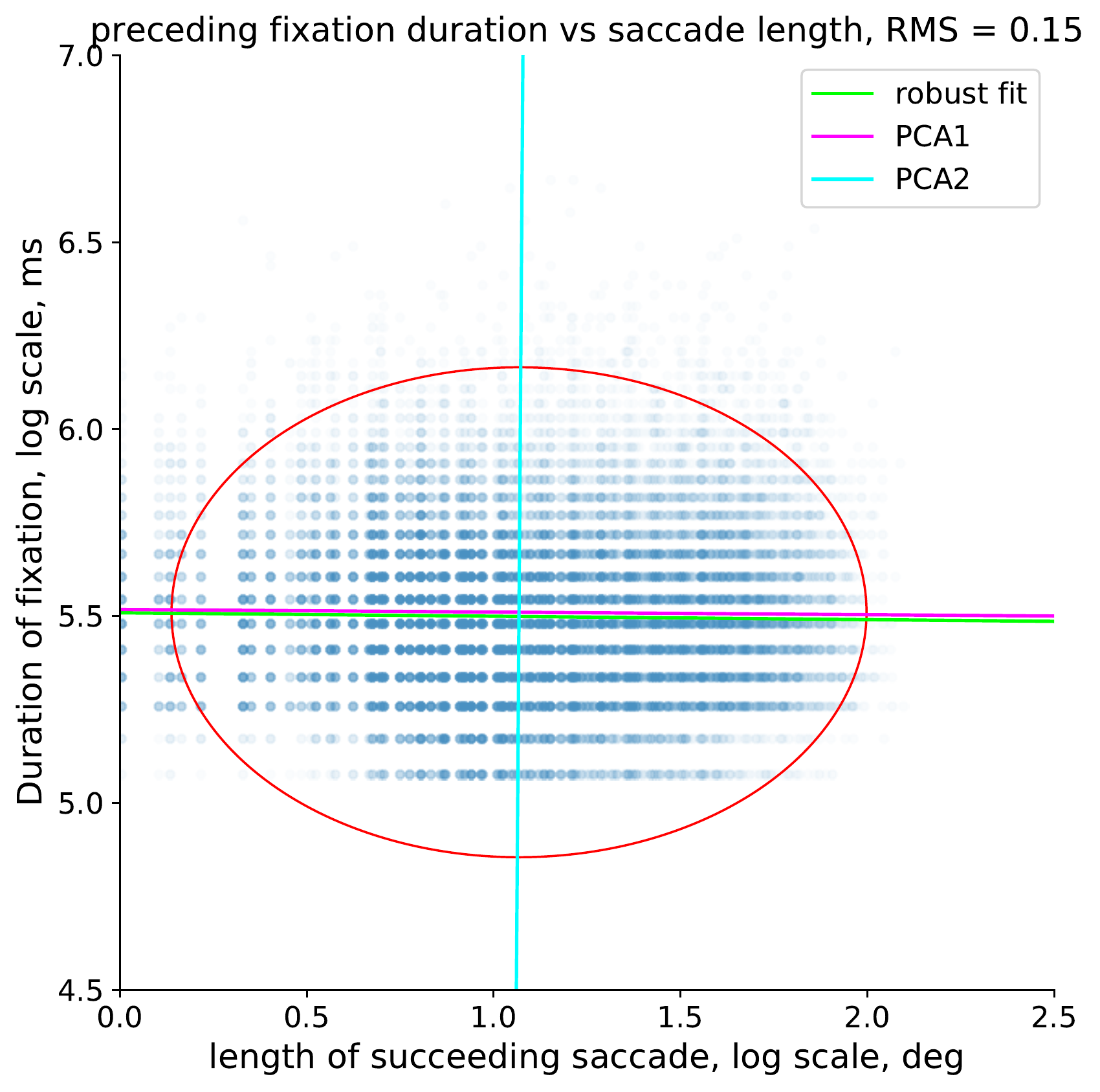}
\includegraphics[scale=0.39]{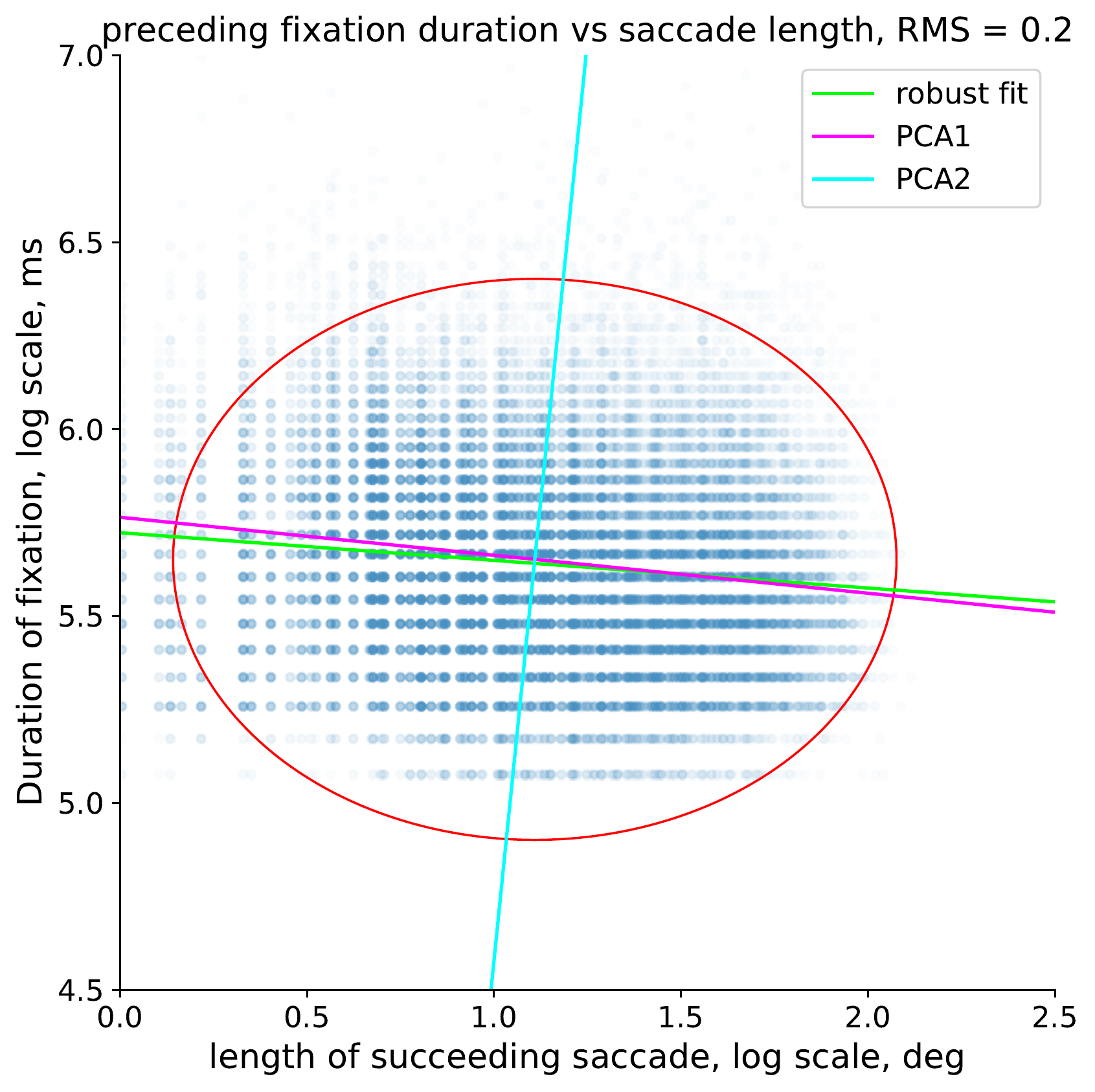}
\includegraphics[scale=0.39]{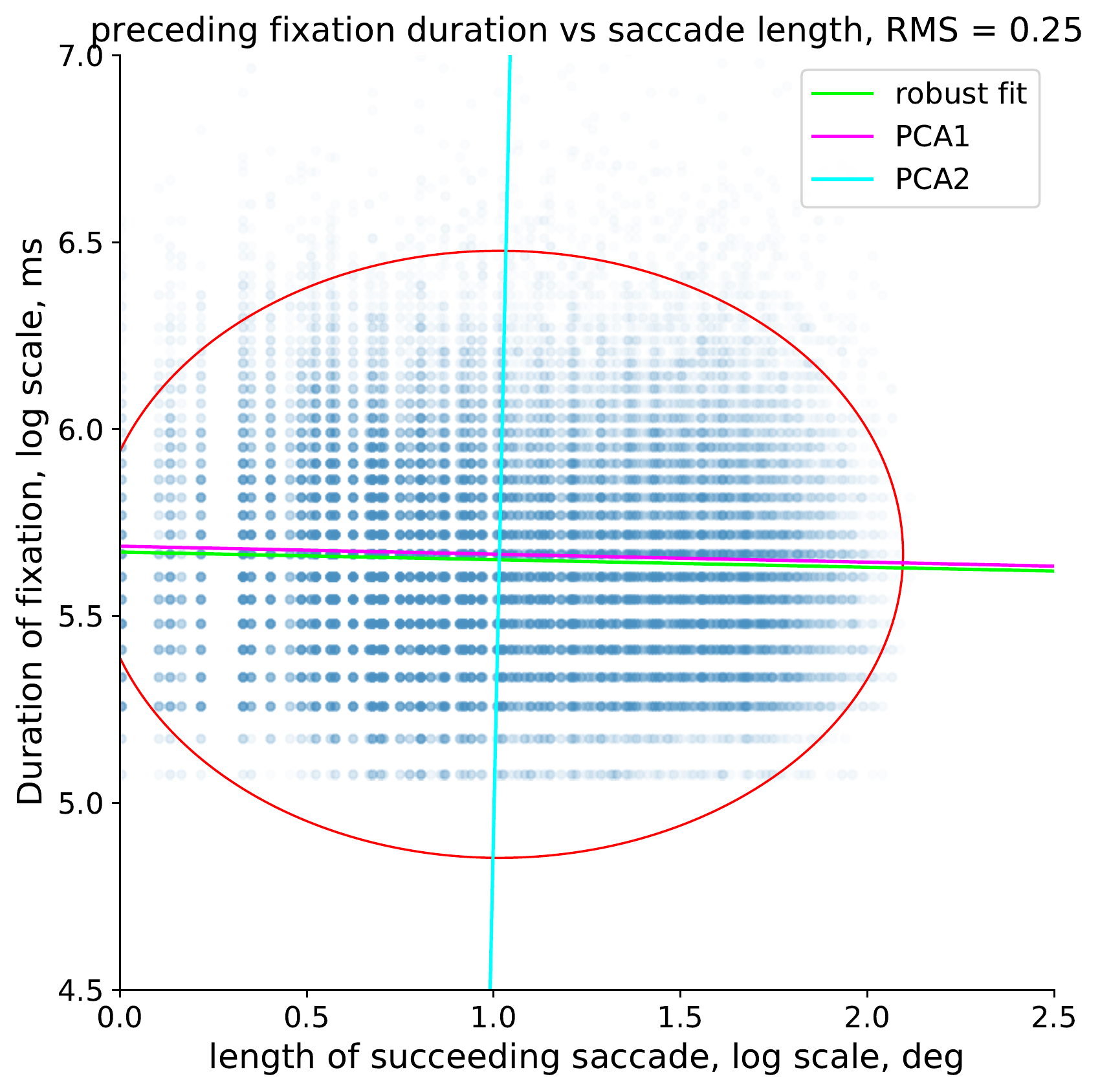}
\caption[Scatter plot of log-transformed data in logarithmic scale for preceding fixation in the case of simulations for synthetic images. ]{Scatter plot of log-transformed data: succeding fixation duration (vertical axis) and saccade length (horizontal axis) in logarithmic scale, principal components and linear approximation by robust fit for preceding fixation in the case of simulations for synthetic images. The regression line of robust fit is close to horizontal and coincides with the first principal component, which explains $58 \%$ of variability in simulations for the case of synthetic images and succeding fixation.}
\label{fig:corrchartsym}
\end{figure*}
\begin{figure*}
\centering
\includegraphics[scale=0.39]{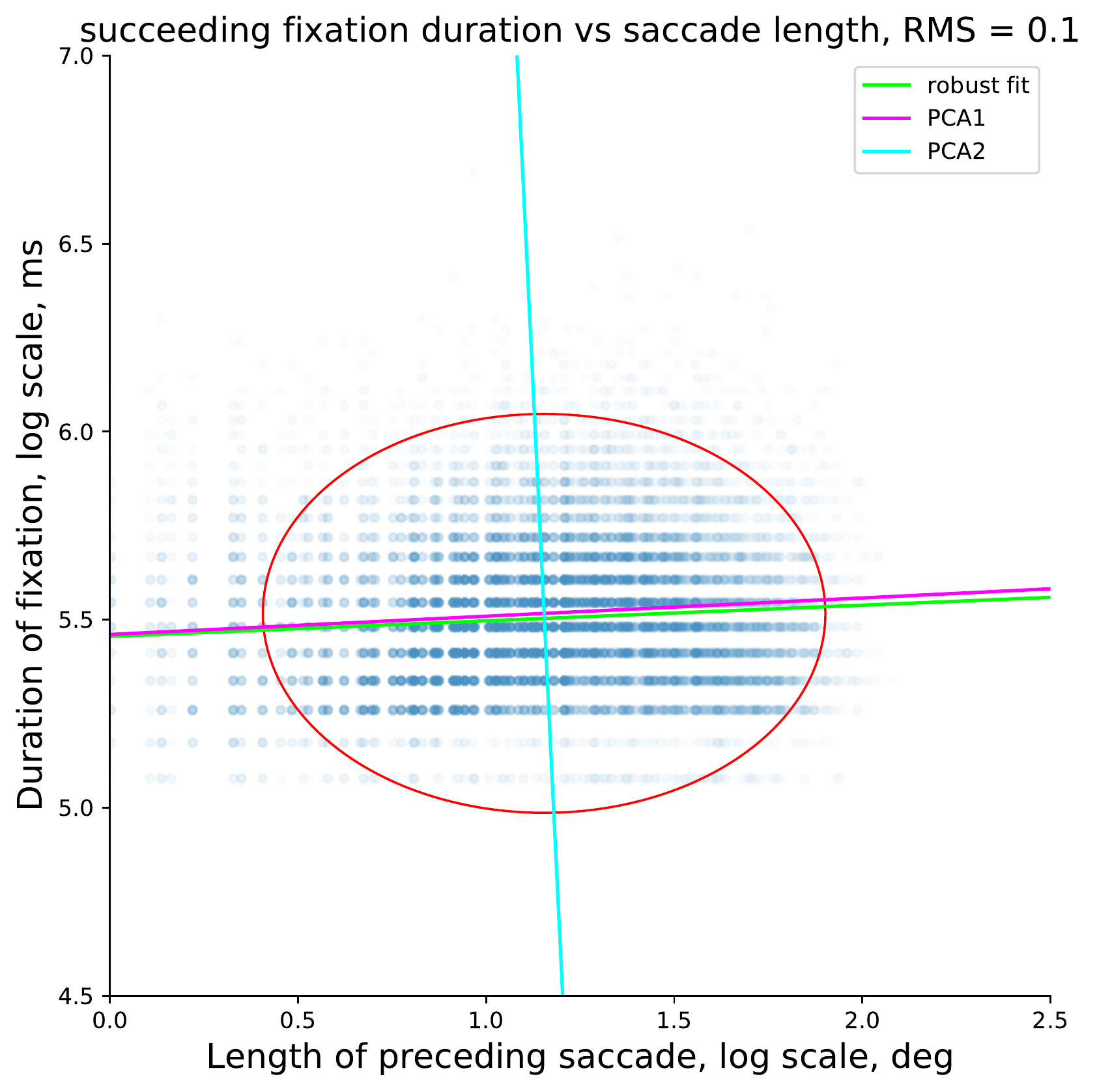}
\includegraphics[scale=0.39]{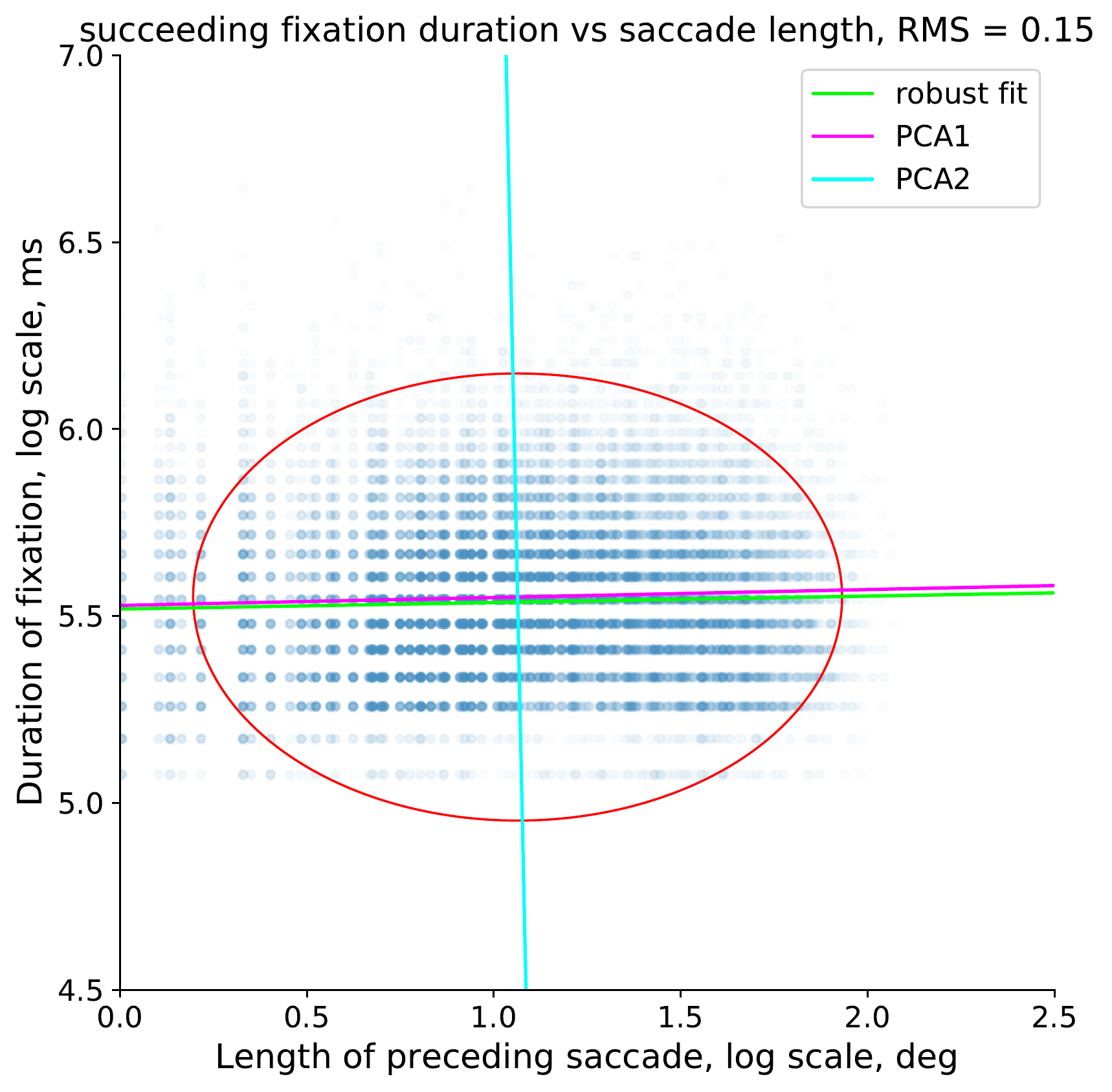}
\includegraphics[scale=0.39]{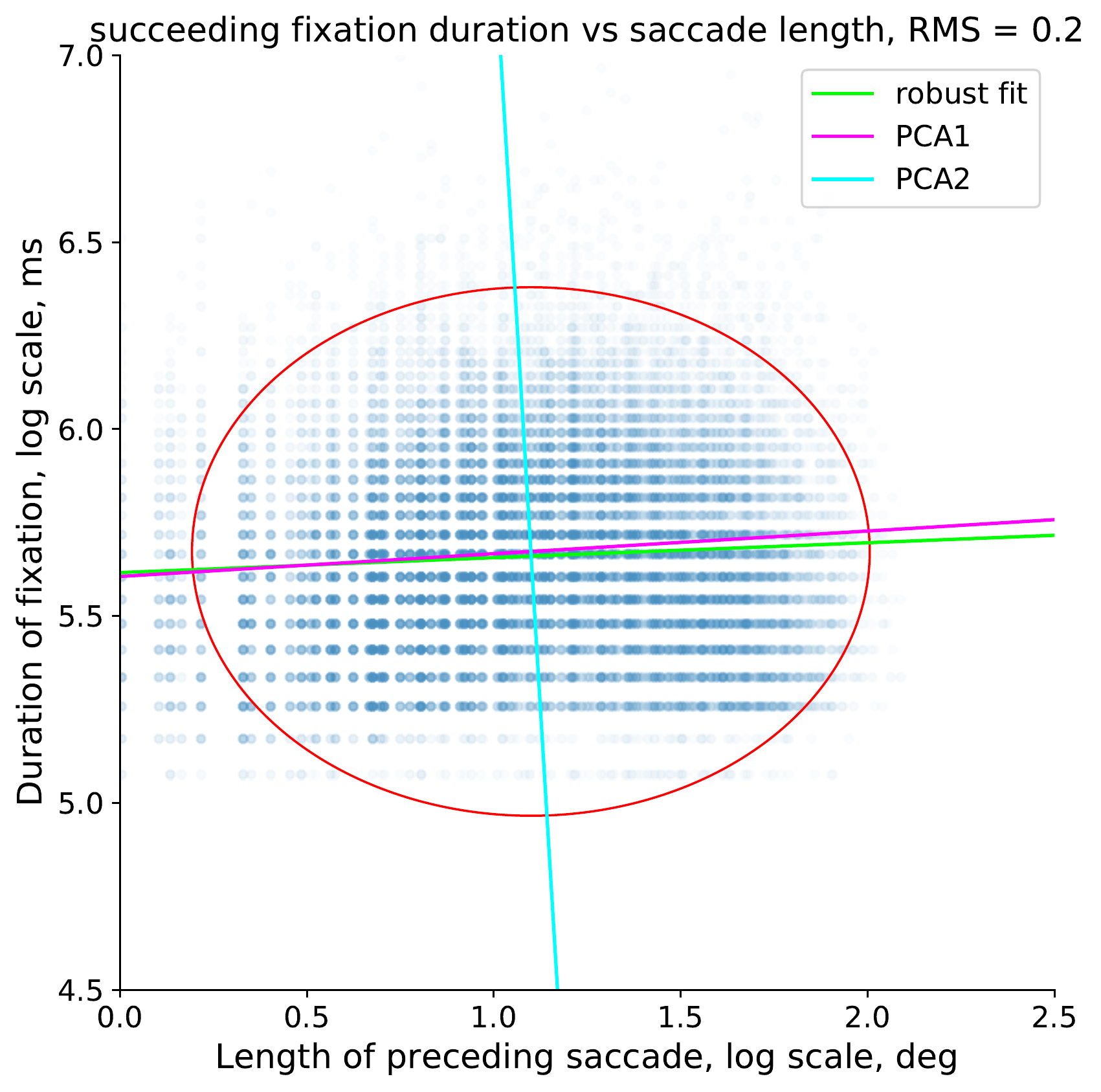}
\includegraphics[scale=0.39]{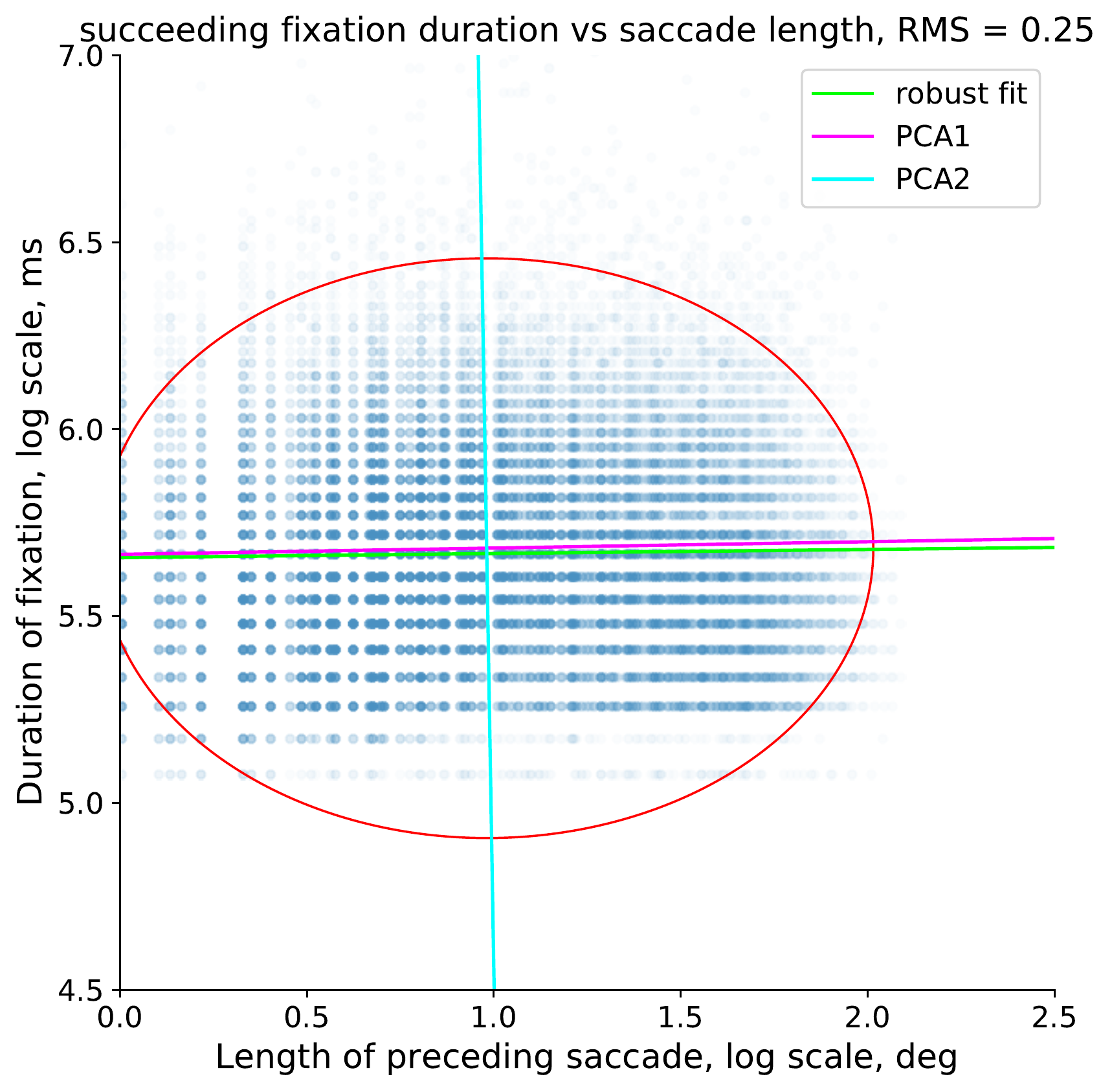}
\caption[Scatter plot of log-transformed data in logarithmic scale for preceding saccade in the case of simulations for synthetic images. ]{Scatter plot of log-transformed data: succeding fixation duration (vertical axis) and saccade length (horizontal axis) in logarithmic scale, principal components and linear approximation by robust fit in the case of simulations for synthetic images. The regression line of robust fit is close to horizontal and coincides with the first principal component, which explains $55 \%$ of variability in simulations for the case of synthetic images and succeding fixation.}
\label{fig:corrchartsym1}
\end{figure*}

\begin{figure*}
\centering
\includegraphics[scale=0.98]{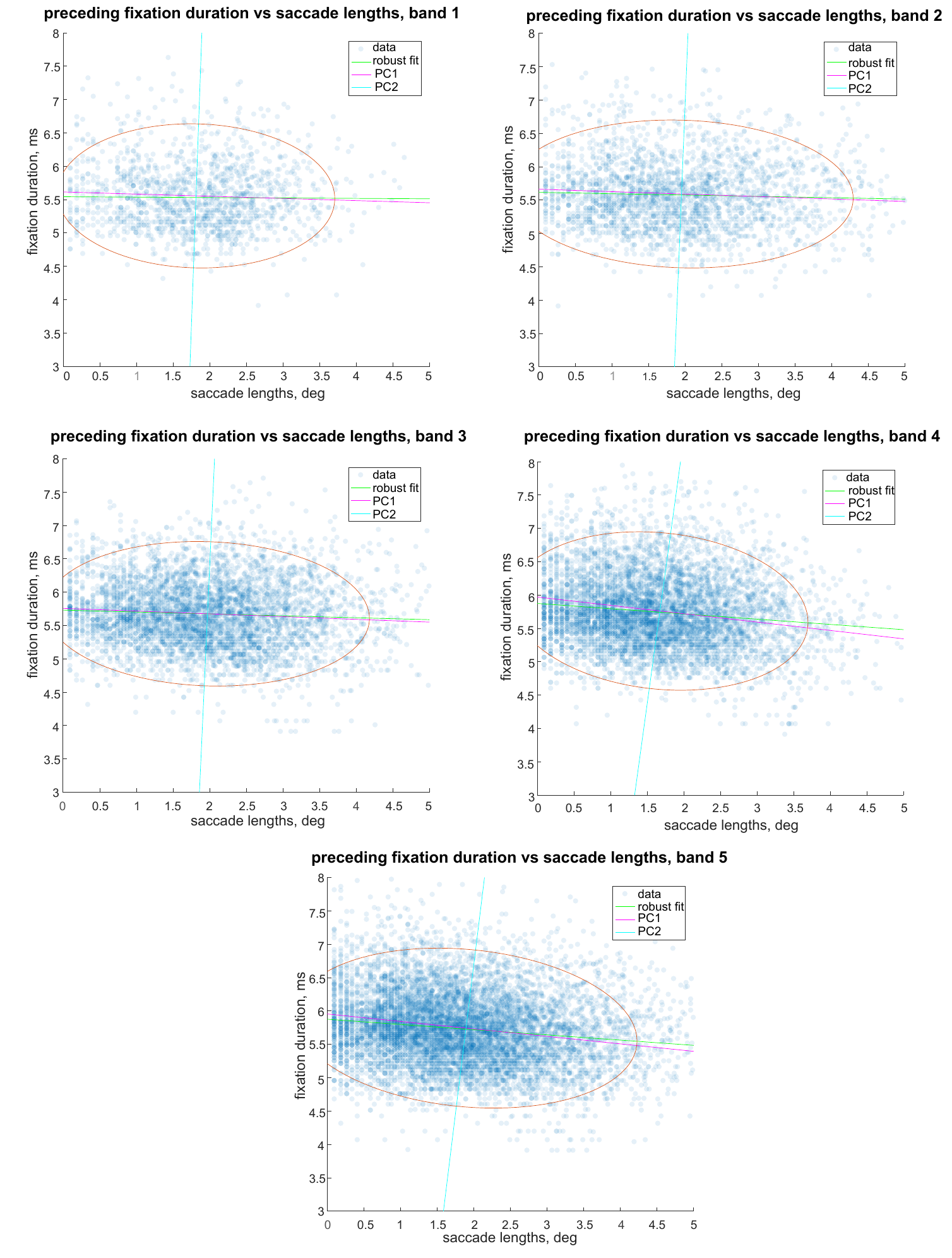}
\caption[Scatter plots of log-transformed data in logarithmic scale for preceding fixation in the case of experiments on natural images.]{Scatter plots of log-transformed data: preceding fixation duration (vertical axis) and saccade lengths (horizontal axis) in logarithmic scale, principal components and linear approximation by robust fit in the case of experiments on natural images. The regression line of robust fit is close to horizontal and coincides with the first principal component, which explains $65 \%$ of variability in experimental data for the case of natural images.}
\label{fig:corrchartnat}
\end{figure*}
\begin{figure*}
\centering
\includegraphics[scale=0.98]{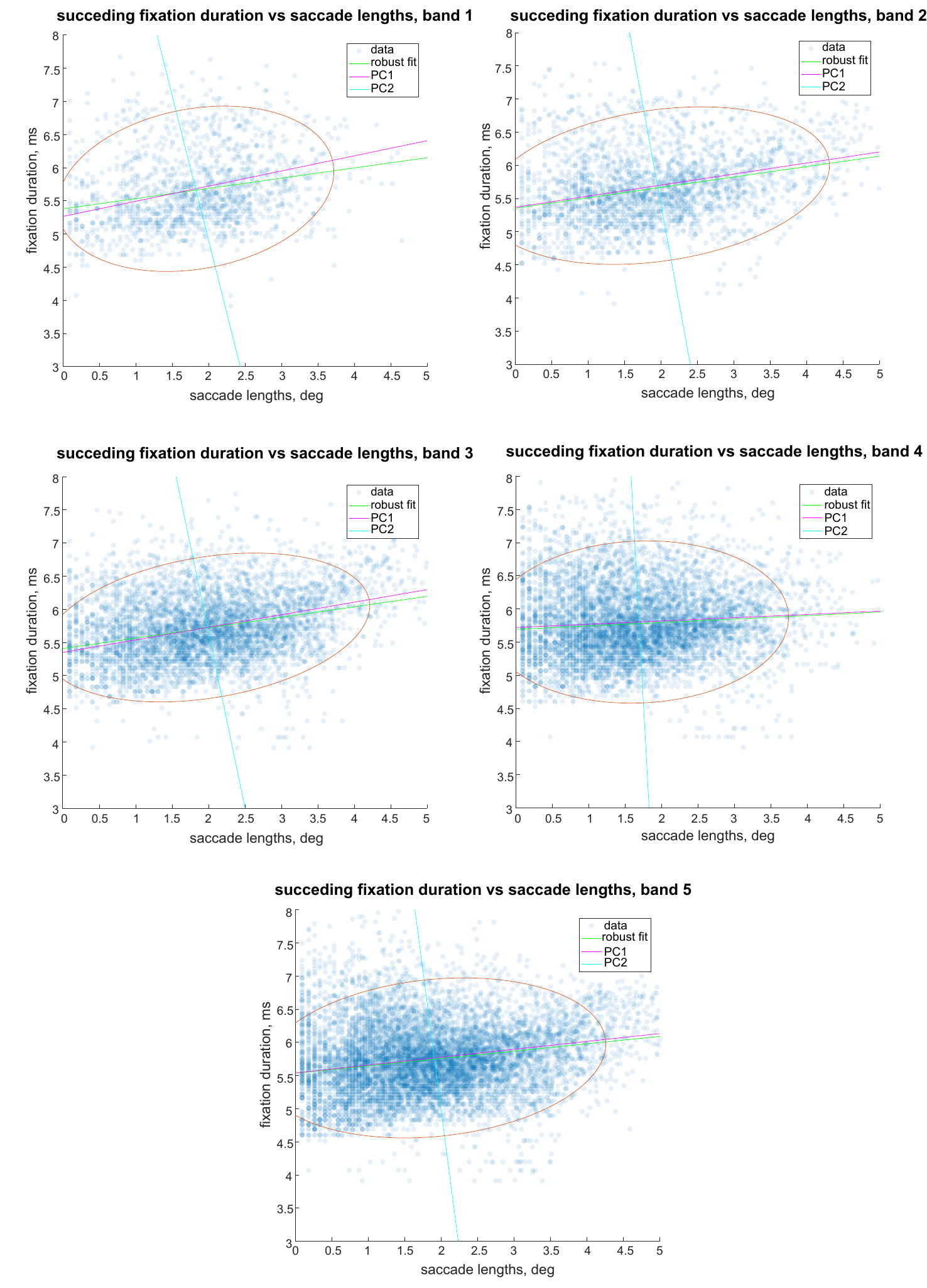}
\caption[Scatter plots of log-transformed data in logarithmic scale for preceding saccade in the case of experiments on natural images.]{Scatter plots of log-transformed data: succeding fixation duration (vertical axis) and saccade lengths (horizontal axis) in logarithmic scalse, principal components and linear approximation by robust fit for preceding saccade in the case of experiments on natural images. The regression line of robust fit is close to horizontal and coincides with the first principal component, which explains $67 \%$ of variability in experimental data for the case of natural images and succeding fixation. }
\label{fig:corrchartnat1}
\end{figure*}
\begin{figure*}
\centering
\includegraphics[scale=0.39]{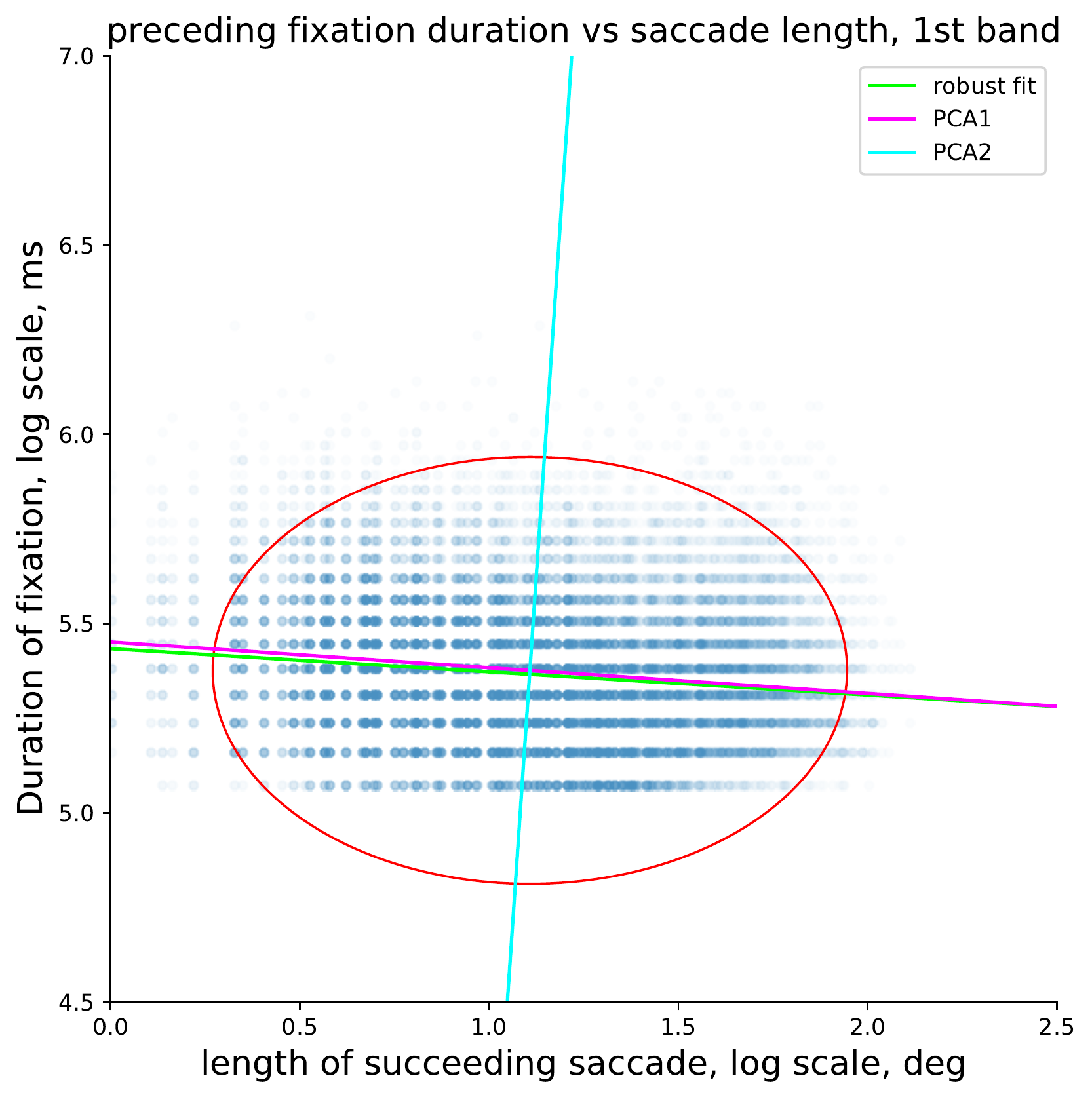}
\includegraphics[scale=0.39]{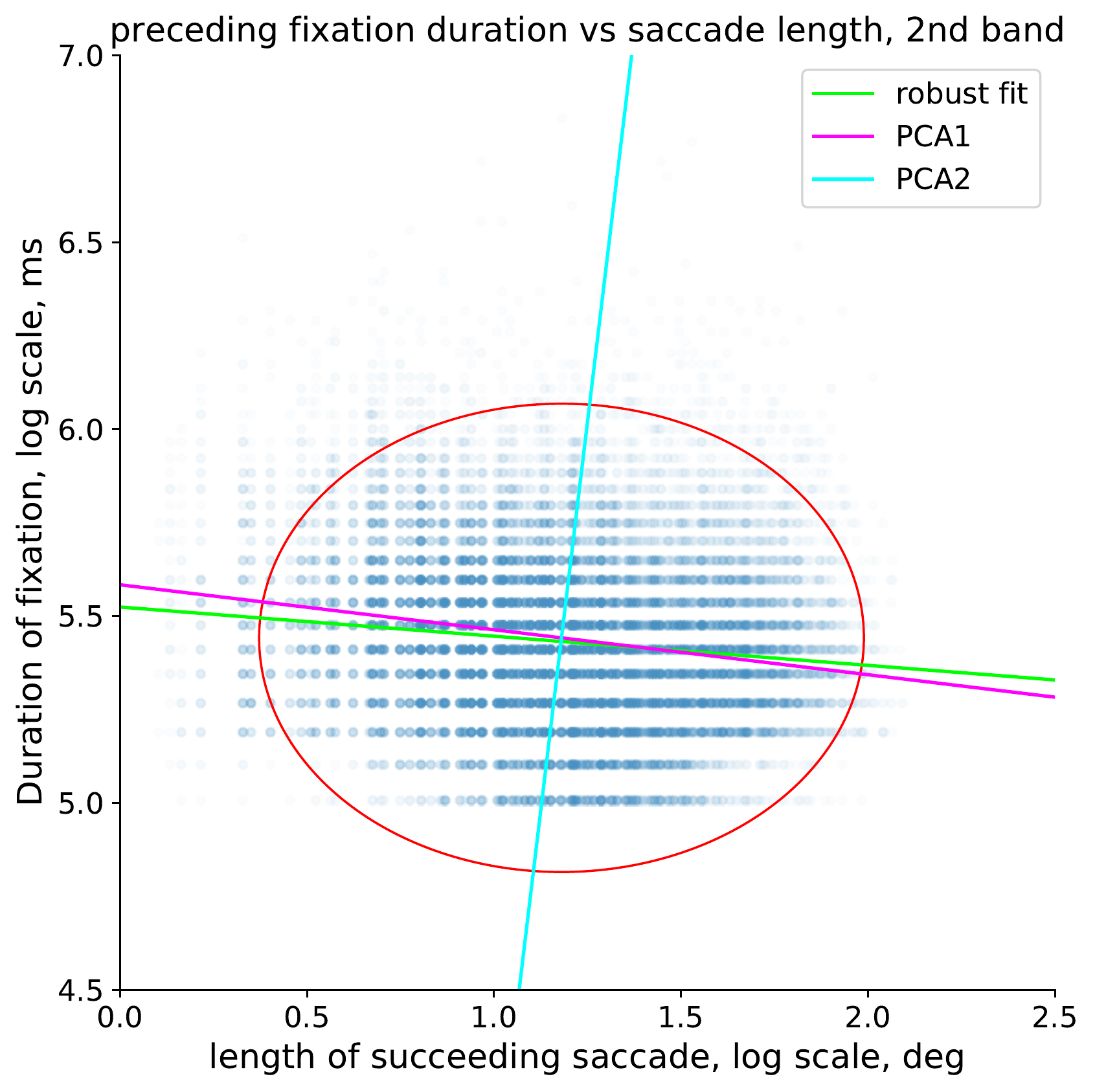}
\includegraphics[scale=0.39]{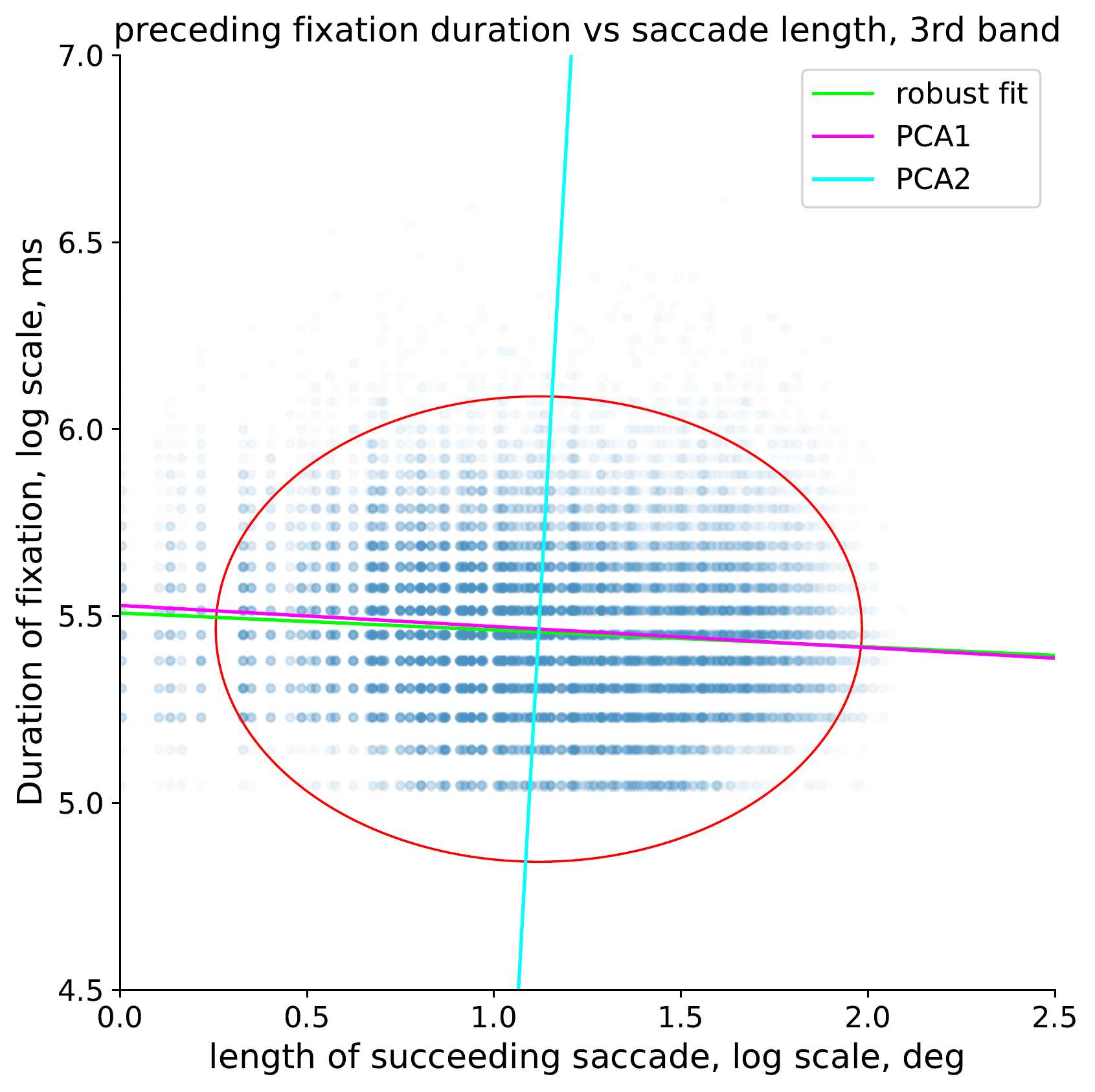}
\includegraphics[scale=0.39]{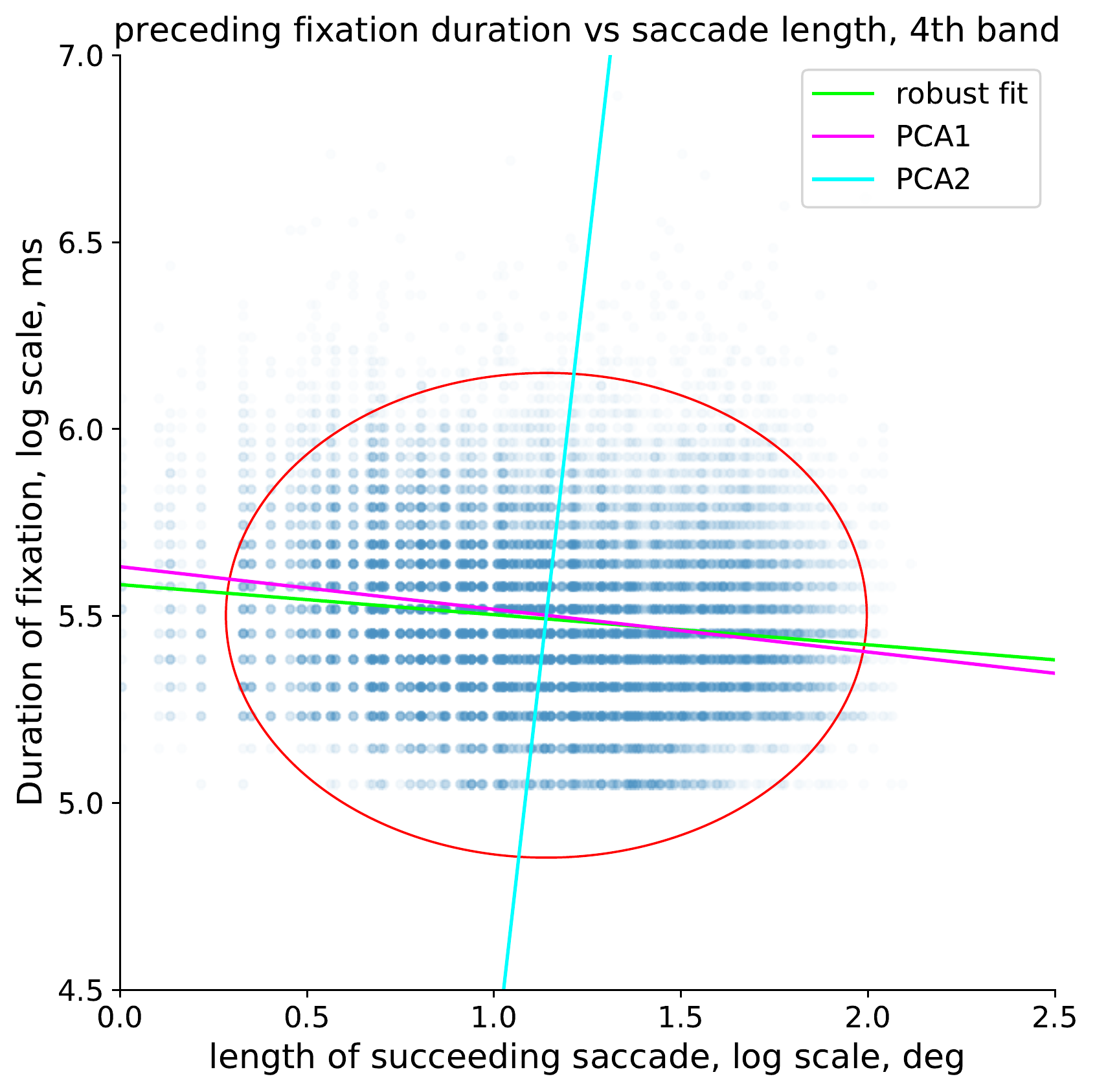}
\includegraphics[scale=0.39]{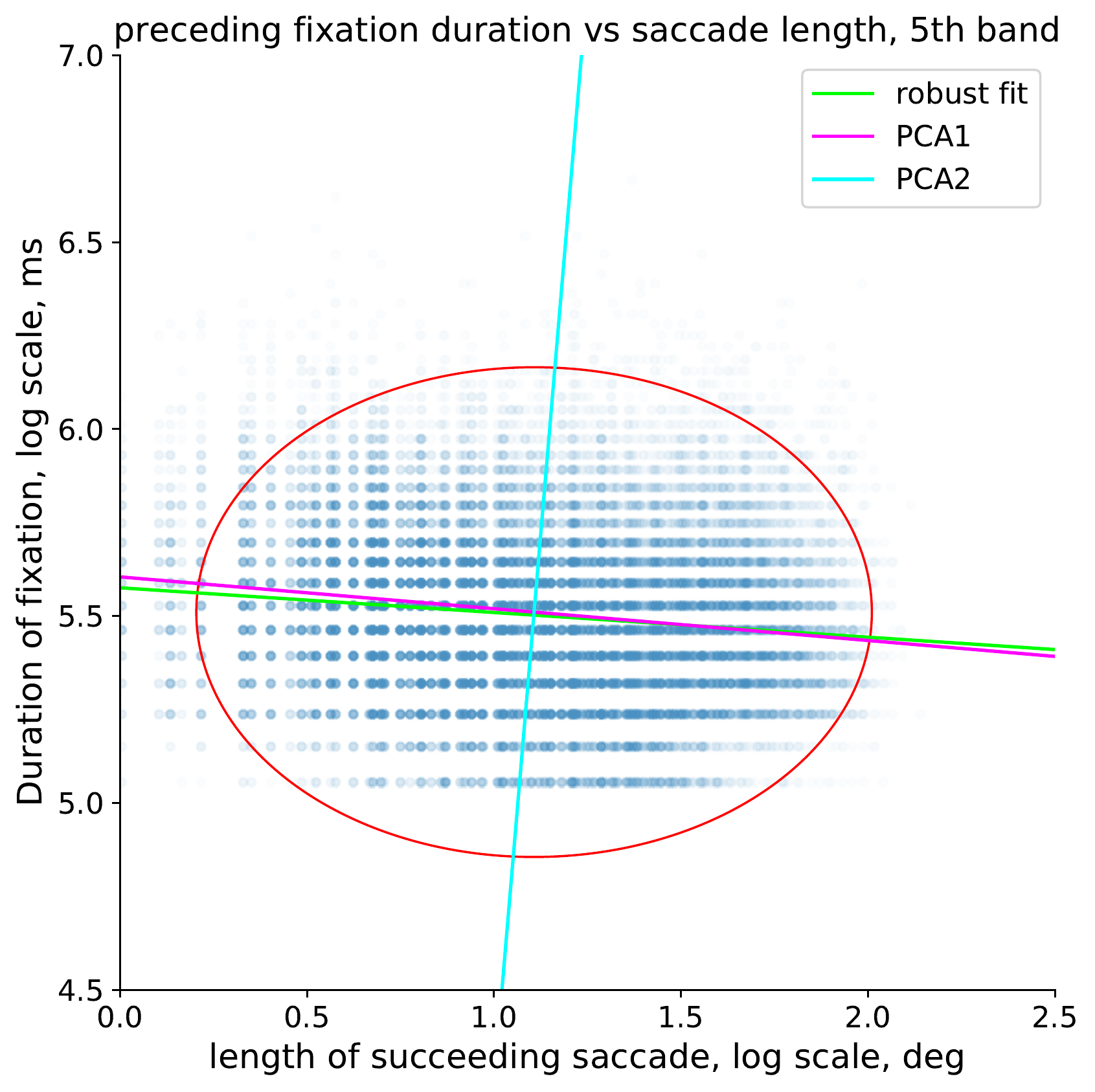}
\caption[Scatter plot of log-transformed data in logarithmic scale for preceding fixation in the case of simulation for natural images.]{Scatter plot of log-transformed data: preceding fixation duration (vertical axis) and saccade lengths (horizontal axis) in logarithmic scale, principal components and linear approximation by robust fit for preceding fixation in the case of simulation for natural images. The regression line of robust fit is close to horizontal and coincides with the first principal component, which explains $56 \%$ of variability in simulations for the case of natural images and preceding fixation.  }
\label{fig:corrchartnatsym}
\end{figure*}
\begin{figure*}
\centering
\includegraphics[scale=0.39]{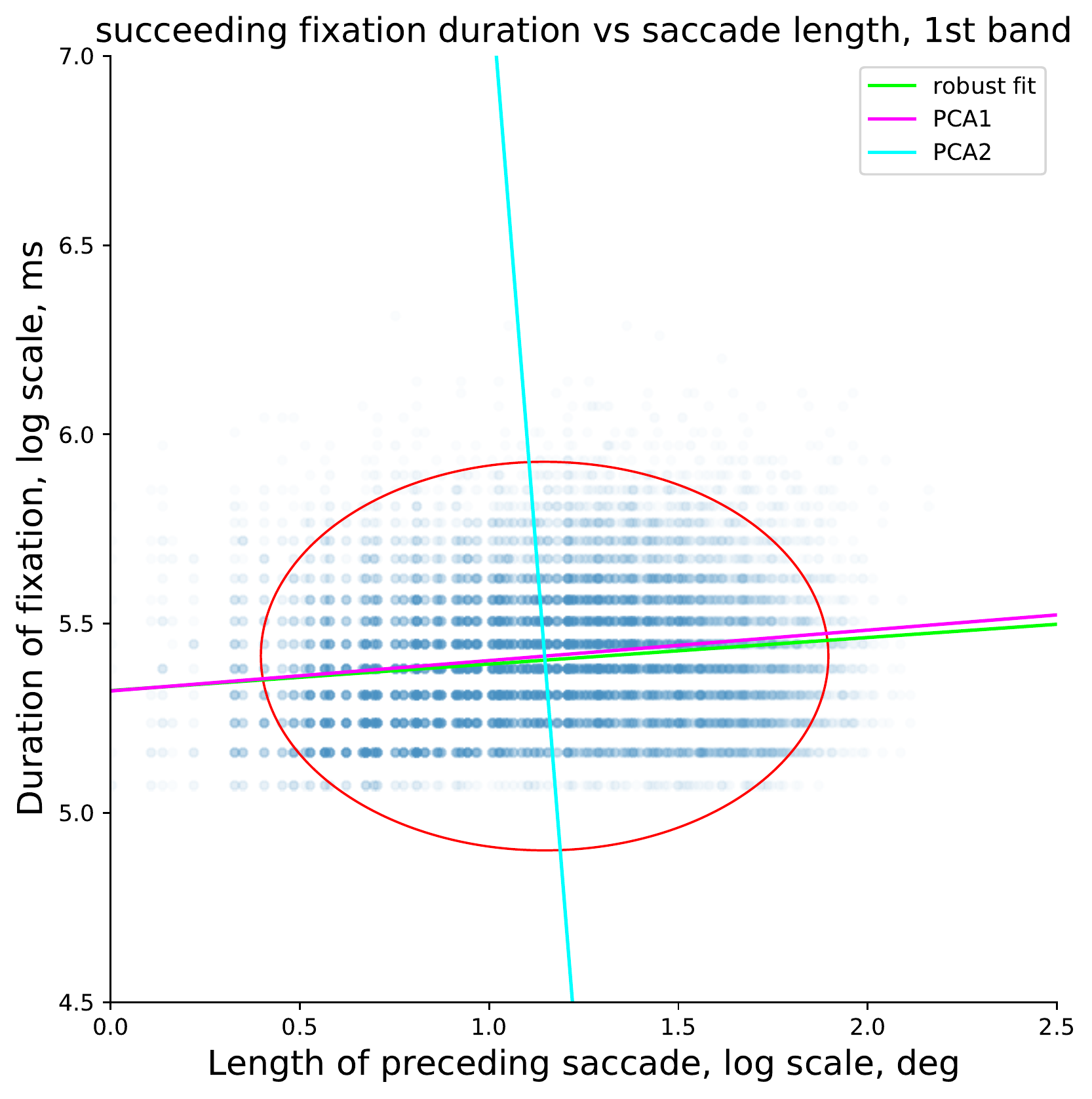}
\includegraphics[scale=0.39]{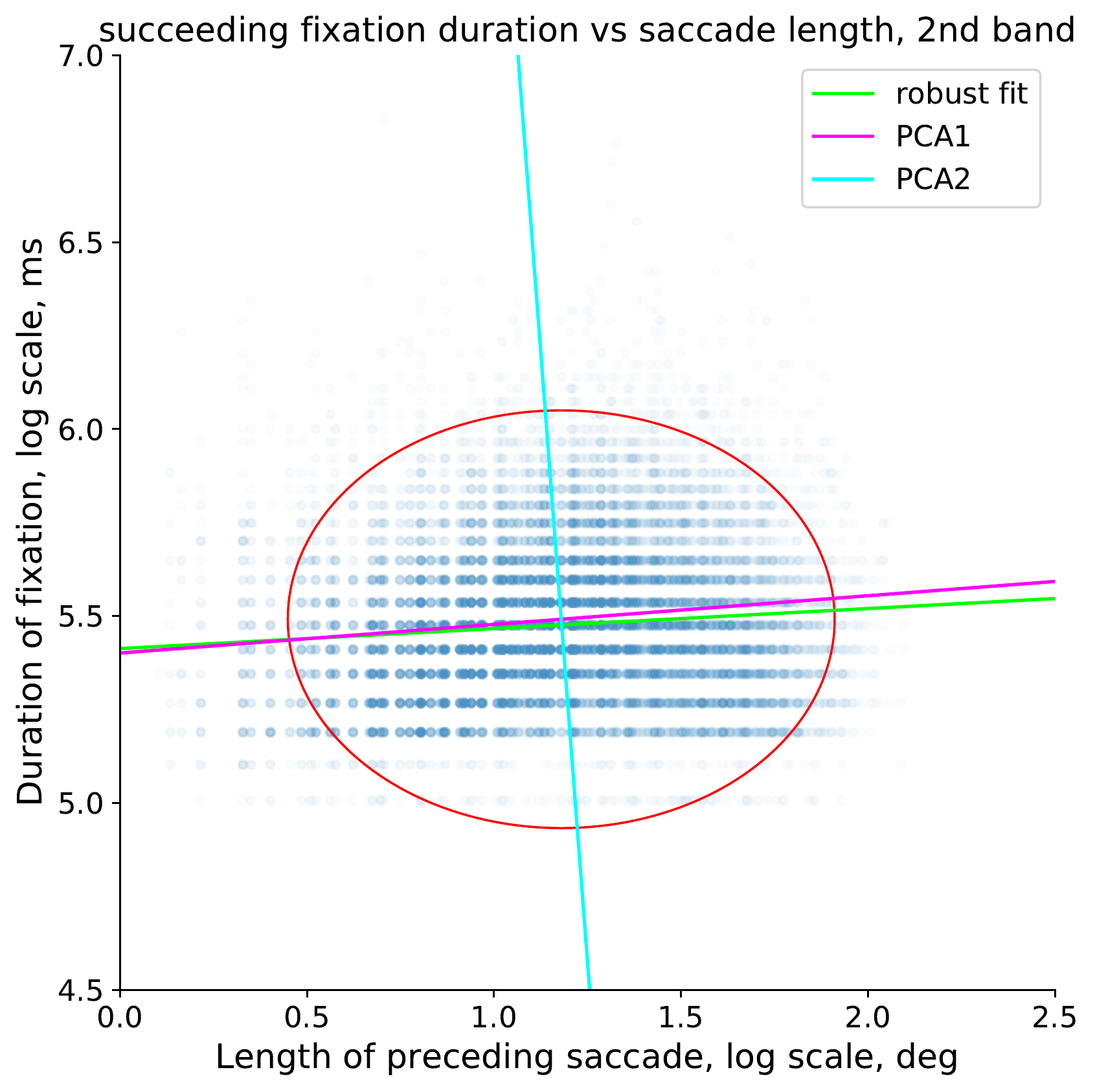}
\includegraphics[scale=0.39]{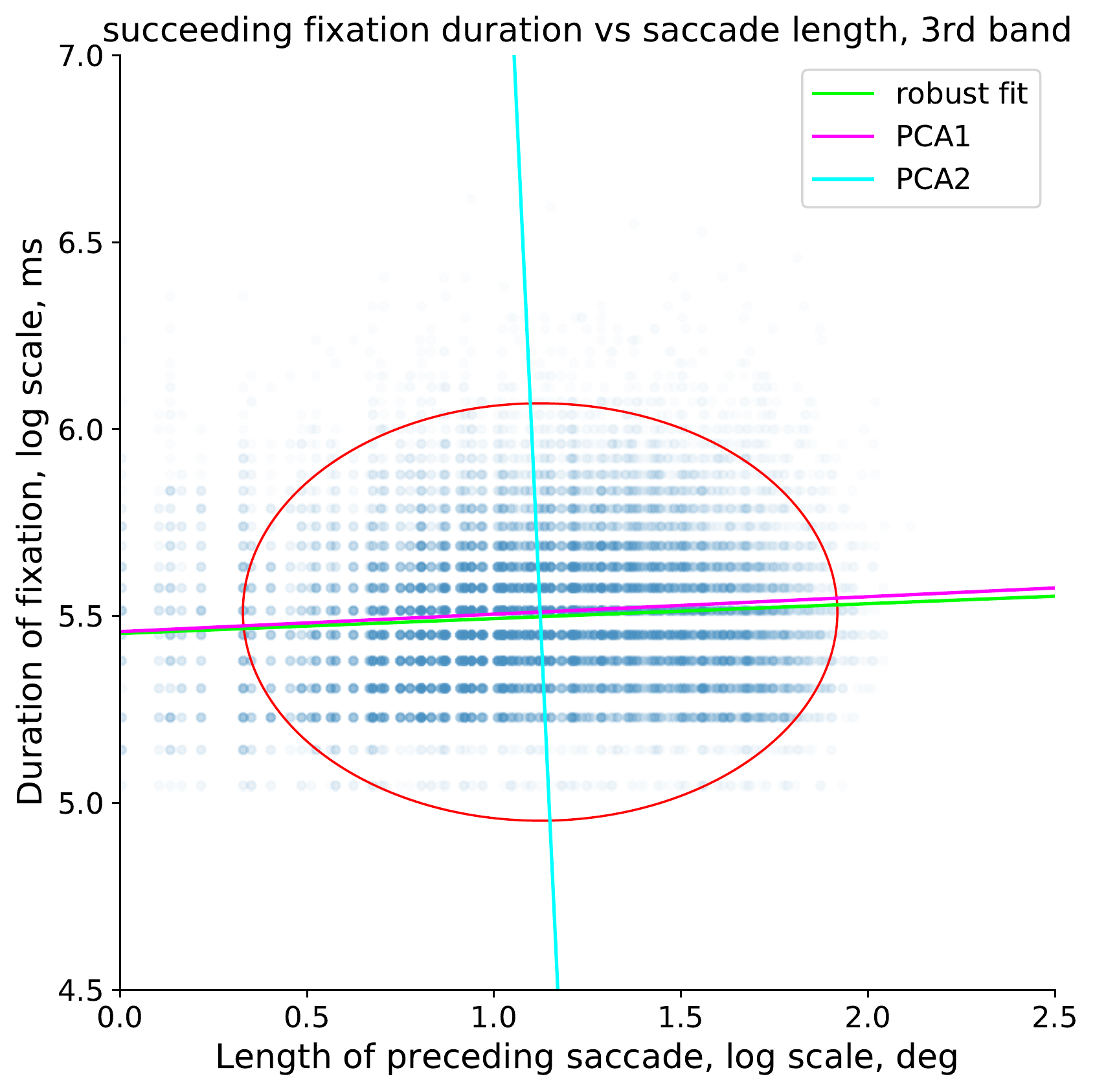}
\includegraphics[scale=0.39]{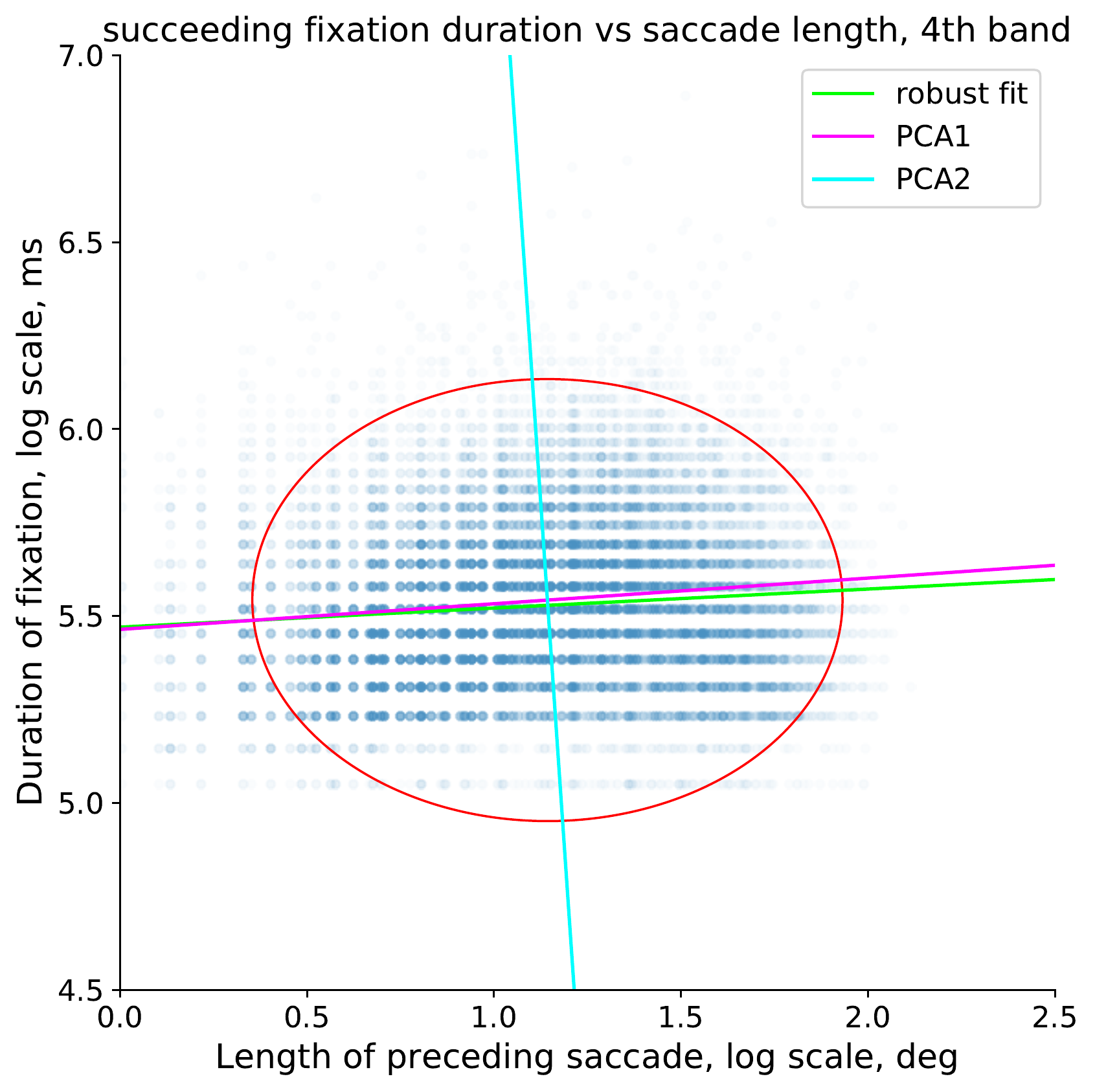}
\includegraphics[scale=0.39]{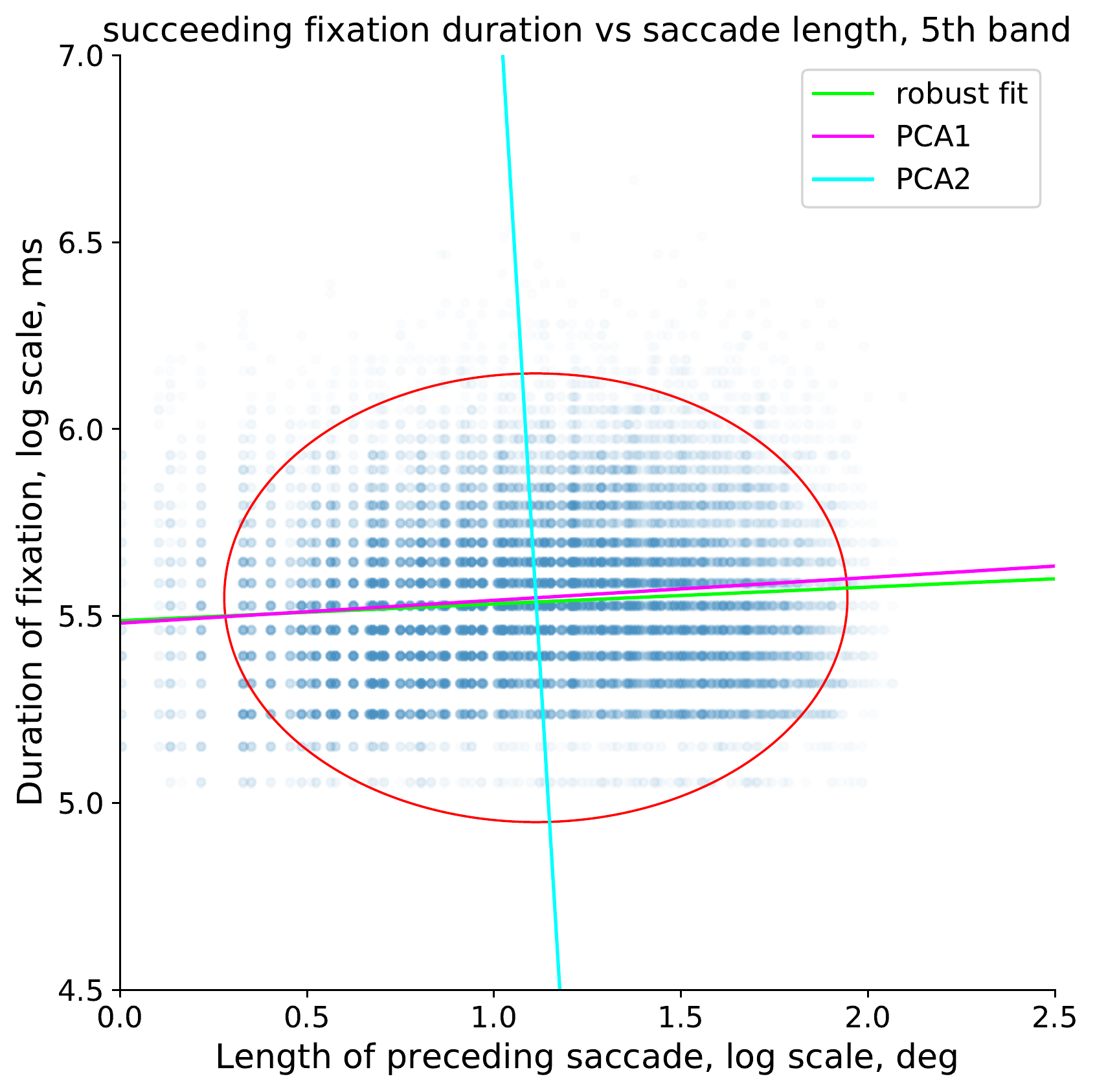}
\caption[Scatter plot of log-transformed data in logarithmic scale preceding saccade in the case of simulation for natural images.]{Scatter plot of log-transformed data: succeding fixation duration (vertical axis) and saccade lengths (horizontal axis) in logarithmic scale, principal components and linear approximation by robust fit for preceding saccade in the case of simulation for natural images. The regression line of robust fit is close to horizontal and coincides with the first principal component, which explains $57 \%$ of variability in simulations for the case of natural images and succeding fixation. }
\label{fig:corrchartnatsym1}
\end{figure*}

\end{document}